\documentclass[11pt]{article}
\usepackage{epsfig} 
\usepackage{amssymb} 
\usepackage{graphics}
\newcommand{\sect}[1]{ \section{#1} \setcounter{equation}{0} } 

\newcommand{\Dslash}{D \! \! \! \! /}

\newcommand{\MSbar}{\overline{\mbox{MS}}} 
 
\newcommand{\RI}{\mbox{RI${}^\prime$}}

\newcommand{\mMOM}{\mbox{mMOM}}

\newcommand{\MOMc}{\mbox{MOMc}}
\newcommand{\MOMcs}{\mbox{\footnotesize{MOMc}}}
\newcommand{\MOMg}{\mbox{MOMg}}
\newcommand{\MOMgs}{\mbox{\footnotesize{MOMg}}}
\newcommand{\MOMq}{\mbox{MOMq}}
\newcommand{\MOMqs}{\mbox{\footnotesize{MOMq}}}
\newcommand{\bare}{\mbox{\footnotesize{o}}}
\newcommand{\Nf}{N_{\!f}}

\newcommand{\NA}{N_{\!A}}
\newcommand{\Nda}{N^d_{\!A}}
\newcommand{\Noda}{N^o_{\!A}}

\newcommand{\zp}{\mbox{\footnotesize{zp}}}

\begin{document}

\begin{centering}

\vfill

%\vspace{-18.2cm}
%\vspace{-18.0cm}
\hspace{14.3cm}
{\bf LTH 1341}

\vspace{1.5cm}
{\LARGE {\bf Scheme and gauge dependence of QCD fixed points at five loops}}

\vspace{1.4cm}
\noindent
{\large J.A. Gracey${}^a$, R.H. Mason${}^a$, Thomas A. Ryttov${}^b$ \& R.M. 
Simms${}^a$}

\vspace{0.7cm}
\noindent
${}^a$ Theoretical Physics Division, Department of Mathematical 
Sciences, University of Liverpool, P.O. Box 147, Liverpool, L69 3BX, 
United Kingdom

\vspace{0.2cm}
${}^b$ CP${}^3$-Origins, University of Southern Denmark, Campusvej 55, 5230
Odense M, Denmark

%\vspace{0.7cm}
%\noindent
%{June, 2023.} 

\end{centering}

\vspace{5cm} 
\noindent 
{\bf Abstract.} We analyse the fixed points of QCD at high loop order in a
variety of renormalization schemes and gauges across the conformal window. We
observe that in the minimal momentum subtraction scheme solutions for the 
Banks-Zaks fixed point persist for values of $\Nf$ below that of the $\MSbar$
scheme in the canonical linear covariant gauge. By treating the parameter of
the linear covariant gauge as a second coupling constant we confirm the 
existence of a second Banks-Zaks twin critical point, which is infrared stable,
to five loops. Moreover a similar and parallel infrared stable fixed point is 
present in the Curci-Ferrari and maximal abelian gauges which persists in 
different schemes including kinematic ones. We verify that with the increased 
available loop order critical exponent estimates show an improvement in 
convergence and agreement in the various schemes.

\newpage 

\sect{Introduction.}

Nonabelian gauge theories with $\Nf$ flavours of quarks are known to possess an
infrared stable fixed point for a range of $\Nf$ from the work of Banks and
Zaks, \cite{1}, as well as that of Caswell, \cite{2}. Known as the conformal 
window the actual range depends on the values of the colour group factors of 
the gluon and quark representations. The fixed point of \cite{1} and \cite{2}, 
which we will refer to as the Banks-Zaks fixed point throughout in keeping with
common usage, was derived from the two loop $\beta$-function of Quantum 
Chromodynamics (QCD), \cite{2,3,4,5}, in the modified minimal subtraction
($\MSbar$) scheme. For the case where the colour group is $SU(3)$ and the 
quarks are in the fundamental representation the conformal window is 
$9$~$\leq$~$\Nf$~$\leq$~$16$. The bounding values are deduced by searching for 
non-zero values of the coupling constant where the $\beta$-function vanishes. 
To ensure the asymptotic freedom property of the underlying theory, \cite{3,4},
is not destroyed requires the one loop term to be negative which determines the
upper bound. That for the lower bound comes from the coefficient of the two 
loop term which has to be positive. Otherwise there is no non-zero critical 
coupling constant. In this senario the Banks-Zaks fixed point is infrared 
stable with the Gaussian critical point at the origin being ultraviolet stable 
as a result of asymptotic freedom. Such conformal windows are not restricted to
QCD itself as they can also occur in supersymmetric gauge theories for 
instance. Indeed the windows have been of interest in general due to their 
potential connection with constructing viable beyond the Standard Model 
candidates. 

One key marker of conformal window studies both perturbatively and
nonperturbatively, in the context of lattice gauge theory, is the critical
exponent of the quark mass. For instance, with the advance in our knowledge of 
all the perturbative renormalization group functions of non-abelian gauge
theories to five loops in the $\MSbar$ scheme
\cite{2,3,4,5,6,7,8,9,10,11,12,13,14,15,16,17,18,19,20,21,22,23} the location
of the Banks-Zaks fixed point not only has been hugely refined but has also for
instance produced accurate quark mass exponent estimates. These are competitive
with lattice measurements at values of $\Nf$ which are on the edge of 
perturbative applicability. Most of these Banks-Zaks fixed point analyses have 
been carried out in the $\MSbar$ scheme. However other schemes have been 
considered for the conformal window such as the minimal momentum subtraction
($\mMOM$) and modified regularization invariant ($\RI$) schemes in 
\cite{24,25,26}. While the renormalization group functions in these schemes are
similar in structure to that of the $\MSbar$ scheme, and constructed to high 
loop order too, \cite{27,28,29,30,31,32,33,34}, kinematic schemes have also 
been studied. These are the momentum subtraction (MOM) schemes of Celmaster and 
Gonsalves, \cite{35,36}. Although there are three schemes, based on the triple 
gluon (MOMg), ghost-gluon (MOMc) and quark-gluon (MOMq) vertices, their 
renormalization group functions are only available to three loops for an
arbitrary linear covariant gauge \cite{35,36,37,38} but to four loops in the 
Landau gauge, \cite{39}. On the whole the values of the quark mass critical
exponent are on a par with the $\MSbar$ estimates in the conformal window. This
is consistent with the underlying properties of critical exponents in that 
since they are physical observables they are renormalization group invariants. 
Moreover in a gauge theory they should have the same value in all gauges. This 
latter property is not trivial for a number of reasons.

It is widely known that in the $\MSbar$ scheme the $\beta$-function is
independent of the gauge parameter of the linear covariant gauge, \cite{40}.
This is not the case in all the other schemes listed earlier except for the
$\RI$ scheme due to the particular prescription used to define the coupling
constant renormalization. Therefore if one wished to study Banks-Zaks fixed 
points as well as the conformal window in these other schemes account has to be
taken of the gauge parameter dependence and its underlying running. This was 
highlighted in \cite{41} for example. To do so one has to solve for the 
critical points of not only the $\beta$-function but also the anomalous 
dimension of the gauge parameter $\alpha$, denoted by $\gamma_\alpha(a,\alpha)$
where $a$ is the coupling constant. Strictly $\alpha$ should be regarded as a 
second coupling constant and its critical values and those for $a$ deduced from
the zeros of $\beta(a,\alpha)$ and $\alpha \gamma_\alpha(a,\alpha)$. The latter 
is the $\beta$-function associated with $\alpha$. In the $\MSbar$ scheme the 
zeros of $\beta(a)$ define the critical coupling. However, it is already known 
in \cite{41} for instance that there is more than one non-zero solution for a 
critical $\alpha$ value in a linear covariant gauge. Indeed \cite{41} examined 
this at length with the then known renormalization group functions in the 
$\MSbar$ and MOM schemes of \cite{35,36}. One interesting observation was that
the Banks-Zaks fixed point, with $\alpha$~$=$~$0$, is a saddle point and there 
is an infrared stable fixed point with $\alpha$~$\neq$~$0$ in the $(a,\alpha)$ 
plane. In fact while the number of such non-zero $\alpha$ solutions increases 
with loop order one solution appears to be robust to the ephemeral ones which 
can disappear at the next loop order. This was a remarkable observation and the 
effect of running to this infrared fixed point was explored in \cite{41}. 

Given there has been a significant advance in the loop expansion of the 
renormalization group functions of QCD in the various schemes mentioned, it 
seems appropriate to revisit the various previous Banks-Zaks fixed point 
analyses and carry out a comprehensive and exhaustive study of the critical 
properties of QCD in the $(a,\alpha)$ plane. This is the purpose of this 
article. In particular we will find the five loop fixed points in the $\RI$ and
$\mMOM$ schemes as well as the three loop ones for the MOM schemes. The five 
loop $\MSbar$ Banks-Zaks fixed point was analysed in \cite{42} but we will 
extend that to the case of $\alpha$~$\neq$~$0$. Several main questions of 
interest are the convergence of exponent estimates in the various schemes as 
well as the robustness of the infrared stable fixed point to higher order loop 
corrections. Since it may be an artefact of the particular gauge fixing in the 
linear covariant gauge we will include several other gauges in our 
investigations. These other gauges are covariant with an associated gauge 
parameter but are non-linear. They are the Curci-Ferrari (CF) gauge, \cite{43},
and the maximal abelian gauge (MAG), \cite{44,45,46}. By including these two
gauges we will be able to comment on whether certain exponents exhibit the
gauge independence property. The evidence for this would be to find the
critical exponents that are common to these gauges have the same value to a
reasonable level of accuracy. It is important to recognize that this is not
the same as saying that exponents are independent of the gauge parameter since
that is a separate coupling constant that takes a value at criticality. Equally
another question is if there is also an infrared stable fixed point in these
other gauges whether or not there is a common value for the gauge parameter.
While this is unlikely various analyses in
\cite{24,41,47,48,49,50,51,52,53,54,55,56} in the linear covariant gauge case 
have also identified $\alpha$~$=$~$-$~$3$ as perhaps being a special case in a
variety of different contexts. It would be interesting to examine whether a
similar negative integer value for the Curci-Ferrari gauge and MAG emerges in
parallel. We finally remark that a complementary approach to finding critical 
exponents based on a scheme independent expansion has also been investigated in
detail in \cite{57,58,59,60,61}.

The article is organized as follows. The field theory background to fixed
points in gauge theories is reviewed in Section $2$ where we discuss the core
methods of our study as well as the properties of the different gauge fixed
QCD Lagrangians we are interested in. Section $3$ details the main results of
the investigation which proceeds in two parts. This centres on determining the
actual fixed points in the $(a,\alpha)$ plane together with their stability
properties prior to numerically analysing the core critical exponents of QCD in
Section $4$. As there were no Banks-Zaks fixed points at five loops in the
$\MSbar$ $\beta$-function, \cite{42}, we reproduce the Pad\'{e} analysis of
\cite{42} in Section $5$ before extending it to the $(a,\alpha)$ plane for not
only the $\MSbar$ scheme but also the $\mMOM$ and $\RI$ schemes at five loops.
Concluding remarks are provided in Section $6$.

\sect{Background.}

As we will be considering QCD fixed in several different gauges it is
instructive to recall the different properties of each. The appropriate way of
viewing this is through the various Lagrangians. First, the most commonly used
covariant gauge is the Lorenz one where the gauge fixing functional is linear 
in the fields and produces the Lagrangian 
\begin{equation}
L ~=~ -~ \frac{1}{4} G^a_{\mu\nu} G^{a \, \mu\nu} ~+~ 
i \bar{\psi} \Dslash \psi ~-~ 
\frac{1}{2\alpha} \left( \partial^\mu A^a_\mu \right)^2 ~+~ 
\bar{c}^a \partial^\mu \partial_\mu c^a ~-~ 
\frac{g}{2} f^{abc} A^a_\mu \partial^\mu \bar{c}^b c^c
\label{laglin}
\end{equation}
where $g$ will be the gauge coupling throughout, $f^{abc}$ are the structure
constants and $\Nf$ will be the number of quarks. The gauge parameter $\alpha$ 
will have a different origin in the three gauges we consider but it will be 
clear from the context in our later discussions which gauge we will be 
referring to. In (\ref{laglin}) the colour index on the gluon and ghost fields 
lie in the range $1$~$\leq$~$a$~$\leq$~$\NA$ where $\NA$ is the dimension of 
adjoint representation. By contrast gauge fixing QCD in the MAG gauge leads to 
a much more involved set of interactions, \cite{44,45,46,62}. This stems partly
from the non-linear nature of the gauge fixing functional but also because the 
gauge field itself is split into two sectors. One sector contains those gluons 
associated with the group generators that commute among themselves and form an 
abelian subgroup. In general there will be $\Nda$ such fields with 
$\Nda$~$=$~$2$ for $SU(3)$. The remaining fields are in what is termed the 
off-diagonal sector and there are $\Noda$ such fields with $\Noda$~$=$~$6$ for 
$SU(3)$. Overall we have $\Nda$~$+$~$\Noda$~$=$~$\NA$. On top of this the 
fields of each sector are gauge fixed differently. Those in the abelian 
subgroup are fixed in the Landau gauge while the off-diagonal ones have a 
non-linear gauge functional. As a consequence the Faddeev-Popov ghosts of each
sector emerge with different interactions. In light of this explanation the MAG
Lagrangian is, \cite{63},
\begin{eqnarray}
L^{\mbox{\footnotesize{MAG}}} &=& 
-~ \frac{1}{4} G^a_{\mu\nu} G^{a \, \mu\nu} ~-~
\frac{1}{4} G^i_{\mu\nu} G^{i \, \mu\nu} ~+~ i \bar{\psi} \Dslash \psi 
- \frac{1}{2\alpha} \left( \partial^\mu A^a_\mu \right)^2
+ \bar{c}^a \partial^\mu \partial_\mu c^a
+ \bar{c}^i \partial^\mu \partial_\mu c^i \nonumber \\
&& +~ g \left[ f^{abk} A^a_\mu \bar{c}^k \partial^\mu c^b
- f^{abc} A^a_\mu \bar{c}^b \partial^\mu c^c
- \frac{1}{\alpha} f^{abk} \partial^\mu A^a_\mu A^b_\nu A^{k \, \nu}
- f^{abk} \partial^\mu A^a_\mu c^b \bar{c}^k
\right. \nonumber \\
&& \left. ~~~~~~
- \frac{1}{2} f^{abc} \partial^\mu A^a_\mu \bar{c}^b c^c
- 2 f^{abk} A^k_\mu \bar{c}^a \partial^\mu \bar{c}^b
- f^{abk} \partial^\mu A^k_\mu \bar{c}^b c^c \right] \nonumber \\
&& +~ g^2 \left[ f^{aci} f^{bdi} A^a_\mu A^{b \, \mu} \bar{c}^c c^d
- \frac{1}{2\alpha} f^{akc} f^{blc} A^a_\mu A^{b \, \mu} A^k_\nu A^{l \, \nu}
+ f^{adb} f^{cjb} A^a_\mu A^{j \, \mu} \bar{c}^c c^d \right.
\nonumber \\
&& \left. ~~~~~~~
- \frac{1}{2} f^{ajb} f^{cdb} A^a_\mu A^{j \, \mu} \bar{c}^c c^d
+ f^{ajb} f^{clb} A^a_\mu A^{j \, \mu} \bar{c}^c c^l
+ f^{alb} f^{cjb} A^a_\mu A^{j \, \mu} \bar{c}^c c^l 
\right. \nonumber \\
&& \left. ~~~~~~~
- f^{cjb} f^{dib} A^i_\mu A^{j \, \mu} \bar{c}^c c^d
- \frac{\alpha}{4} f^{abi} f^{cdi} \bar{c}^a \bar{c}^b c^c c^d
- \frac{\alpha}{8} f^{abe} f^{cde} \bar{c}^a \bar{c}^b c^c c^d
\right. \nonumber \\
&& \left. ~~~~~~~
+ \frac{\alpha}{8} f^{ace} f^{bde} \bar{c}^a \bar{c}^b c^c c^d 
- \frac{\alpha}{4} f^{abe} f^{cle} \bar{c}^a \bar{c}^b c^c c^l
+ \frac{\alpha}{4} f^{ace} f^{ble} \bar{c}^a \bar{c}^b c^c c^l
\right. \nonumber \\
&& \left. ~~~~~~~
- \frac{\alpha}{4} f^{ale} f^{bce} \bar{c}^a \bar{c}^b c^c c^l
+ \frac{\alpha}{2} f^{ake} f^{ble} \bar{c}^a \bar{c}^b c^k c^l \right] 
\label{lagmag}
\end{eqnarray}
where the indices $a$ to $e$ label the off-diagonal sector and $i$ to $l$ label
the fields of the abelian subgroup. For completeness we note that if $A^i_\mu$ 
was gauge fixed in the full linear covariant gauge, rather than the Landau 
gauge specifically, then the term
$\frac{1}{2\bar{\alpha}} \left( \partial^\mu A^i_\mu \right)^2$ would be
appended to (\ref{lagmag}) where $\bar{\alpha}$ is the sector's gauge 
parameter. Although the indices of the off-diagional sector have a different 
range from that of the linear gauge we retain that notation as the fields of 
the $\Nda$-dimensional subgroup in effect act as a background field. This is 
due to the fact that there is a Slavnov-Taylor identity in the MAG which 
implies that the wave function renormalization of the commuting gluon fields 
equates to the coupling constant renormalization. A similar property holds in 
the background field gauge, \cite{64,65,66,67}. The other reason why we retain 
this index labelling for the off-diagonal sector is when the formal limit where
$\Nda$~$\to$~$0$ is taken. In (\ref{lagmag}) this equates to deleting terms 
with abelian subgroup fields or products of structure constants where there is 
a summation over an index of the abelian subgroup. The resulting Lagrangian is
\begin{eqnarray}
L^{\mbox{\footnotesize{CF}}} &=& 
-~ \frac{1}{4} G^a_{\mu\nu} G^{a \, \mu\nu} ~+~ i \bar{\psi} \Dslash \psi ~-~ 
\frac{1}{2\alpha} \left( \partial^\mu A^a_\mu \right)^2 ~+~ 
\bar{c}^a \partial^\mu \partial_\mu c^a
\nonumber \\
&& -~ \frac{g}{2} \left[ 
f^{abc} A^a_\mu \bar{c}^b \partial^\mu c^c
- f^{abc} A^a_\mu \partial^\mu \bar{c}^b c^c
\right]
+~ \frac{1}{8} \alpha g^2 \left[ 
f^{acbd} \bar{c}^a \bar{c}^b c^c c^d 
- f^{abcd} \bar{c}^a \bar{c}^b c^c c^d
\right] 
\label{lagcf}
\end{eqnarray}
which corresponds to the non-linear Curci-Ferrari gauge originally discussed in
\cite{43}. In (\ref{lagcf}) the gluon and ghost colour indices have the same 
range as (\ref{laglin}). While the terms of each gauge fixed Lagrangian that
are derived from the square of the field strength are the same due to gauge
invariance, the essential difference in the remaining cubic interaction terms
concern the ghost-gluon ones. In the Curci-Ferrari gauge there are two such
terms distinguished by which ghost field the spacetime derivative acts on.
There is only one such term in (\ref{laglin}). Both the MAG and Curci-Ferrari
gauges have quartic ghost interactions directly as a result of their gauge
fixing functional depending on a ghost antighost bilinear term. Such quartic
ghost terms do not make the respective Lagrangians non-renormalizable. On the
contrary the Lagrangians have been shown to be renormalizable through the
construction of the Slavnov-Taylor identities \cite{43,45,46,62}. 

Another difference with regard to the renormalization in the various gauges 
rests in the renormalization of the respective gauge parameters. If one defines
the renormalization of $\alpha$ with the convention
\begin{equation}
\alpha_{\bare} ~=~ Z^{-1}_\alpha Z_A \, \alpha
\end{equation}
with the subscripts denoting a bare parameter then the associated anomalous
dimension is given by
\begin{equation}
\gamma_\alpha(a,\alpha) ~=~ \left[ \beta(a,\alpha) \frac{\partial}{\partial a}
\ln Z_\alpha ~-~ \gamma_A(a,\alpha) \right] \left[ 1 ~-~ \alpha
\frac{\partial}{\partial \alpha} \ln Z_\alpha \right]^{-1} ~.
\label{gammaaldef}
\end{equation}
This is the general relation between $Z_\alpha$ and $\gamma_\alpha(a,\alpha)$ 
but when $Z_\alpha$~$=$~$1$, as is the case in the linear covariant gauge, it
reduces to the more familiar relation between $\gamma_A(a,\alpha)$ and 
$\gamma_\alpha(a,\alpha)$. However in both the non-linear gauges considered
here $Z_\alpha$~$\neq$~$1$ and so 
\begin{equation}
\gamma_A(a,\alpha) ~+~ \gamma_\alpha(a,\alpha) ~\neq~ 0 ~.
\end{equation}
Although this is an unusual property of the renormalization group functions in
a gauge theory it is important in the context of the present analysis. The 
explicit form of $\gamma_\alpha(a,\alpha)$ will be used to determine the
critical properties of QCD since $\alpha$ is interpreted as a second coupling
constant. This is partly the reason why we included $\alpha$ as a second
argument of the $\beta$-function in (\ref{gammaaldef}). The other reason is
that in general the $\beta$-function is gauge dependent. It is only independent
of a covariant gauge parameter in the $\MSbar$ scheme, \cite{40}, as well as 
the $\RI$ scheme among the suite of schemes we consider. In kinematic schemes 
such as the MOM ones of \cite{35,36} the explicit gauge parameter dependence 
has been determined for instance to several loop orders. 

Indeed it is this gauge parameter dependence of the $\beta$-function that 
formed the foundation for the critical point analysis of \cite{41} that we
extend to several loop orders, schemes and gauges. Therefore it is instructive
to recall the essence of that approach. The two key renormalization group
functions that determine the critical behaviour in the $(a,\alpha)$ plane are
defined by
\begin{equation}
\mu \frac{d a}{d\mu} ~=~ \beta(a,\alpha) ~~~,~~~
\mu \frac{d \alpha}{d\mu} ~=~ \alpha \gamma_\alpha(a,\alpha)
\label{betgamdef}
\end{equation}
where $\mu$ is the mass scale associated with the renormalization group
equation where the functions appear together. Equally the order by order
solution of the coupled differential equations of (\ref{betgamdef}) determines
how $a$ and $\alpha$ depend on $\mu$. In particular at a critical point a
system is scale free which means in the renormalization group context that 
$\beta(a,\alpha)$ and $\gamma_\alpha(a,\alpha)$ are key with
the solutions of
\begin{equation}
\beta(a,\alpha) ~=~ 0 ~~~,~~~
\alpha \gamma_\alpha(a,\alpha) ~=~ 0
\label{critpteqn}
\end{equation}
determining the fixed points of the system. While the properties of the first
equation have been examined at depth since the discovery of the Banks-Zaks
fixed point, \cite{1,2}, the scheme that was focused on was the $\MSbar$ one in
a linear covariant gauge. In \cite{24,41} it was observed that the second 
equation could not be ignored. In particular it was noted in \cite{24,41} that 
there was a second fixed point with a non-zero gauge parameter at criticality. 
This arose from the solution of $\gamma_\alpha(a,\alpha)$ vanishing giving a 
non-zero critical value for $\alpha$. In other words a non-Landau gauge 
solution. One might be tempted to assume that there is always a gauge parameter
fixed point that produces the Landau gauge. This is only the case if 
$\gamma_\alpha(a,\alpha)$ is not singular at $\alpha$~$=$~$0$. By contrast in 
the MAG $\gamma_\alpha(a,\alpha)$ has a singularity at $\alpha$~$=$~$0$,
\cite{62}.

In order to gain as large a viewpoint as possible of the critical parameter
plane we will determine the critical couplings in the three gauges of interest
in a variety of schemes. These will be the $\MSbar$, $\RI$ and $\mMOM$ schemes
to five loops in the linear covariant gauges and to three loops in the other 
two gauges. The three MOM kinematic schemes of \cite{35,36} will also be 
considered for all three gauges but only to three loops given the difficulty of
computing the underlying master integrals for the non-exceptional momentum
configuration of the vertex functions necessary for the MOM scheme 
prescription. In this respect we will use the results derived over a number of 
years from 
\cite{2,3,4,5,6,7,8,9,10,11,12,13,14,15,16,17,18,19,20,21,22,23,63,68}
Although the four loop MOM scheme renormalization group functions are available
in \cite{39} they were only determined in the Landau gauge rather than a 
general linear covariant gauge. One reason for carrying out a high loop order 
fixed point analysis rests in the issue of convergence. For instance evaluating
the anomalous dimensions at a fixed point produces critical exponents which are
renormalization group invariants. In other words the exponents are physical 
quantities but in estimating them perturbatively one has to have a measure of 
their convergence. This can be studied not only by considering values at 
successive orders in perturbation theory but also by computing the same 
exponents in a different renormalization scheme. In addition the exponents 
ought to be independent of the choice of gauge. So evaluating anomalous 
dimensions that are common to all gauges at criticality are an additional 
measure of consistency. Though for completeness we have analysed the anomalous 
dimensions of all the fields as well as the quark mass operator. The remaining 
exponents which are important to our analysis are those dealing with the 
stability properties of the fixed point. In general these properties are 
derived from the eigenvalues of the matrix
\begin{equation}
\beta_{ij}(g_k) ~=~ \frac{\partial \beta_i}{\partial g_j}
\label{hessdef}
\end{equation}
for a theory of $n$ couplings where $1$~$\leq$~$i,j,k$~$\leq$~$n$. Here we have
$n$~$=$~$2$ with $g_1$~$=$~$a$, $g_2$~$=$~$\alpha$ and
\begin{equation}
\beta_1(a,\alpha) ~\equiv~ \beta(a,\alpha) ~~~,~~~
\beta_2(a,\alpha) ~\equiv~ \alpha \gamma_\alpha(a,\alpha) ~.
\label{beta12}
\end{equation}
The main aim is to ascertain whether there are critical points that are fully
stable in the infrared rather than saddle points. We will denote the two
critical eigen-exponents of (\ref{hessdef}) by $\omega_i$.
 
\sect{Fixed point analysis.}

While perturbative results are available at various loop orders in different
schemes and gauges for a general gauge group, our results will focus on the
particular group of $SU(3)$ as it is the strong sector of the Standard Model.
Equally we will concentrate on fermions in the fundamental representation of
$SU(3)$. One issue that we will consider is the properties of the conformal 
window and whether the interrelation of real critical points at various values 
of $\Nf$, as determined perturbatively, provide a clue as to when it ceases to
exists for low $\Nf$. First we make several observations concerning our 
analysis method. As a mathematical problem finding the solutions to 
(\ref{critpteqn}) is a straightforward exercise numerically. We have used 
various tools such as {\sc Maple} to do this. Consequently as the loop order 
increases one finds a large number of zeros for the critical coupling and gauge
parameter. However given the origin of the two renormalization group functions 
and the underlying theory they represent we have filtered out solutions that 
are not physical. For instance we have ignored solutions that are complex as
well as those where $a$~$<$~$0$ since the coupling constant has to be positive.
Equally cases where the critical coupling is large can be discounted as it 
would exceed the limits of the perturbative approximation. Although for values 
of $\Nf$ towards the lower boundary of the conformal window this latter 
scenario arises, we have retained these in order to have an overall perspective
similar to the original work of \cite{1,2}. 

What remains after this general sieving is a handful of critical points at 
three and higher loops. However as we are dealing with polynomials in the 
coupling with increasing order some fixed points arise at a particular order 
which have no candidate partner at the next order in the neighbourhood of the 
critical coupling and gauge parameter. Instead it arises at the order 
subsequent to that one. This is because at the intermediate order there is a 
complex fixed point and so we regard such critical points as artefacts. Indeed 
they tend to be associated with saddle points or unstable fixed points. After 
the general filtering of the fixed points our next stage was to determine their
stability properties from the eigenvalues of (\ref{hessdef}). These are either 
ultraviolet stable, ultraviolet unstable or saddle points. One of our general 
observations was that the Banks-Zaks fixed point, \cite{1,2}, in each of the 
gauges and schemes was infrared stable in the coupling constant direction but 
infrared unstable in the gauge parameter direction making it a saddle point. 
However in the running away from the infrared stable direction there was an 
infrared stable fixed point for both non-zero coupling and gauge parameter 
confirming earlier observations of \cite{41}. Moreover the value of the 
critical coupling at this infrared stable point was in general the {\em same} 
value of the coupling at the Banks-Zaks itself. Therefore we will invariably 
refer to the fully infrared stable fixed point as the mirror or twin point.

Having provided an overview of the analysis it is instructive to look at the
specific situation and for the moment we will concentrate on the $SU(3)$ linear
covariant gauge fixed points in the $\MSbar$ scheme. We have recorded the
location of the critical points at two, three, four and five loops in Tables 
\ref{2lsu3cfw} to \ref{5lsu3cfw}\footnote{More detailed tables upon which all
the Tables are based are available in a data file associated with the arXiv 
version of this paper.}. These are banked into groups with the same $\Nf$ 
values. The quantities $a_\infty$ and $\alpha_\infty$ are the critical coupling
values in the same notation as \cite{41} with $\omega_i$ being the values of 
the eigenvalues of (\ref{hessdef}). These determine the stability properties in
the infrared limit which is indicated in the final column. In all tables of 
fixed points we omit the trivial Gaussian one at the origin which is infrared 
unstable. The data relating to the original Banks-Zaks fixed point agrees with 
earlier analyses \cite{24,25,26,41,42}. The mirror point is clearly evident in 
all five tables. For the lower values of $\Nf$ in the conformal window the 
perturbative approximation ceases to be reliable. Indeed it was an initial 
surprise when the five loop $\MSbar$ QCD $\beta$-function became available in 
\cite{17} that the Banks-Zaks fixed point did not appear to exist for 
$\Nf$~$<$~$13$ when the standard solution method was employed, \cite{42}. This 
was clearly resolved in the same article, \cite{42}, where the five loop 
conformal window was accessed via Pad\'{e} resummation methods. At four loops a
second set of connected solutions arises in Tables \ref{4lsu3cfw912} and 
\ref{4lsu3cfw1316} that has no relation to a lower or five loop solution. These
can be regarded as artefacts and moreover have no physical importance given 
that none relate to a stable point. Equally they have a large critical coupling
and might have suggested a type of asymptotically safe solution being 
ultraviolet stable but the lack of a connection with lower loop order rules 
that out. One other interesting feature is the value of the critical gauge 
parameter at the mirror fixed point. For $\Nf$ values close to the top of the 
conformal window the value is always in the neighbourhood of 
$\alpha_\infty$~$\approx$~$-$~$3$. This value has been observed before in 
different contexts, \cite{24,41,47,48,49,50,51,52,53,54,55,56}, and in 
particular several instances related to infrared issues in QCD. We qualify this 
observation by noting that $\alpha_\infty$~$=$~$-$~$3$ arose in the particular 
case of the linear covariant gauge. There is no reason to expect this property 
to arise in other gauges let alone for the same value of the critical gauge 
parameter. 

In order to visualize the renormalization group flow we have constructed flow 
plots in the coupling constant and gauge parameter plane. These are provided in
Figures \ref{flow3msn12} and \ref{flow3msn16} for $SU(3)$ at two, three, four 
and five loops for $\Nf$~$=$~$12$ and $16$ respectively in the case of the 
linear covariant gauge. The flow arrow is towards the infrared away from the 
Gaussian fixed point at the origin. Our notation is that a fixed point is 
indicated by various markers where their shapes indicate their stability 
property. In particular an infrared stable fixed point is indicated by 
$\times$, the Banks-Zaks critical point is denoted by $\blacksquare$ where the 
other saddle points are marked by $\blacklozenge$. The remaining shape, 
$\bigstar$, denotes an infrared unstable fixed point and therefore an 
ultraviolet stable one. Examining Figure \ref{flow3msn12} by way of example one
sees the Banks-Zaks fixed point in line with its mirror partner at each loop 
order and $\Nf$ value. By contrast infrared unstable fixed points are present 
at three and higher loops order but their location varies by a significant 
amount at successive orders and have no physical significance. The general 
locations of the Banks-Zaks and its stable mirror partner are clearly 
unaffected by the increasing loop order. The five loop plot of the flow for 
$\Nf$~$=$~$12$ reflects the absence of the Banks-Zaks fixed point noted in 
\cite{42}. 

We have repeated the same analysis for the Curci-Ferrari gauge and MAG but only
to three loop order as no renormalization group functions are available beyond 
that. While this may appear to limit what can be extracted from a fixed point 
analysis, it transpires that although a similar general picture emerges the 
details are necessarily different. The fixed point values at two and three 
loops together with their stability for the $\MSbar$ scheme are provided in 
Tables \ref{2lcfsu3cfw} and \ref{3lcfsu3cfw} for the Curci-Ferrari gauge with 
the parallel data for the MAG presented in Tables \ref{2lmagsu3cfw} and 
\ref{3lmagsu3cfw}. Various flow plots for both gauges at two and three loops
are provided in Figures \ref{flow3mscfn12} to \ref{flow3msmagn16}. For both 
gauges we include three loop results for $\Nf$ less than the perturbatively 
accepted value of $8$ for the lower boundary of the conformal window. This is 
to illustrate the changing nature of the Banks-Zaks fixed point in both gauges 
as it ceases being a saddle point and transforms into an infrared stable point
when $\Nf$~$\leq$~$7$. However for that range the analysis cannot be regarded 
as reliable as one is clearly beyond the region of perturbative validity. We 
need to qualify what we mean by a Banks-Zaks fixed point in the case of the 
MAG. In a linear gauge it corresponds to a non-trivial $\beta$-function zero in
the $\MSbar$ scheme which is gauge parameter independent and so lies on the 
$\alpha$-axis in an $(a,\alpha)$ flow plot. Given the nature of the MAG gauge 
parameter anomalous dimension there is no fixed point when $\alpha$~$=$~$0$ but
as is evident from Tables \ref{2lmagsu3cfw} and \ref{3lmagsu3cfw} there is a 
fixed point with a value of $\alpha$ close to zero. This is the one we refer to
as the MAG Banks-Zaks fixed point. We have also checked that when the 
$\Nda$~$\to$~$0$ limit is taken this fixed point smoothly tends to the 
Banks-Zaks fixed point of the Curci-Ferrari model. The flows for both gauges 
are given in Figures \ref{flow3mscfn12} and \ref{flow3mscfn16} for the 
Curci-Ferrari gauge and the corresponding ones for the MAG can be seen in 
Figures \ref{flow3msmagn12} and \ref{flow3msmagn16} where again we focus on 
$\Nf$~$=$~$12$ and $16$. What is also apparent in the tables for the MAG is 
that an additional stable fixed point arises when $\Nf$~$\leq$~$10$. Whether 
such extra solutions are an artefact of the loop order is not clear but they 
are present in an $\Nf$ range where perturbative reliability could be 
questioned. One feature that differs from the linear covariant gauge is the 
value of the critical gauge parameter at the stable fixed point in both gauges.
In the linear gauge the value was around $(-3)$ but in the two non-linear 
gauges the value appears to be in the neighbourhood of $(-5)$ for $\Nf$ near 
the top end of the conformal window which has not been noted previously. Of 
course the parameters are not the same in each gauge and such parameters are 
not physically measureable. Their main application is in the determination of 
critical exponents which are renormalization group invariants and discussed 
later.

To this point the focus has been on the $\MSbar$ scheme for different gauges.
It is instructive to analyse the fixed points in another scheme. As the
renormalization group functions are available to five loops in the $\mMOM$
scheme \cite{30,31,32,33} for an arbitrary linear gauge parameter we have 
solved for the critical values of $a$ and $\alpha$ in that scheme. In this
scheme the $\beta$-function depends on $\alpha$ in contrast to the $\MSbar$
scheme. The results for two, three, four and five loops are recorded in Tables 
\ref{2lmmsu3cfw}, \ref{3lmmsu3cfw}, \ref{4lmmsu3cfw} and \ref{5lmmsu3cfw} 
respectively. In general terms the tables reflect very similar properties to 
those observed in the $\MSbar$ scheme. The Banks-Zaks fixed point is evident 
across all loop orders and is a saddle point. The associated infrared stable 
fixed point is also a feature but with the caveat that it does not have the 
same critical coupling as the Banks-Zaks one. Instead its critical coupling 
value is very close to it. The critical gauge parameter value of the infrared 
stable points are all again in the neighbourhood of $\alpha$~$=$~$-$~$3$. 
For low loop order towards the lower end of the conformal window there is a 
larger deviation of this scenario. However at five loops there is a marked 
reduction in the discrepancy from $(-3)$ which is apparent at low $\Nf$ values.
This reinforces the $\MSbar$ observation that this gauge parameter choice may 
be a deeper property of the underlying theory. While the $\mMOM$ and the 
$\MSbar$ tables display a degree of similarity there are clearly several major 
differences. The first is that the eigenvalues of (\ref{hessdef}) are complex 
conjugates at several fixed points. The stability property is determined by the
sign of the real part of the eigenvalues and the non-zero imaginary part 
indicates a spiral flow to or away from criticality. However as the loop order 
increases the spiral flow to the infrared stable fixed point appears to 
disappear. Perhaps the most significant aspect of the $\mMOM$ analysis is that 
there are real solutions for both $a$ and $\alpha$ for $\Nf$~$\leq$~$12$ unlike
the $\MSbar$ scheme. This therefore supports the five loop $\MSbar$ analysis of
the Banks-Zaks critical point of \cite{42} which required a Pad\'{e} analysis 
to access fixed points for $\Nf$~$\leq$~$12$. We have included all the real 
solutions to (\ref{critpteqn}) partly for completeness but also for background 
when viewing the associated $\mMOM$ flow plots. These are given in Figures 
\ref{flow3mmn12} and \ref{flow3mmn16} for $\Nf$~$=$~$12$ and $16$ respectively.
Clearly some of the extra fixed points in the $\mMOM$ tables are beyond the 
region of perturbative validity. However their presence are responsible for the
flow in the regions immediately outside the boundaries of the plots.

We complete this section by recording the situation with the kinematic scheme
fixed points in the three gauges of interest. Unlike the $\MSbar$ and $\mMOM$
schemes the full renormalization group functions for the MOM schemes of
\cite{35,36} are only available at three loops for arbitrary gauge parameter.
In order to compare the fixed point properties with other schemes we have
provided the $\Nf$~$=$~$12$ and $16$ three loop flow plots for the linear
covariant and Curci-Ferrari gauges as well as the MAG in Figures
\ref{flow3mom3n1216}, \ref{flow3momcf3n1216} and \ref{flow3mommag3n1216}
respectively. In all three gauges and MOM schemes the Banks-Zaks and the
infrared stable fixed points are clearly evident for $\Nf$~$=$~$16$. For
$\Nf$~$=$~$12$ the situation is similar except in one or two schemes the
infrared stable fixed point is not present in the flow plane. While the
value of the critical gauge parameter of the infrared stable fixed point is 
consistently around the values of $(-3)$ for the linear gauge and $(-5)$ for 
the Curci-Ferrari gauge and MAG the associated critical coupling is actually
large. While it is outside the range plotted it is also outside the domain 
of perturbative reliability. As an example Tables \ref{2lsu3momcmagcfw} and 
\ref{3lsu3momcmagcfw} record the fixed point data for the two and three loop 
MAG fixed points in the MOMc scheme. Given the improvement with convergence 
at $\Nf$~$=$~$12$ that was apparent in the higher order $\MSbar$ and $\mMOM$
scheme fixed points we would expect that situation to improve if the full four 
loop renormalization group functions with $\alpha$~$\neq$~$0$ were available. 

\sect{Critical exponents.}

While the location of fixed points is important for understanding the 
renormalization group flow in QCD, the quantities of physical relevance are the
critical exponents. Their values define the properties of the critical theory
and are important for discerning the underlying conformal field theory at a
fixed point. In our case values or estimates for the exponents are derived from
the renormalization group functions of QCD by evaluating them at the various 
fixed points. In this section we will discuss the critical exponents derived 
from the anomalous dimensions of the gluon, ghost and quark fields as well as 
the quark mass dimension in the various gauges and schemes. The exponent 
derived from the quark mass anomalous dimension is gauge independent. So one
aspect of our analysis will be to examine this in the various schemes and 
gauges. Moreover by comparing their values for a variety of schemes and gauges
the issue of perturbative convergence can be examined. In addition the 
exponents $\omega_1$ and $\omega_2$ are derived from the underlying 
$\beta$-functions (\ref{beta12}) and therefore should also be independent of 
the gauge fixing procedure. We qualify this outline of the analysis by 
recalling that data for the MOM kinematic schemes will be restricted to three 
loops unlike the $\MSbar$, $\mMOM$ and $\RI$ schemes where five loop data are 
available. Previous exponent studies of the QCD critical points in various 
schemes was restricted to the Banks-Zaks fixed point in the $\MSbar$, $\RI$ and
$\mMOM$ schemes, \cite{24,25,69,70}, and the kinematic schemes, \cite{71}. 
Since we have examined the $(a,\alpha)$ flow plane in depth here our main focus
will be on the exponents of the infrared stable fixed point. In the linear 
covariant gauge the value of the critical coupling for the Banks-Zaks and 
infrared stable fixed points are the same. Therefore there is a natural 
question as to whether the corresponding exponents are similar. If so this 
would reinforce the notion that the infrared stable fixed point is a mirror of 
the Banks-Zaks one. 

First we concentrate on the four and five loop estimates for the $\MSbar$,
$\mMOM$ and $\RI$ schemes with values recorded in Tables \ref{4lmssu3exp} to
\ref{5lrisu3exp}. In these and other such tables the syntax is that BZ 
indicates Banks-Zaks and IRS denotes the infrared stable fixed point which is 
closest to the $\alpha$ axis on the flow plane. In addition type (IRS) in a 
table denotes another infrared stable fixed point that was identified in the 
corresponding earlier list of fixed points. Only Banks-Zaks and infrared stable
fixed point exponents are recorded in the tables as exponents at saddle or 
infrared unstable fixed points are not of physical relevance. We recall that it
is not always the case that values for exponents are available for each fixed 
point type across each scheme. Indeed this is not the case at five loops in the
$\MSbar$ scheme as noted in \cite{42} with the same situation for the $\RI$ 
scheme since its $\beta$-function is formally the same as the $\MSbar$ one. The
quantities $\gamma_\phi$ where $\phi$~$\in$~$\{A,c,\psi\}$ in the tables denote
the critical exponents for the gluon, ghost and quark fields. The final column 
of each of our exponent tables will be the exponent $\rho_m$ which is related 
to the quark mass anomalous dimension and is given by
\begin{equation}
\rho_m ~=~ -~ 2 \gamma_{\bar{\psi}\psi}(a_\infty,\alpha_\infty) ~.
\end{equation}
In focusing on the $\MSbar$, $\mMOM$ and $\RI$ schemes in the first instance
the four loop tables are provided for orientation with the main task of 
studying the effect of the next order correction. We note that exponent 
estimates for these and the MOM schemes had been recorded earlier in \cite{72}
for the Banks-Zaks critical point. That study together with \cite{24} included 
results for $SU(2)$ and $SU(4)$ as well as for the quark in a variety of 
representations. For the $\rho_m$ exponent one obvious property that is evident 
here in the $\MSbar$ scheme at all loop orders is that the BZ and IRS values of
$\rho_m$ are the same at each loop order. This follows simply from the fact 
that the anomalous dimension of the quark mass operator is gauge parameter 
independent in the $\MSbar$ scheme. Therefore the value of the critical gauge 
parameter is irrelevant for the exponent. In the $\mMOM$ and $\RI$ schemes the 
quark mass operator does depend on the gauge parameter. So one would expect 
some deviation of the values for $\rho_m$ at the BZ and IRS fixed points. This 
is quite clearly the case at four loops for both schemes. Although at each 
fixed point for $\Nf$~$=$~$15$ and $16$ $\rho_m$ is virtually identical which 
is the part of the conformal window where perturbative reliability is best. 
What is striking is that at five loops the value of $\rho_m$ at the BZ and IRS 
fixed points appear to show a marked convergence to a common value even down to
$\Nf$~$=$~$12$ for $\mMOM$ and $\Nf$~$=$~$13$ for $\RI$. Below that $\Nf$ value
for the $\RI$ scheme there are no five loop solutions of (\ref{critpteqn}). 
This strongly suggests that $\MSbar$ property of $\rho_m$ at the BZ and IRS 
fixed points will become a scheme independent property with more accuracy.

The convergence for the gluon, ghost and quark exponents is not as accurate
except of course for the gluon at the IRS fixed point in the linear covariant
gauge. This is because in that gauge the gluon anomalous and gauge parameter
anomalous dimensions are equal and opposite and the vanishing of the latter 
is used to find the critical point values. For the BZ fixed point the gluon
exponent is reasonably consistent down to $\Nf$~$=$~$14$ for the three schemes
at four and five loops. For $\Nf$~$\leq$~$13$ there is a marked distinction
between the value for $\RI$ compared to the other two schemes. This can simply
be put down to the lack of fixed points at five loops for $\Nf$~$\leq$~$12$
in the $\MSbar$ and $\RI$ schemes. The deviation is apparent at four loops. The
picture for $\gamma_c$ is somewhat similar but with the value for the BZ fixed 
point is in better agreement across the three schemes even down to
$\Nf$~$=$~$12$ at five loops. By contrast the critical value of $\gamma_\psi$ 
at the IRS fixed point appears to be more consistent in the three schemes down 
to the $\Nf$~$=$~$12$ demarcation. More importantly there is a clear 
improvement in convergence comparing the values of $\gamma_\psi$ at four loops 
with their five loop counterparts. This reinforces the earlier observation that
the extra loop order is pointing to the emergence of a more accurate picture 
deeper into the conformal window. It is worth remarking at this point that a
similar observation has been made recently in this respect but in a different 
context in \cite{73}. There dimensionless ratios involving meson decay 
constants were studied in the conformal window using the perturbative expansion
method about the Banks-Zaks fixed point introduced in \cite{1}. Using a fourth 
order expansion $\Nf$~$=$~$12$ was identified as a boundary where the 
perturbative approach was reliable, \cite{73}.

The situation with the kinematic schemes and non-linear gauges is not as clear
cut. The values of exponents at three loops in the three MOM schemes and the 
gauges of interest are given in Tables \ref{3lmgsu3exp} to \ref{3lmqmagsu3exp}.
For the tables with MAG results the type (BZ) indicates the saddle point that 
would be in the neighbourhood of the Banks-Zaks fixed point of the other two
gauges. In terms of relevance the gluon and ghost critical exponents can only
be compared for convergence within each gauge and not in any other gauge. This
is because these fields are not strictly the same object in different gauges. 
The only exponents that can be compared across gauges are $\rho_m$ and 
$\omega_i$. Examining $\rho_m$ for the MOM schemes shows up several patterns of
consistency. First for $\Nf$ close to the top of the conformal window the value
of $\rho_m$ at both the Banks-Zaks and infrared stable fixed points are 
generally in very good agreement to three decimal places in the three gauges 
and MOM schemes. There is one exception though which is MOMg value at the 
infrared stable point in the Curci-Ferrari gauge. Similar exceptions to the 
general pattern for other exponents in the various gauges and schemes are 
apparent in the various tables. They can usually be attributed to accidental 
cancellations in evaluating the perturbative expansion. This actually lends 
weight to ensuring exponent estimates are derived in as many different ways as 
possible in order to the general trends. At lower values of $\Nf$ in the 
conformal window the MOM scheme exponents are not as accurate as one would 
expect. For instance examining the $\Nf$~$=$~$12$ case there is reasonable 
consistency at the Banks-Zaks point across the three gauges for each scheme. 
However there is a discrepancy across schemes within each gauge. For the 
infrared stable fixed point only the MOMq results have a degree of accuracy but
the values for both fixed points undershoot the three loop $\MSbar$ values. We 
recall that at $\Nf$~$=$~$12$ there is no infrared stable fixed point in the 
MOMg scheme. Clearly what is lacking for these lower values of $\Nf$ are the 
higher order corrections in the perturbative series.

While the only four loop MOM scheme renormalization group functions that are
available are in the Landau gauge, \cite{39}, we can try this information to
probe whether the exponent convergence improves albeit at the Banks-Zaks fixed
point. The results of this exercise are presented in Table \ref{momilan4} in
the conformal window. What is evident from comparing the values of $\omega$ in
the three schemes is that there is a degree of agreement for down to 
$\Nf$~$=$~$14$ at which point the estimates in the $\MOMg$ scheme cease to be
commensurate. In the $\MOMq$ case the values of $\omega$ are in reasonable
agreement with those of $\MOMc$ down to $\Nf$~$=$~$11$. What is interesting is
that comparing the $\omega$ estimates for $\MOMc$ with the four loop $\mMOM$
ones there is remarkable agreement down to $\Nf$~$=$~$9$. Although both schemes
are based on the properties of the same vertex function the momentum
configuration where the subtraction is carried out is different. One is for a
non-exceptional momentum setup while the other is exceptional respectively. In 
this instance it would seem to reflect that there is an improvement in 
convergence. Comparing with the same region of the conformal window for the 
$\MSbar$ scheme is probably not a reliable exercise given that there was no 
five loop Banks-Zaks fixed point solution for low $\Nf$. Regarding the $\MOMg$ 
scheme the drop off in the $\omega$ estimate as $\Nf$ reduces might be 
indicative of a slow convergence or an indication that the Banks-Zaks fixed 
point might disappear at five loops. As such an eventuality arises in one 
scheme the same behaviour in another cannot be excluded.
 
The other exponents that should have a degree of consistency across schemes and
gauges are those connected with corrections to scaling which are $\omega_1$ and
$\omega_2$. First examining the $\MSbar$ values of $\omega_i$ in the linear
covariant gauge in Tables \ref{4lsu3cfw912}, \ref{4lsu3cfw1316} and 
\ref{5lsu3cfw} several observations emerge that are also evident in other 
schemes and gauges. The obvious one is that one of the eigenvalues of 
(\ref{hessdef}) is the same for both the Banks-Zaks and infrared stable fixed 
point. Curiously the other stability eigenvalues are roughly equal in magnitude
which is a feature of all $\Nf$ values in the conformal window. Whether their
equality will emerge with higher precision is an open question. Moreover the 
convergence from four to five loops is present down to $\Nf$~$=$~$14$ but the 
loss of a solution for the conformal below $\Nf$~$=$~$13$ is again reflected in
the values for $\omega_i$ at this latter $\Nf$ value. To gauge the structure at
five loops better it is convenient to examine the $\omega_i$ for the $\mMOM$ 
scheme which are given in Tables \ref{4lmmsu3cfw} and \ref{5lmmsu3cfw} as the 
latter covers the conformal window more fully\footnote{Our tables for the 
stability exponents were generated and written automatically using {\sc Maple}.
So the values of $\omega_1$ and $\omega_2$ for the $\mMOM$ scheme tables need 
to be swapped in order to compare with their $\MSbar$ counterparts.}. Comparing
the four loop values there is good agreement at the top of the conformal window
down to $\Nf$~$=$~$13$ but the positive exponent that is supposed to be the 
same at both the Banks-Zaks and infrared stable fixed points begin to differ at
this $\Nf$ value consistent with the appearance of the complex values for 
$\omega_i$ for $\Nf$~$\leq$~$12$ and the switch of the stable infrared fixed 
point away from a critical gauge parameter value in the neighbourhood of 
$\alpha$~$=$~$-$~$3$. At five loops for both schemes the consistency for the 
top end of the window improves as expected from the increased perturbative 
accuracy. What is more striking though can be seen from the $\Nf$~$=$~$13$ 
values of the four loop $\MSbar$ and five loop $\mMOM$ $\omega_i$ estimates. 
These are much more in keeping with each other indicating the importance of 
studying the properties in more than one scheme. 

While these comments concern the linear covariant gauge in particular it is
instructive to look at the situation in the two non-linear gauges. However for 
a fair comparison we will focus on the three loop values as nothing is
available beyond beyond that order for the Curci-Ferrari gauge and MAG. First 
examining the values of $\omega_i$ in the linear covariant and Curci-Ferrari 
gauges given in Tables \ref{3lsu3cfw} and \ref{3lcfsu3cfw} there are several 
points to note. First while the $\omega_2$ values are the same for both fixed 
points in $9$~$\leq$~$\Nf$~$\leq$~$16$ this reflects the $\alpha$ independence 
of the common $\beta$-function in both gauges in the $\MSbar$ scheme. What is 
more interesting is that the values $\omega_1$ are remarkably consistent for 
the same range of $\Nf$ with a slight discrepancy in value at the lower end. 
What has to be remembered is that the critical gauge parameter for each $\Nf$ 
value is not the same unlike the critical coupling. This lends weight to the 
observation that the infrared stable fixed point is a Banks-Zaks twin point.
This pattern is less conclusive for the MOM schemes in the Curci-Ferrari gauge
with only the MOMq data for $\omega_i$ producing comparable values with those
in the $\MSbar$ scheme for $\Nf$ down to $13$ flavours at three loops. In the
MOMg scheme only in the $\Nf$~$=$~$16$ case is there a stable infrared fixed 
point and then the $\omega_i$ values are not in keeping with the other schemes 
and gauges. This is again perhaps related to the absence of a similar critical
point for lower values of $\Nf$ in this scheme for the Curci-Ferrari gauge. The
values for $\omega_i$ are not as accurate at the Banks-Zaks point either aside 
from $\omega_2$ down to $\Nf$~$=$~$11$. For this fixed point in the MOMc scheme
the situation at three loops is almost on a par with the MOMq results. However 
aside from $\Nf$~$=$~$16$ the values of $\omega_i$ in the MOMc scheme are not 
reliable. In light of this one natural question concerns why the picture in the
MOMq scheme is better than those in the MOMg and MOMc ones. Perhaps this can be
explained by the position in the respective two loop cases. For MOMg there are 
no infrared stable fixed point solutions and only one appears at three loops 
unlike the linear covariant gauge. In the MOMc case at two loops there are 
infrared stable fixed points but aside from $\Nf$~$=$~$16$ the $\omega_i$ are 
complex conjugates which become real at three loops. So it seems clear that for
these two schemes the existence and properties of this infrared stable fixed 
point requires several more loop orders to become manifest on a comparable 
level to other schemes. 

To complete our commentary on the situation with the corrections to scaling
exponents we turn to the MAG results where the two and three loop $\MSbar$
results are given in Tables \ref{2lmagsu3cfw} and \ref{3lmagsu3cfw}. Focusing 
for the moment on three loops it is clear that the $\omega_2$ value is in good 
agreement for a wide range of $\Nf$ with the other two gauges. For the other 
exponent there is a wide disparity even for the fixed point close to the 
origin. This seems particularly peculiar given that the Curci-Ferrari gauge and
MAG are intimately connected but the $\MSbar$ $\omega_i$ values are better 
aligned with those of the linear covariant gauge. This is perhaps indicative of
a specific property of the MAG itself rather than a particular breakdown in the
connection with the Curci-Ferrari gauge. In this respect we recall that the MAG
treats the diagonal and off-diagonal gluons differently in the gauge fixing 
functional. The parameter $\alpha$ in this gauge relates to the non-linear 
gauge fixing for the off-diagonal gluons but the diagonal ones are fixed in the 
Landau gauge. In the same way that we explored the $(a,\alpha)$ plane for fixed
points in the linear covariant and Curci-Ferrari gauge from the point of view 
of a critical point analysis of the renormalization group functions full gauge 
fixing a separate gauge parameter should be considered for the diagonal gluons.
In other words the infrared stable fixed point associated with the gauge 
parameter in the neighbourhood of $\alpha$~$=$~$-$~$5$ may not be stable in the
direction towards that extra gauge parameter. There may be an infrared stable 
fixed point in the $(a,\alpha,\bar{\alpha})$ hyperspace. If so then it would be
the one for comparing the values of $\rho_m$ and two of the three $\omega_i$ 
exponents. Trying to explore this is beyond the scope of the present work. The 
reason for this is we would have to renormalize the QCD in a maximal abelian 
gauge fixing with one or more extra parameters. By this we mean that an 
interpolating gauge was constructed in \cite{74} which involved six additional 
gauge parameters to ensure renormalizability. Taking separate limits of these 
parameters produces the usual linear covariant and maximal abelian gauges. Once
these limits were verified at three loops one would then have to examine the 
fixed point structure to ascertain whether there was a stable infrared fixed 
point. If there were more than one such solution the absolute minimum should be
the one that is the natural partner of the infrared stable fixed point of the 
other two gauges. 

\sect{Pad\'{e} analysis.}

The absence of a Banks-Zaks fixed point at five loops in the $\MSbar$ scheme
for the full range of the conformal window, \cite{42}, was unexpected. This was
especially the case since perturbative estimates of the $\rho_m$ exponent are 
of importance to compare with lattice methods. Consequently an additional tool 
was employed to explore the lower reaches of the conformal window in this 
scheme which was Pad\'{e} approximants, \cite{42}. The approach was to 
establish a more accurate fixed point and thence a better estimate of $\rho_m$.
Given that our five loop $\mMOM$ analysis has revealed that the absence of
the $\MSbar$ five loop Banks-Zaks fixed point can be attributed to a
scheme artefact it is a worthwhile exercise to repeat the Pad\'{e} analysis of 
\cite{42} for the $\mMOM$ scheme but also for the $\RI$ one. By doing so we 
will be able to see if a consensus emerges for exponent values at the lower end
of the conformal window below $\Nf$~$\leq$~$12$. Moreover we will not restrict 
the analysis to the Banks-Zaks case but include its twin partner. To facilitate
such a study therefore requires incorporating the gauge parameter into the 
rational polynomials of the Pad\'{e} approximants. 

By way of establishing our approach we focus first on the $\MSbar$ scheme with
the aim of reproducing the data of \cite{42} for the Banks-Zaks case which will
also play the role of a check on the setup as it will include $\alpha$
dependence. This will be of importance for the mirror fixed point since 
ultimately exponent estimates should be independent of the gauge. As before 
searching for stationary running means searching for solutions 
$(a_\infty,\alpha_\infty)$ to $\beta(a_\infty,\alpha_\infty)$~$=$~$0$ and
$\alpha_\infty\gamma_\alpha(a_\infty,\alpha_\infty)$~$=$~$0$. However both
$\beta(a,\alpha)$ and $\gamma_\alpha(a,\alpha)$ will be supplied by Pad\'{e}
approximants such that
\begin{eqnarray}
\beta^{[p,q]}(a,\alpha) &=& \beta_0 a^2 \
\frac{\left[1+\sum_{j=1}^p u^{[p,q]}_j(\alpha)a^j\right]}
{\left[1+\sum_{k=1}^q v^{[p,q]}_k(\alpha)a^k\right]} \nonumber \\
\gamma_\alpha^{[p,q]}(a,\alpha) &=& \gamma_1 a 
\frac{\left[1+\sum_{j=1}^p w^{[p,q]}_j(\alpha)a^j\right]}
{\left[1+\sum_{k=1}^q x^{[p,q]}_k(\alpha)a^k\right]}
\end{eqnarray}
where
\begin{equation}
\beta_0 ~=~ -~ \frac{1}{3} [ 11 C_A - 4 T_F \Nf ] ~~~,~~~
\gamma_1 ~=~ \frac{1}{6} [ 8 T_F \Nf - 13 C_A + 3 \alpha C_A ]
\end{equation}
and only $a$ acts as the expansion parameter in the approximant with $\alpha$
entering in the coefficients. While we could have approximated 
$\beta(a,\alpha)$ and $\alpha\gamma_\alpha(a,\alpha)$ with different $[p,q]$
approximants, since any approximation will be correct to the order in 
perturbation theory we consider here, for simplicity we have used the same 
$[p,q]$ structure for both functions. Although we use the same method as 
Section $3$ to search for fixed point solutions we need to ensure they are in 
the region of validity of the approximation. By this we mean the approximant to
both renormalization group functions has no poles or denominator zeros that are
closer to the axis than the critical point when the gauge parameter is fixed to
$\alpha_\infty$. The fixed points will be classified into three categories. The
first class is the valid fixed points which are those for which no zero of the 
denominator has a smaller absolute coupling constant value than the fixed 
point. The second class is termed the zero pole (zp) fixed points which are 
those where there exist zeros in the denominator closer to the origin than the 
fixed point but these are also zeros of the numerator. Thus these zero pole 
zero numerator pair will cancel each other. The final group are critical points
with a pole closer to $a$~$=$~$0$ but these will be discarded from further 
analysis. With regard to using the critical point data for the first two 
classes the last step will be to determine the quark mass anomalous dimension. 
We do this by evaluating the regular polynomial perturbative series for the 
renormalization group function at the fixed point.

The results of our Pad\'{e} analysis for the $\MSbar$ scheme are recorded in
Table \ref{padems}. Since the values of the coupling constant and quark mass 
exponent have already appeared in \cite{42} it is satisfying to note that we 
find full agreement allowing for the difference in convention for the coupling 
constant. Given that reassuring check we now restrict the discussion to the 
sector with non-zero gauge parameter values. Since the $\beta$-function and 
mass anomalous dimension are gauge parameter independent in the $\MSbar$ scheme
we again see that their values are consistent across multiple gauge parameter 
values. As all approximations presented are accurate to an order in 
perturbation theory within their region of validity, if we can identify the 
Banks-Zaks' twin then we can use the values from different approximations to 
provide a range for the location of the fixed point. We note the fixed point we
label as the Banks-Zaks twin is chosen on the basis of consistency across 
schemes, loop order, anomalous dimensions and $\Nf$ but still its 
identification is not absolutely defined. As one would expect of something 
whose value is accurate to the truncation order under consideration we see that
the critical gauge parameter values are more congruous with like values in the 
other approximations at high $\Nf$ where the corresponding $a_\infty$ is lower 
than at smaller $\Nf$. For example, at $\Nf$~$=$~$16$ the difference provided 
by these values is $5\times10^{-6}$, whereas at $\Nf$~$=$~$13$ this difference 
is $0.03$. While there were more fixed points outside of the region of validity
it is of interest to note that in this case the fixed points that were retained 
for the different Pad\'{e} approximants are, in the case of $[2,2]$ and 
$[1,3]$, only the Banks-Zaks fixed point and its partner. For $\Nf$~$=$~$12$ we
can compare the non-zero gauge parameter for the $[3,1]$ approximant with those
of the other approximants in order to ask whether the zp fixed points provide 
an accurate representation of the fixed point. In this case we see the 
difference between the $[3,1]$ fixed point and the $[1,3]$ fixed point is 
roughly ten times the difference between $[2,2]$ and $[1,3]$ in both the 
coupling constant and the gauge parameter. While this is not necessarily 
promising, we point out that the quark mass exponent values are not 
considerably more consistent between the valid fixed points than between the zp
and valid fixed points. We will not draw any firm conclusions about this before
examining other schemes.

With regard to this the critical coupling constants are the same in the 
$\MSbar$ and $\RI$ scheme because their $\beta$-functions are formally
equivalent. However the two schemes are unique and the gauge parameter running 
of both is different. This therefore allows for a simple point of comparison in
relation to their Pad\'{e} approximations. Table \ref{paderi} gives the $\RI$
scheme results that is the parallel to Table \ref{padems} and is accurate to 
the five loop level. First we note that the location of the secondary 
$\alpha$~$=$~$0$ fixed point changes drastically between $[4,0]$ and $[3,1]$
approximants even at $\Nf$~$=$~$16$. This suggests this value is not 
perturbatively reliable. By contrast the Banks-Zaks fixed point value itself is 
stable for the entire conformal window when compared between different Pad\'{e}
approximants. This is as expected since the critical coupling constant values 
are the same as in the $\MSbar$ scheme as they ought to be given the way the 
two scheme $\beta$-functions are related. While the mass anomalous dimension is
not gauge parameter independent in the $\RI$ scheme, we will identify the 
Banks-Zaks twin as the fixed point whose $\rho_m$ and coupling constant best 
matches the Banks-Zaks values. At $\Nf$~$=$~$16$ the Banks-Zaks twin is 
approximately in the same place in all Pad\'{e} approximants with a range of 
$0.000030$ between the maximum and minimum values. It is less clear what would 
be referred to as the Banks-Zaks twin at lower $\Nf$ values since the 
truncation errors affect both the mass exponent and the gauge parameter 
position. At $\Nf$~$=$~$14$ this value is at $\alpha$~$\approx$~$-$~$4$ with a 
range of $0.2$ but at $\Nf$~$=$~$13$ this range is $2$. Below this we cannot 
identify any with this on more than one approximant. At the upper end of the 
conformal window the value of $\rho_m$ is consistent for the different 
Banks-Zaks pairs across the different approximations. This decreases as we move
towards the lower end. At $\Nf$~$=$~$13$ the values are compatible with values 
of approximately $0.19$ and a secondary value of $0.3$ appears for some of the 
other fixed points. As only the $[3,1]$ approximant has a positive $\rho_m$ as 
$\Nf$ is reduced, further comparison is difficult beyond this point. However 
considering the four loop approximation we do see agreement between fixed point
values at $\Nf$~$=$~$12$ but not really lower $\Nf$.

We close this section by extending this analysis to schemes with a gauge 
parameter dependent $\beta$-function which means the $\mMOM$ scheme as it is
the only one available at five loops. The results are given in Table
\ref{pademmom}. Examining $\Nf$~$=$~$16$ we again see good agreement between 
the two critical point parameters and the quark mass exponent calculated at the
fixed points for the different Pad\'{e} approximants. The case $\Nf$~$=$~$14$ 
provides us with the first zero pole pair fixed point which has a close 
analogue in the set of fixed points in the canonical perturbative expansion. 
However at $\Nf$~$=$~$10$ this is no longer the case with a range of values 
from $0.056$ for the $[3,1]$ zp case to $0.024$ for the $[1,3]$ fixed point 
with the $[2,2]$ approximant best matching the fixed point from the regular 
expansion. At $\Nf$~$=$~$11$ the critical coupling constant value for the 
$[3,1]$ zp approximant is congruous with that of the Banks-Zaks one as well as 
showing agreement with the mass exponent suggesting that the zp fixed points 
should be taken into consideration. Examining the range of gauge parameter 
values for the Banks-Zaks twin we find it is $3\times10^{-6}$ when 
$\Nf$~$=$~$16$, $0.004$ at $\Nf$~$=$~$13$ but $2$ at $\Nf$~$=$~$10$ if we 
include the $\alpha$~$=$~$-$~$1.6$ fixed point for the $[1,3]$ approximant. In 
this case the critical coupling constant has a difference of $0.01$ which is at
the same order as the coupling constant itself. For the quark mass exponent of 
the Banks-Zaks fixed points we see again a decreasing degree of consistency as 
$\Nf$ decreases. For $\Nf$~$=$~$14$ where we find a zp there is still a good 
compatibility between this value and those calculated for the Banks-Zaks fixed
point of the other approximants. Even at $\Nf$~$=$~$12$ we see good agreement 
with a range of values from $0.306$ to $0.284$.

\sect{Discussion.}

We have provided a comprehensive analysis of the fixed point structure of the
renormalization group functions in QCD where the gauge parameter is treated as
an additional coupling constant. While there have been earlier studies in this
respect, \cite{41}, which revealed interesting structure given the progress in
determining the renormalization group functions to much higher order in recent
years it was important to revisit that work. This was not just in the context
of one scheme and gauge. Instead our analysis not only included the canonical
linear covariant gauge fixing as well as the $\MSbar$ and MOM schemes of
\cite{35,36} we incorporated the $\RI$ and $\mMOM$ schemes in addition to two
nonlinear covariant gauges. Although five loop results are not yet available
for the latter two gauges it was important to study the extent to which 
critical exponent estimates of observables were not only scheme independent 
but also independent of the choice of gauge. This is because when solving for 
the fixed points numerically, as with any perturbative truncation, there will 
always be a degree of tolerance in final estimates. Within the conformal window
at the centre of our investigation one of the main observations was that in the
regime where the perturbative approximation is reliable good agreement with 
exponent estimates across schemes and gauges emerged. Indeed where five loop 
results were available it was clear that these higher orders increased the 
range of perturbative applicability. In addition at five loops the absence of 
a solution for the Banks-Zaks fixed point in the $\MSbar$ scheme for 
$\Nf$~$\leq$~$12$ was shown to be a scheme artefact. Instead solving for the 
zeros of the $\beta$-functions in the $\mMOM$ scheme produced solutions at and 
below $\Nf$~$=$~$12$ at five loops. With data from these two schemes as well as
the $\RI$ one Pad\'{e} approximants were used to study if convergence could be 
improved away from the top end of the conformal window. 

One observation of \cite{41} was the existence of a fixed point in the interior
of the $(a,\alpha)$ plane away from either axis in addition to the Gaussian and
Banks-Zaks one. As its critical coupling in the linear covariant gauge was the 
same as the latter fixed point we referred to this as its mirror or twin fixed 
point. Moreover we verified that it was infrared stable unlike the Banks-Zaks 
one which is a saddle point on the $(a,\alpha)$ plane. To answer the obvious 
question as to whether this was an artefact of the linear covariant gauge we 
searched for a similar infrared stable fixed point in the two non-linear 
covariant gauges. Intriguingly such a similar critical point is present in both
cases. While this is qualified by noting that the critical gauge parameter 
value differ in each of the three gauges since this is not an observable what 
did emerge was the consistency of the estimate for the critical exponent 
$\rho_m$ in the region of perturbative reliability at this infrared stable
point. Moreover in the MAG the Banks-Zaks fixed point is absent in the 
conventional sense since there is no point on the coupling constant axis 
although there is a saddle point in its neighbourhood. Given that the existence
of such a stable infrared critical point in QCD appears to be independent of 
the covariant gauge fixing procedure it reinforces the observation of \cite{41}
that retaining the gauge parameter in studies might assist infrared analyses of
colour confinement. This would be an interesting topic to pursue but would
clearly require a non-perturbative approach. Finally we remark that we have 
completely focused on the $SU(3)$ colour group with quarks in the fundamental 
representation. There is no a priori reason why one should restrict gauge 
theory fixed point analyses to this particular Lie group or matter 
representation. Rather it might be of interest to embed the $SU(3)$ analysis in 
the Standard Model as well as other gauge theories that seek to explore beyond 
that fundamental theory. Indeed in this context it is worth noting that the 
choice of $\alpha$~$=$~$-$~$3$ for the linear covariant gauge fixing in the 
electroweak sector was singled out as a special case in \cite{75}. In 
particular at this value the $Z$ boson is renormalized multiplicatively.

\vspace{1cm}
\noindent
{\bf Acknowledgements.} This work was carried out with the support of the STFC
Consolidated Grant ST/T000988/1 (JAG), an EPSRC Studentship EP/R513271/1 (RHM)
and an STFC Studentship ST/M503629/1 (RMS). For the purpose of open access, the
authors have applied a Creative Commons Attribution (CC-BY) licence to any 
Author Accepted Manuscript version arising. The data representing the full 
fixed point and critical exponent analysis of the work presented here are 
accessible in electronic form from the arXiv ancillary directory associated 
with the article.

{\begin{figure}[ht]
\includegraphics[width=7.80cm,height=6cm]{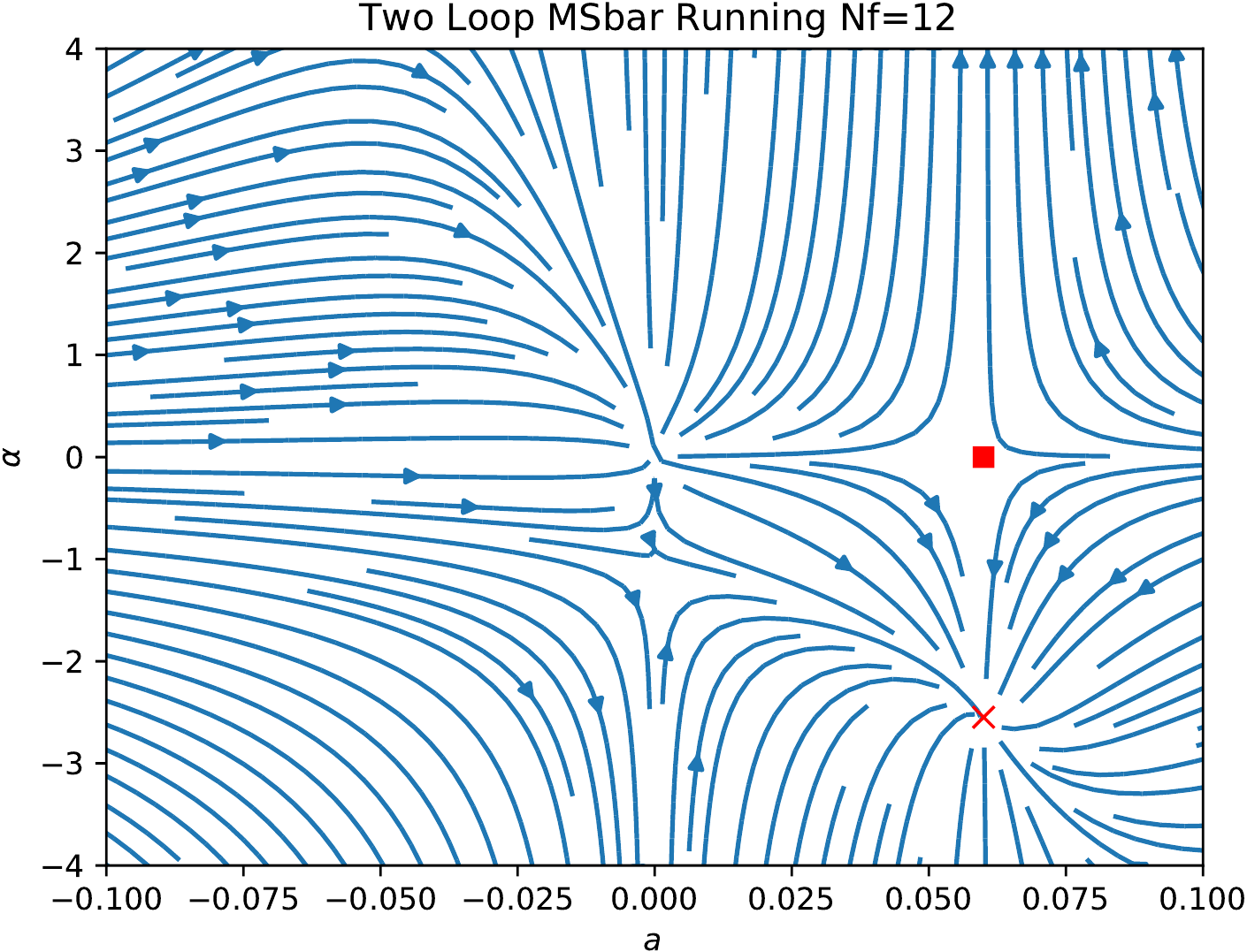}
\quad
\quad
\includegraphics[width=7.80cm,height=6cm]{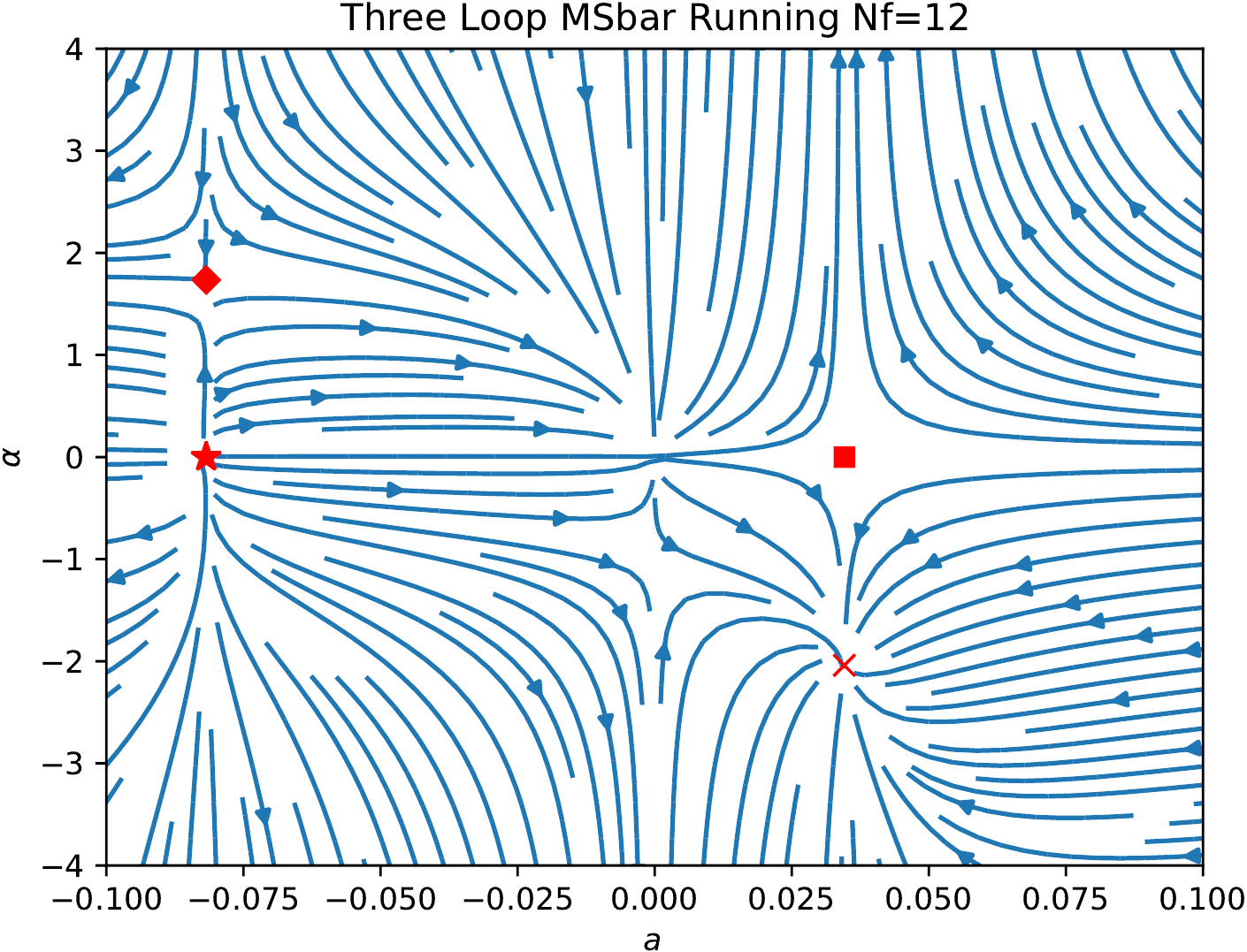}
\\ \\ \\ \\
\includegraphics[width=7.80cm,height=6cm]{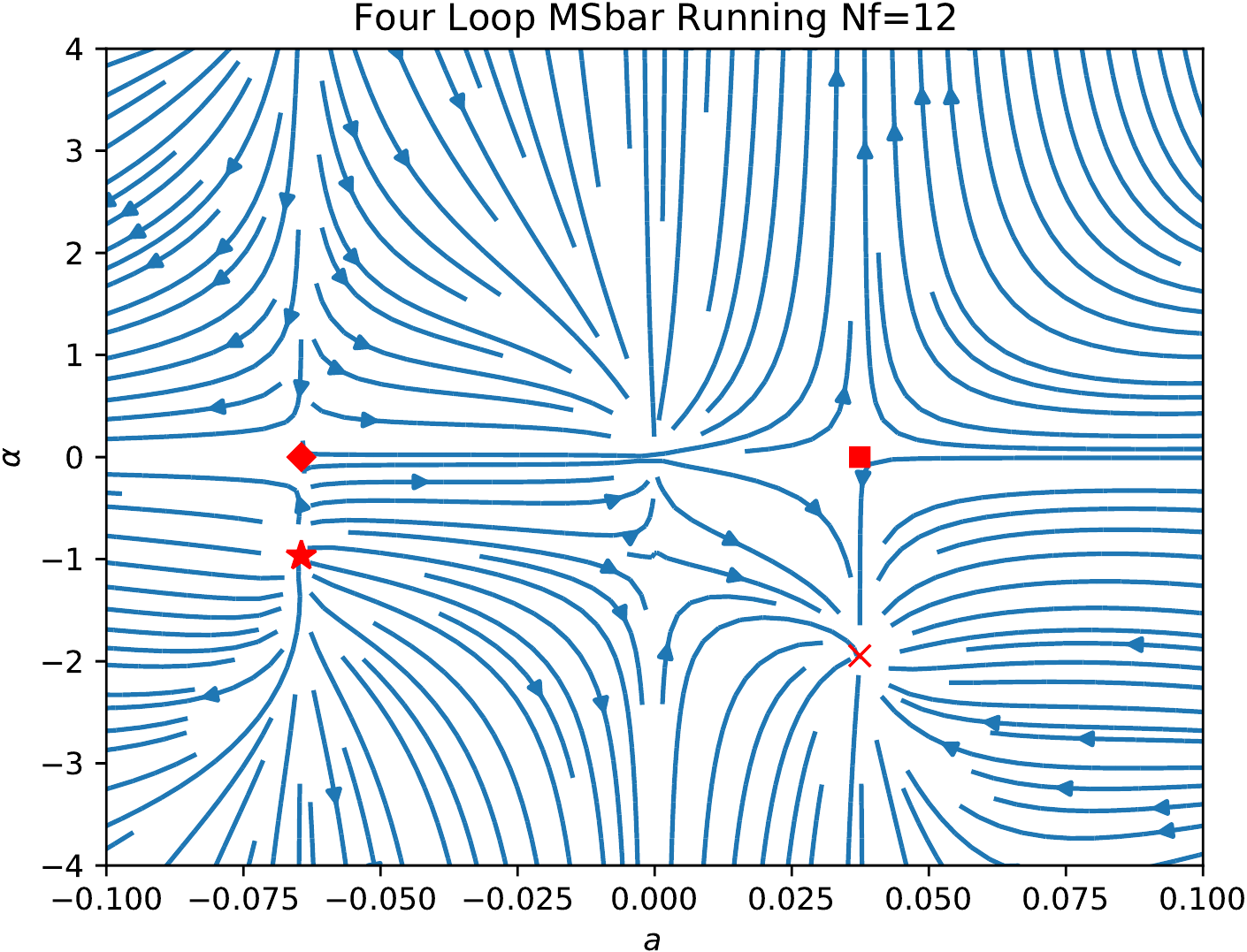}
\quad
\quad
\includegraphics[width=7.80cm,height=6cm]{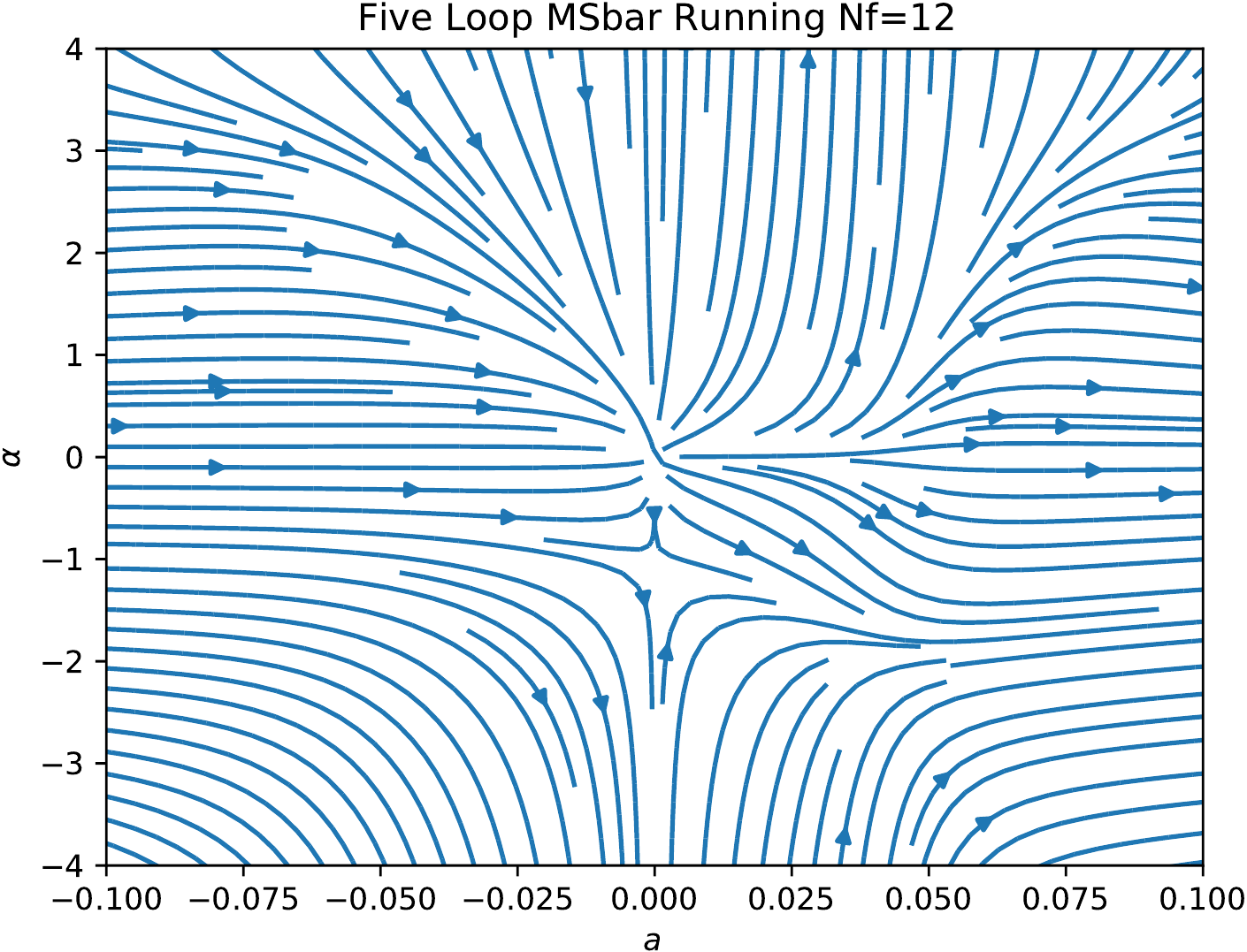}
\caption{Flow plane for the $\MSbar$ scheme $SU(3)$ linear gauge at two, three,
four and five loops when $\Nf$~$=$~$12$}
\label{flow3msn12}
\end{figure}}

\clearpage 

{\begin{figure}[ht]
\includegraphics[width=7.80cm,height=6cm]{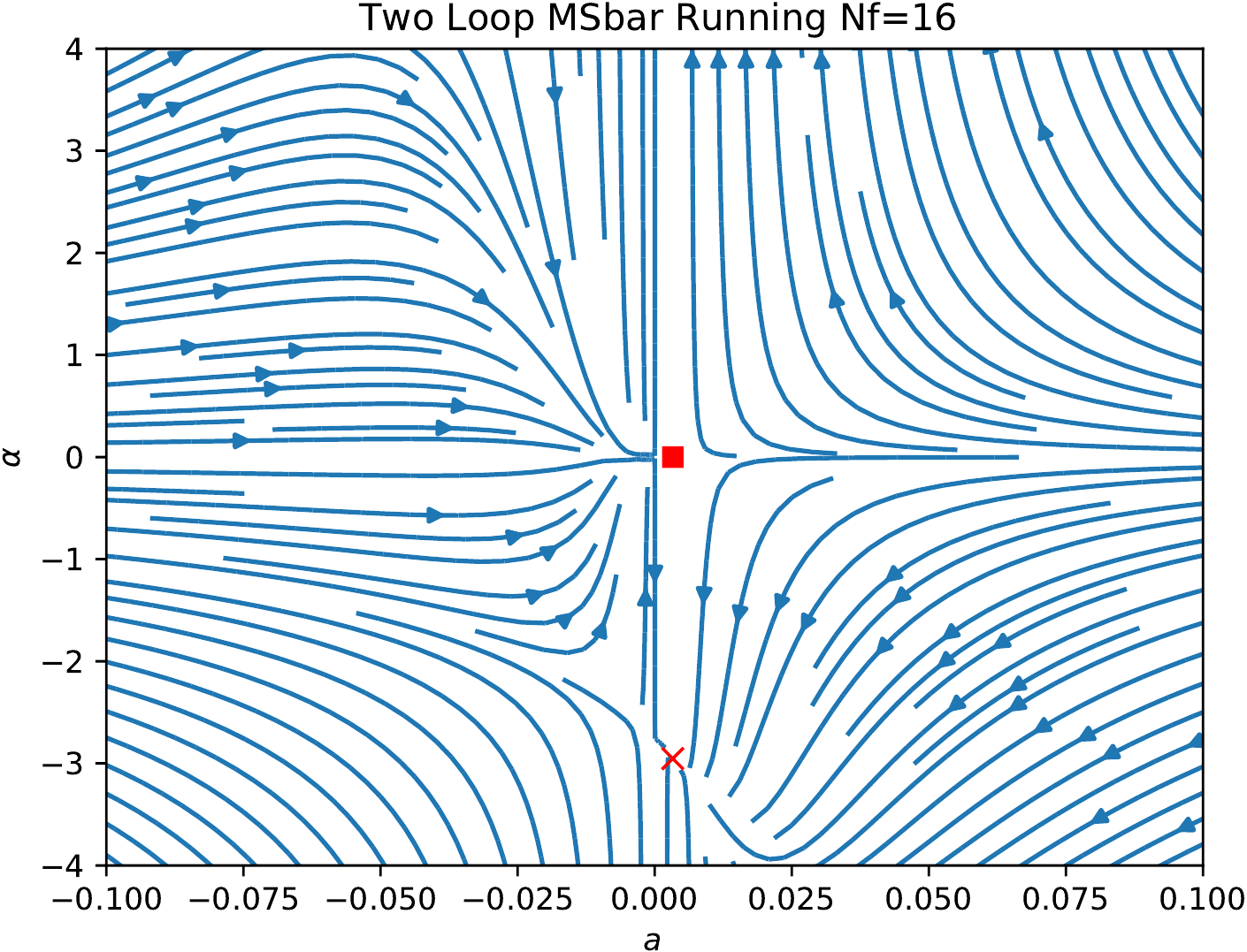}
\quad
\quad
\includegraphics[width=7.80cm,height=6cm]{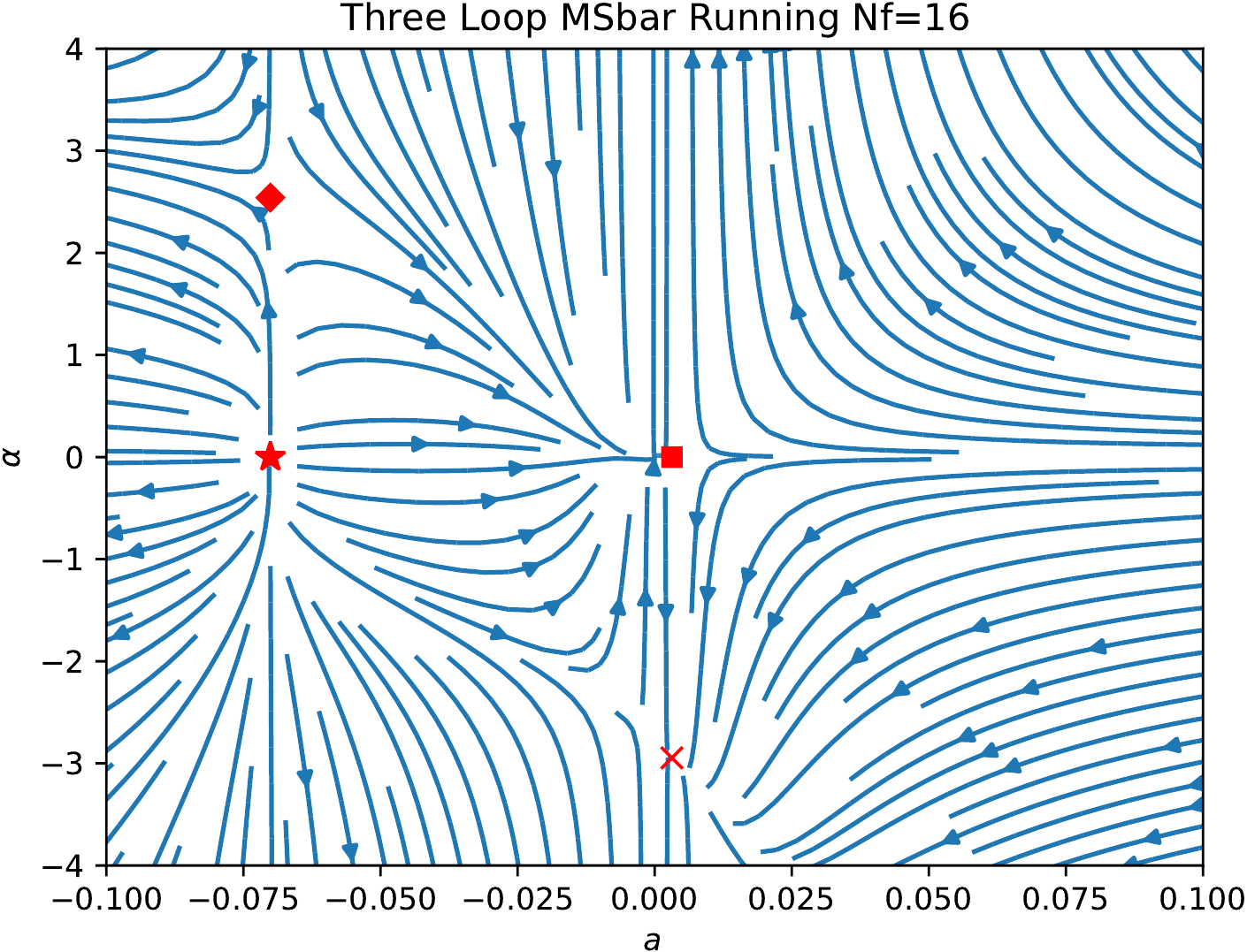}
\\ \\ \\ \\
\includegraphics[width=7.80cm,height=6cm]{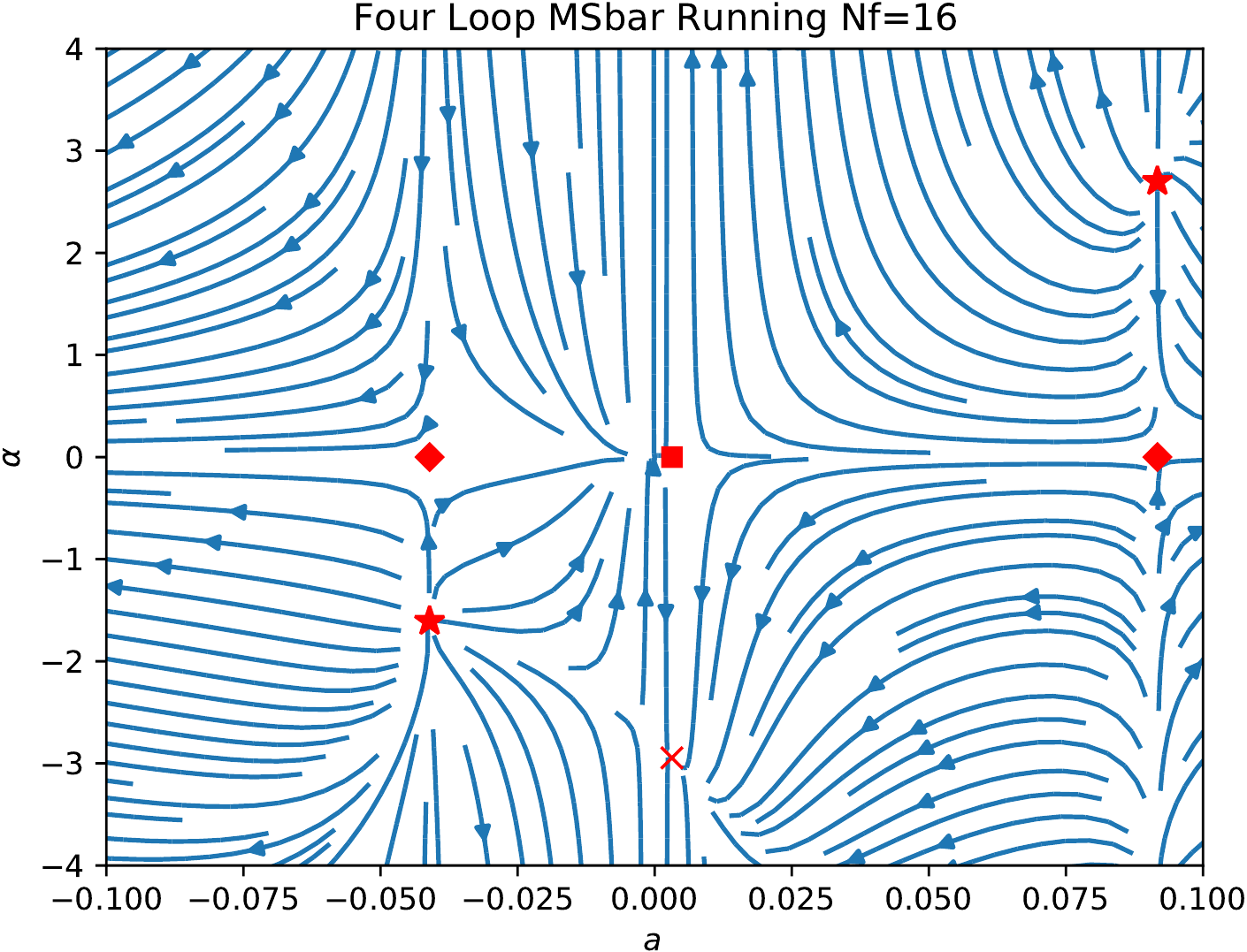}
\quad
\quad
\includegraphics[width=7.80cm,height=6cm]{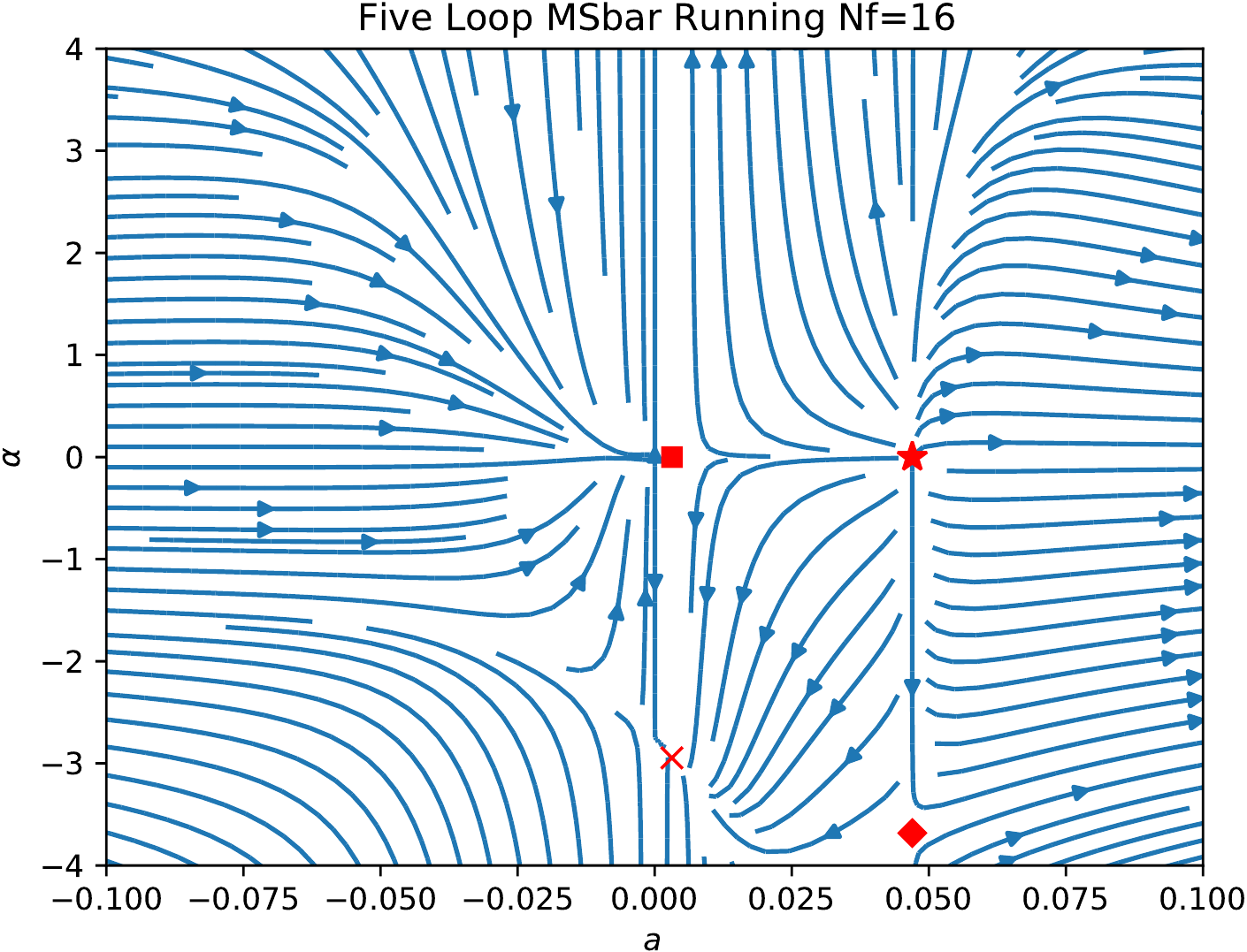}
\caption{Flow plane for the $\MSbar$ scheme $SU(3)$ linear gauge at two, three,
four and five loops when $\Nf$~$=$~$16$}
\label{flow3msn16}
\end{figure}}

{\begin{figure}[ht]
\includegraphics[width=7.80cm,height=6cm]{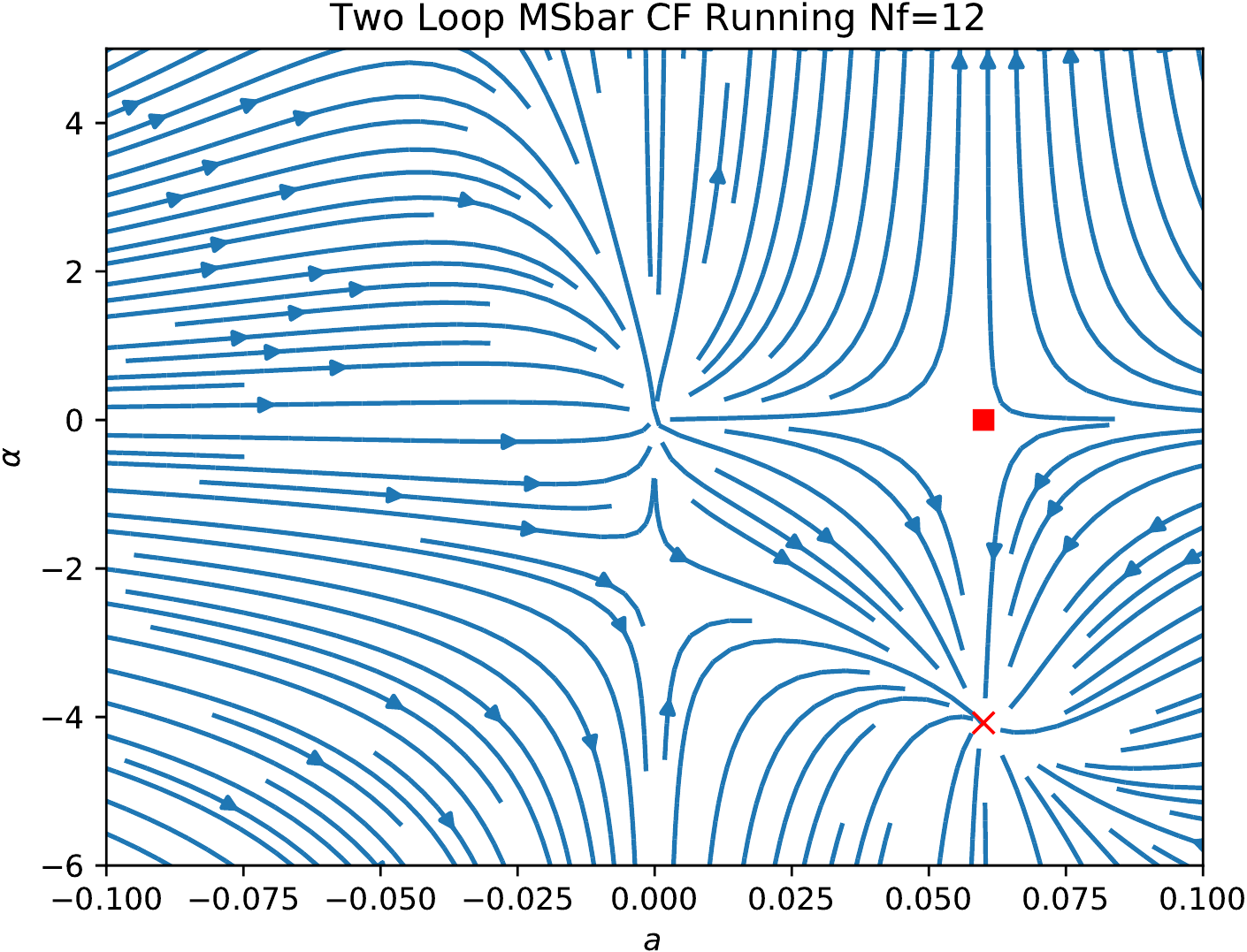}
\quad
\quad
\includegraphics[width=7.80cm,height=6cm]{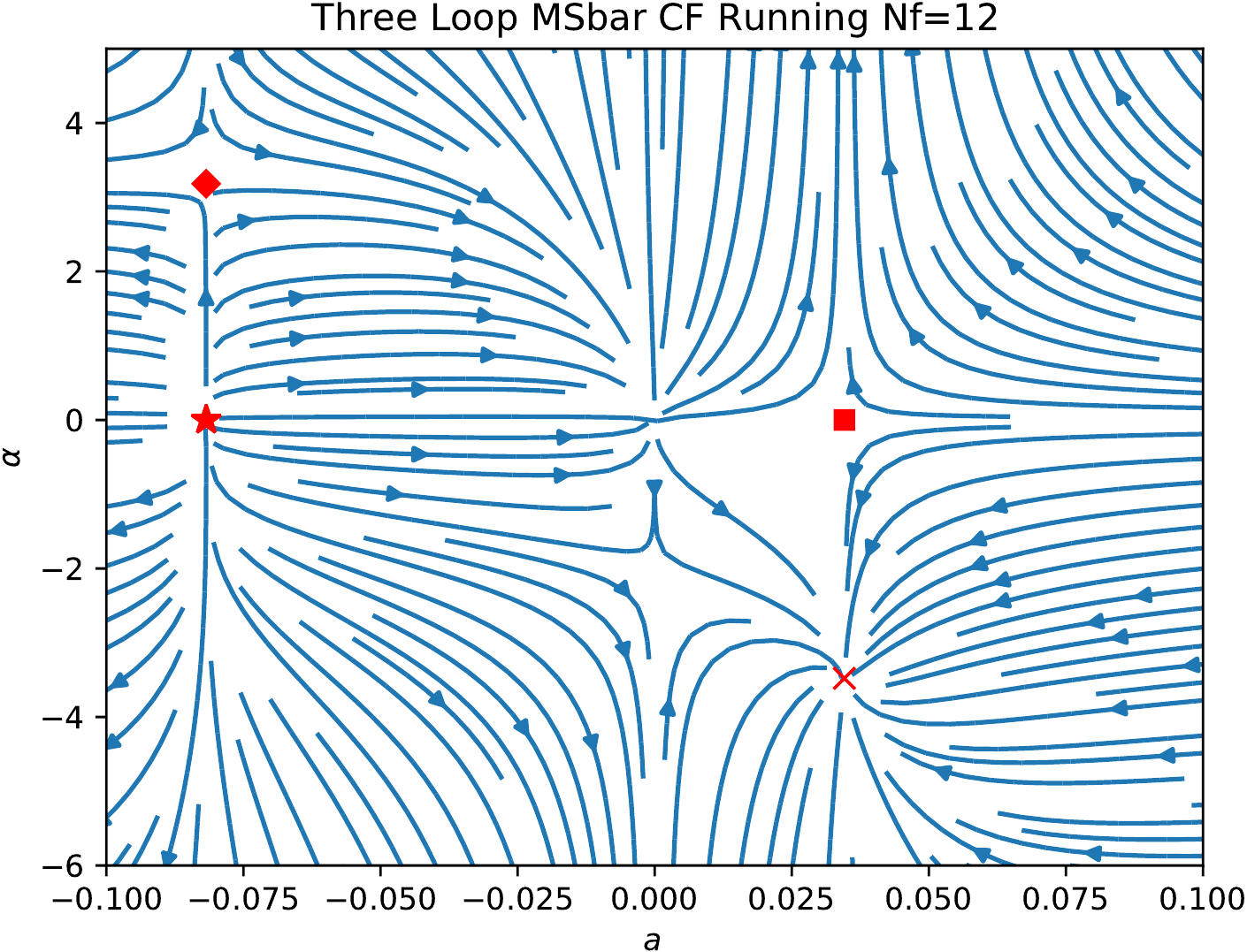}
\caption{Flow plane for the $\MSbar$ scheme $SU(3)$ Curci-Ferrari gauge at two 
and three loops 
when $\Nf$~$=$~$12$}
\label{flow3mscfn12}
\end{figure}}

{\begin{figure}[ht]
\includegraphics[width=7.80cm,height=6cm]{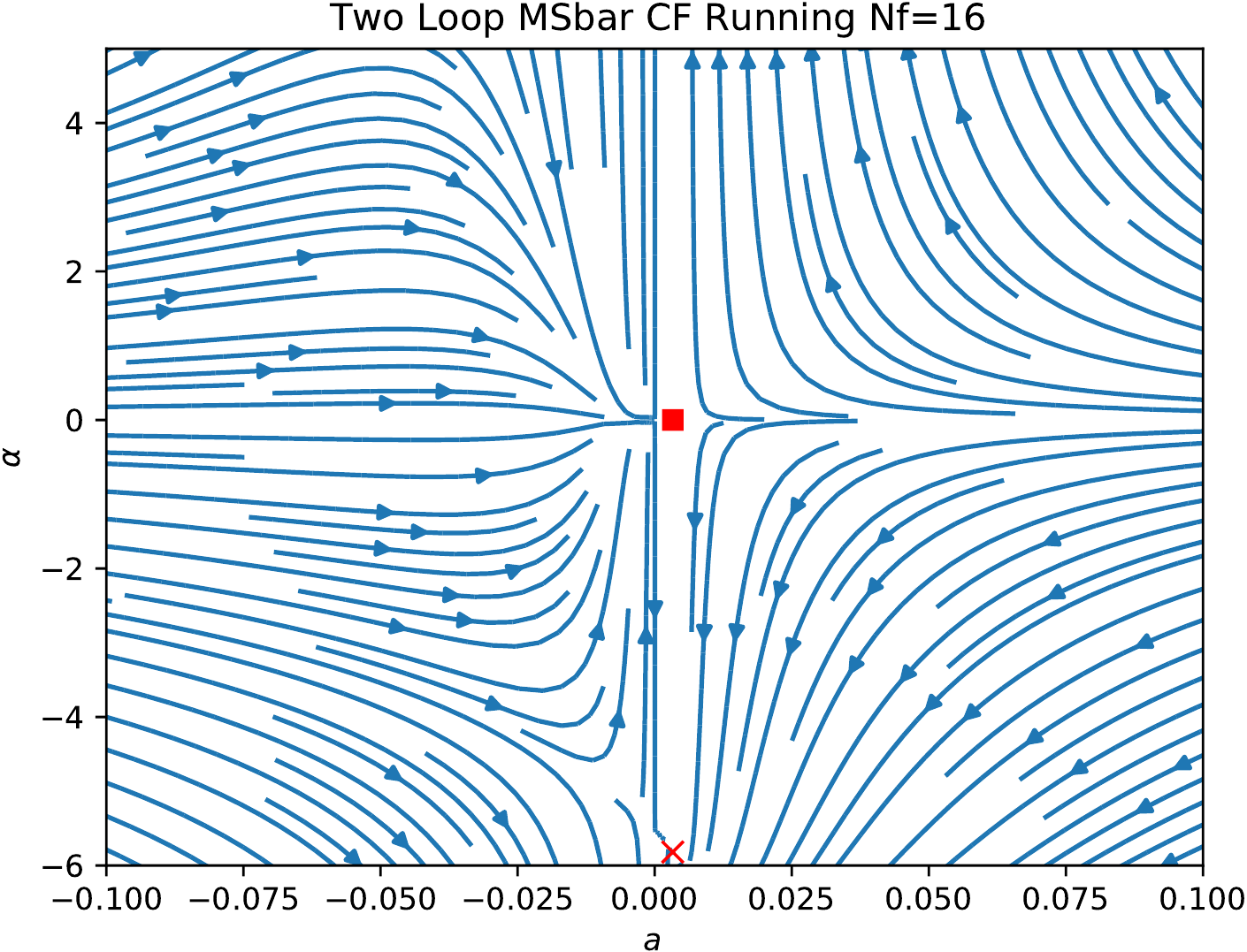}
\quad
\quad
\includegraphics[width=7.80cm,height=6cm]{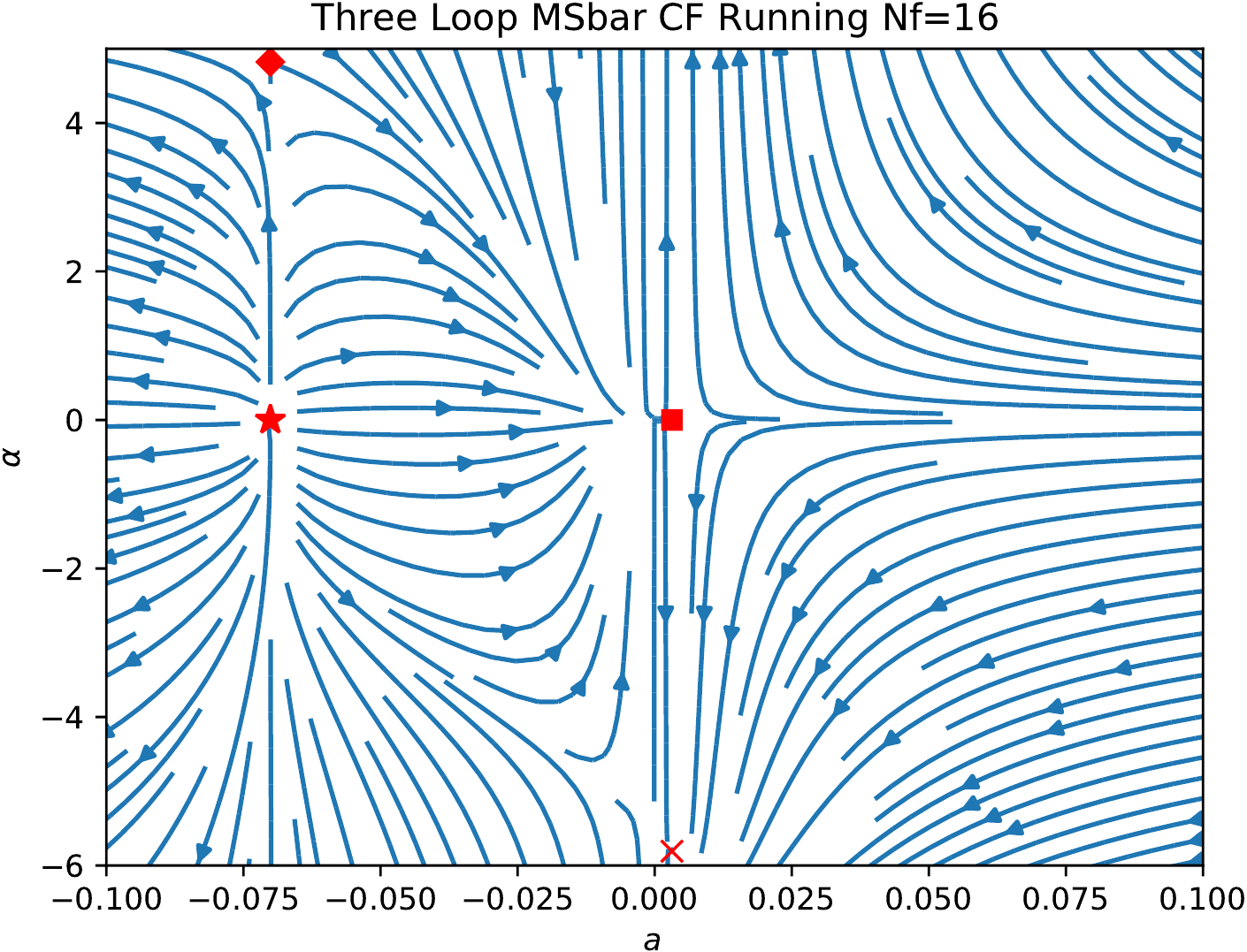}
\caption{Flow plane for the $\MSbar$ scheme $SU(3)$ Curci-Ferrari gauge at two
and three loops 
when $\Nf$~$=$~$16$}
\label{flow3mscfn16}
\end{figure}}

{\begin{figure}[ht]
\includegraphics[width=7.80cm,height=6cm]{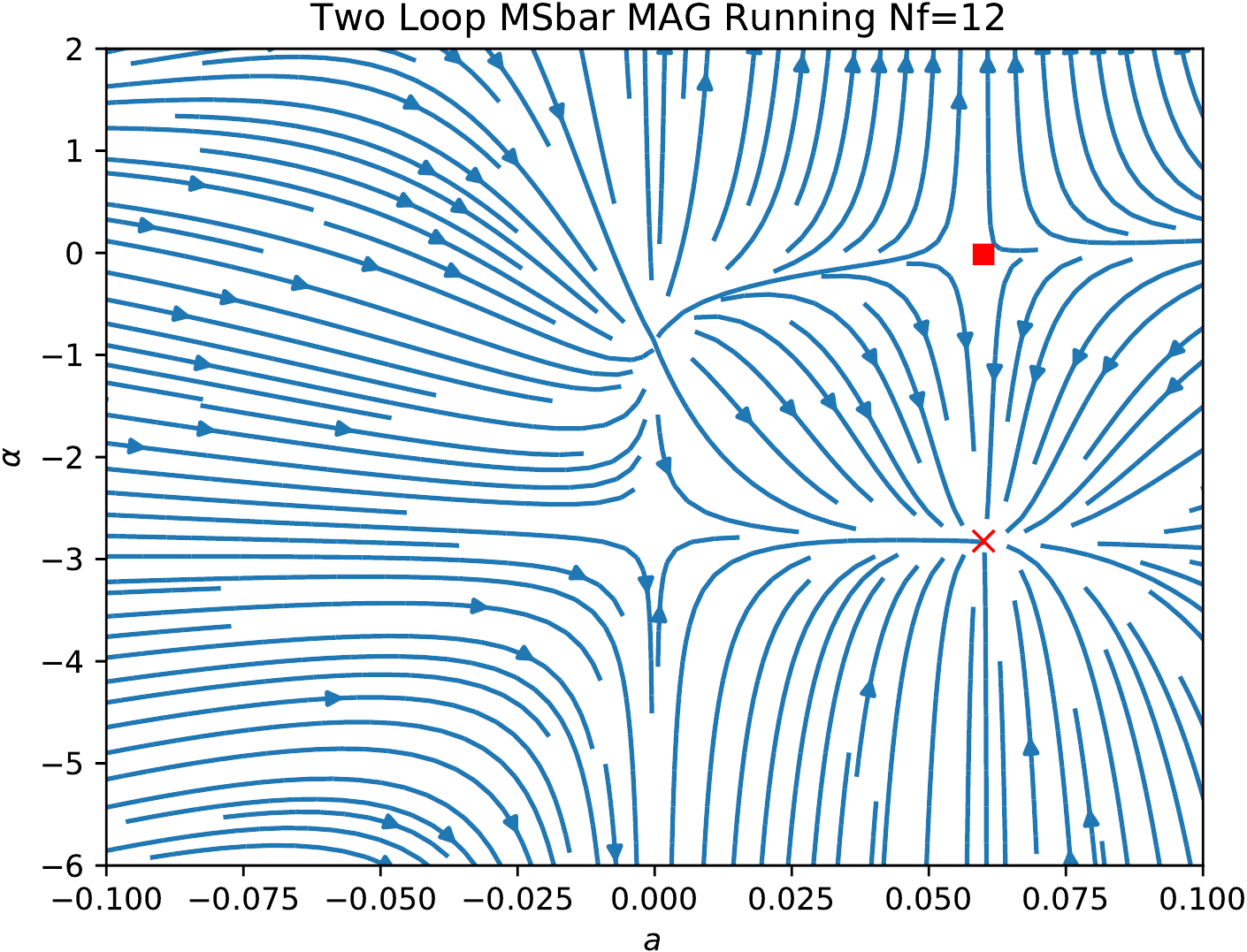}
\quad
\quad
\includegraphics[width=7.80cm,height=6cm]{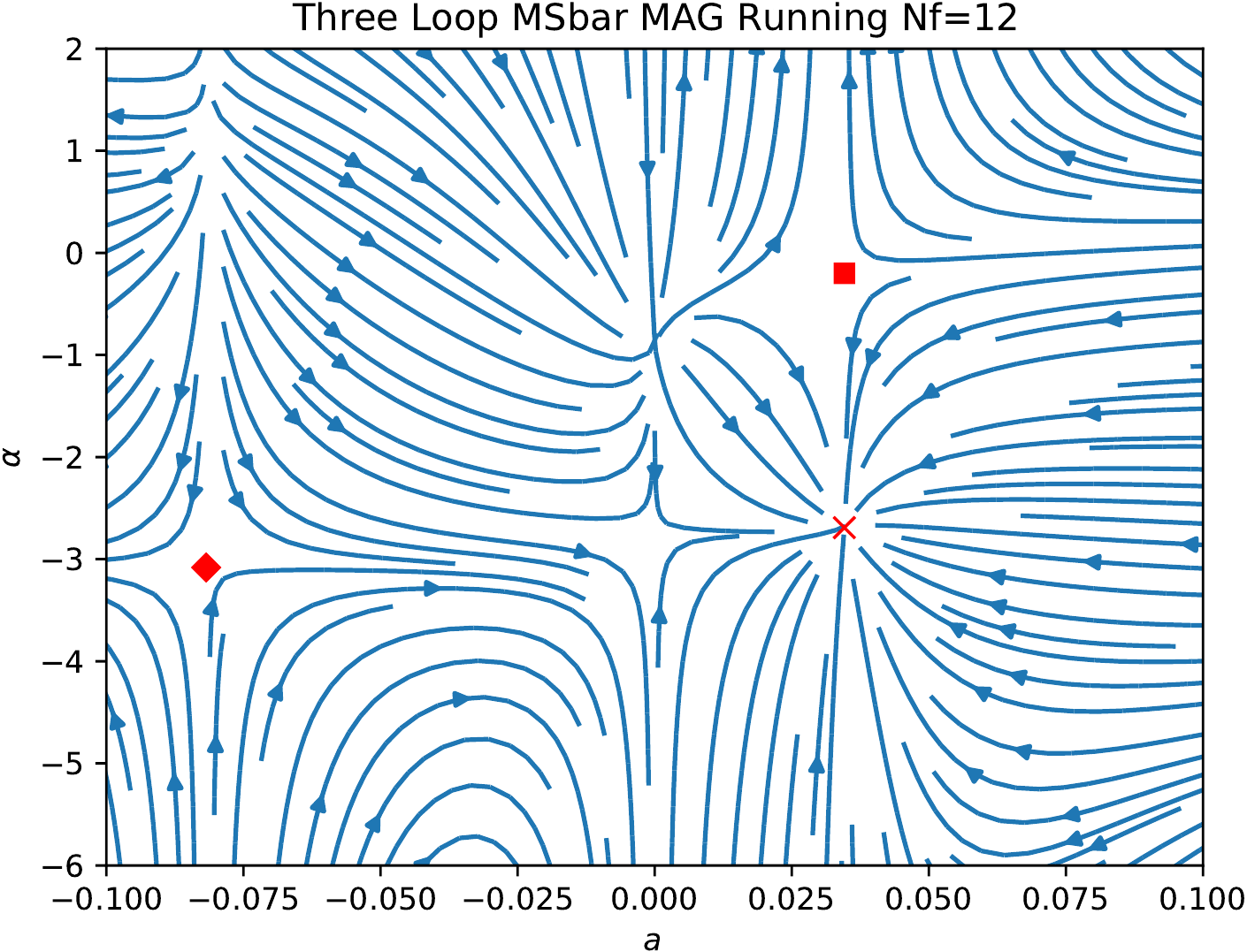}
\caption{Flow plane for the $\MSbar$ scheme $SU(3)$ MAG at two and three loops 
when $\Nf$~$=$~$12$}
\label{flow3msmagn12}
\end{figure}}

{\begin{figure}[ht]
\includegraphics[width=7.80cm,height=6cm]{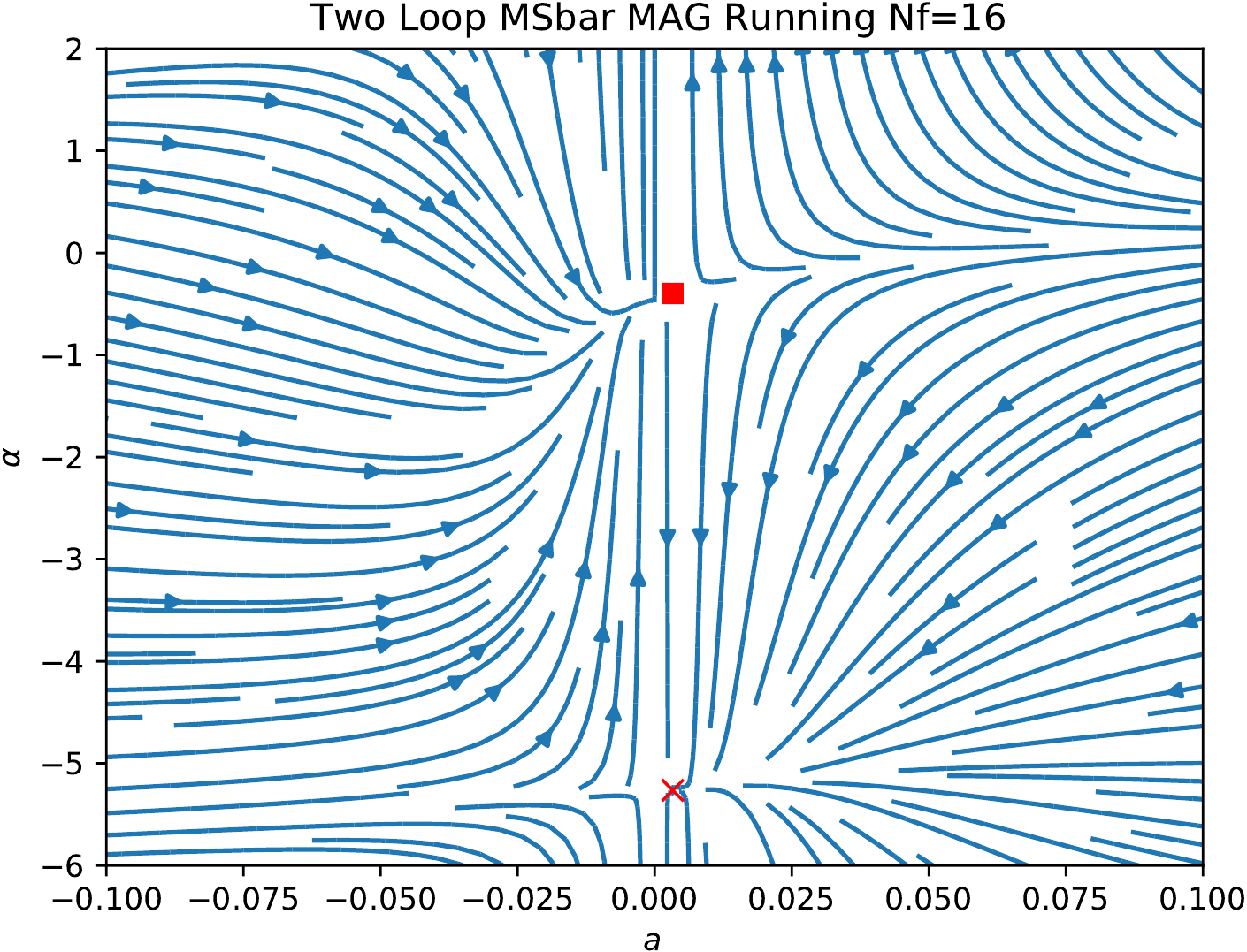}
\quad
\quad
\includegraphics[width=7.80cm,height=6cm]{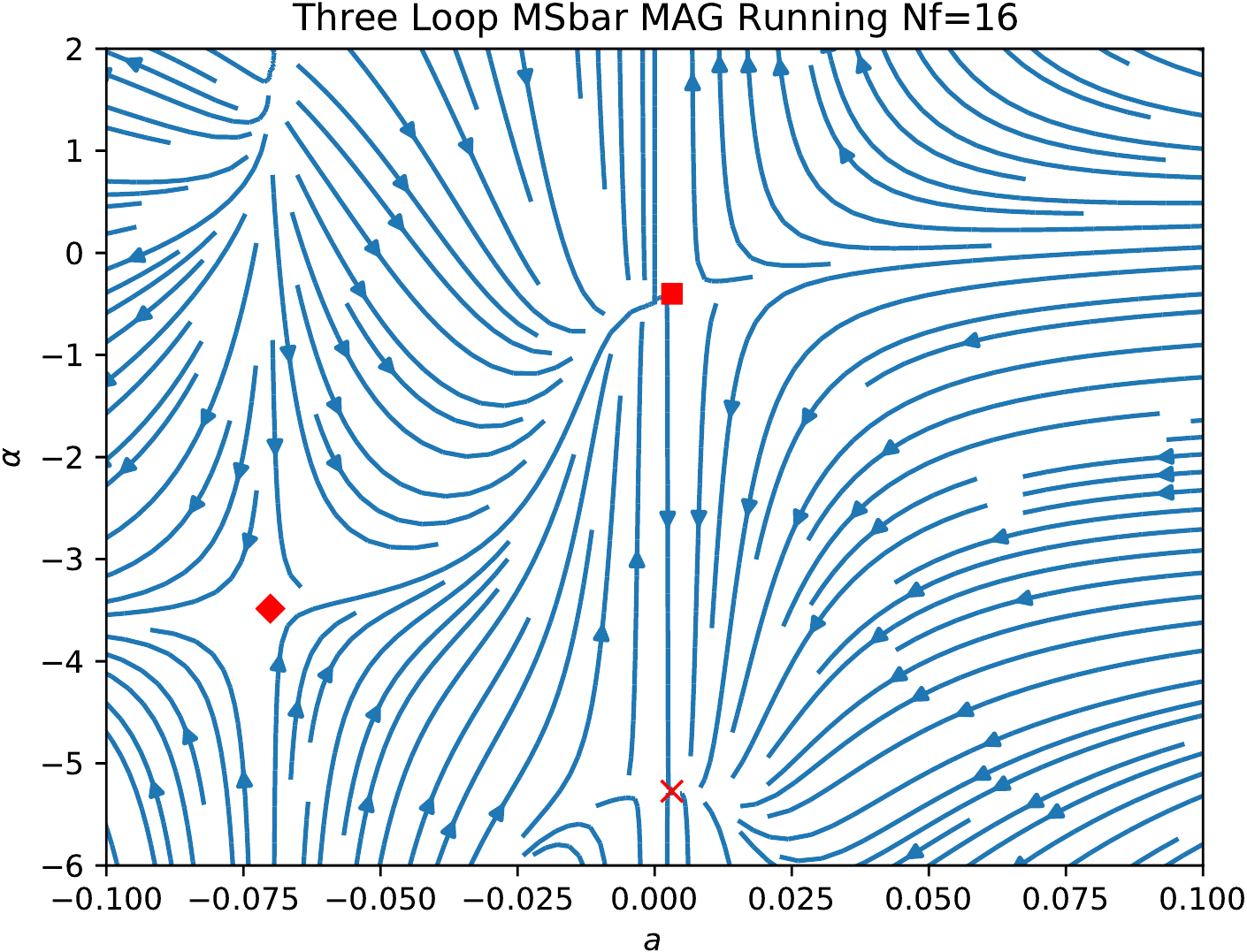}
\caption{Flow plane for the $\MSbar$ scheme $SU(3)$ MAG at two and three loops 
when $\Nf$~$=$~$16$}
\label{flow3msmagn16}
\end{figure}}

{\begin{figure}[ht]
\includegraphics[width=7.80cm,height=6cm]{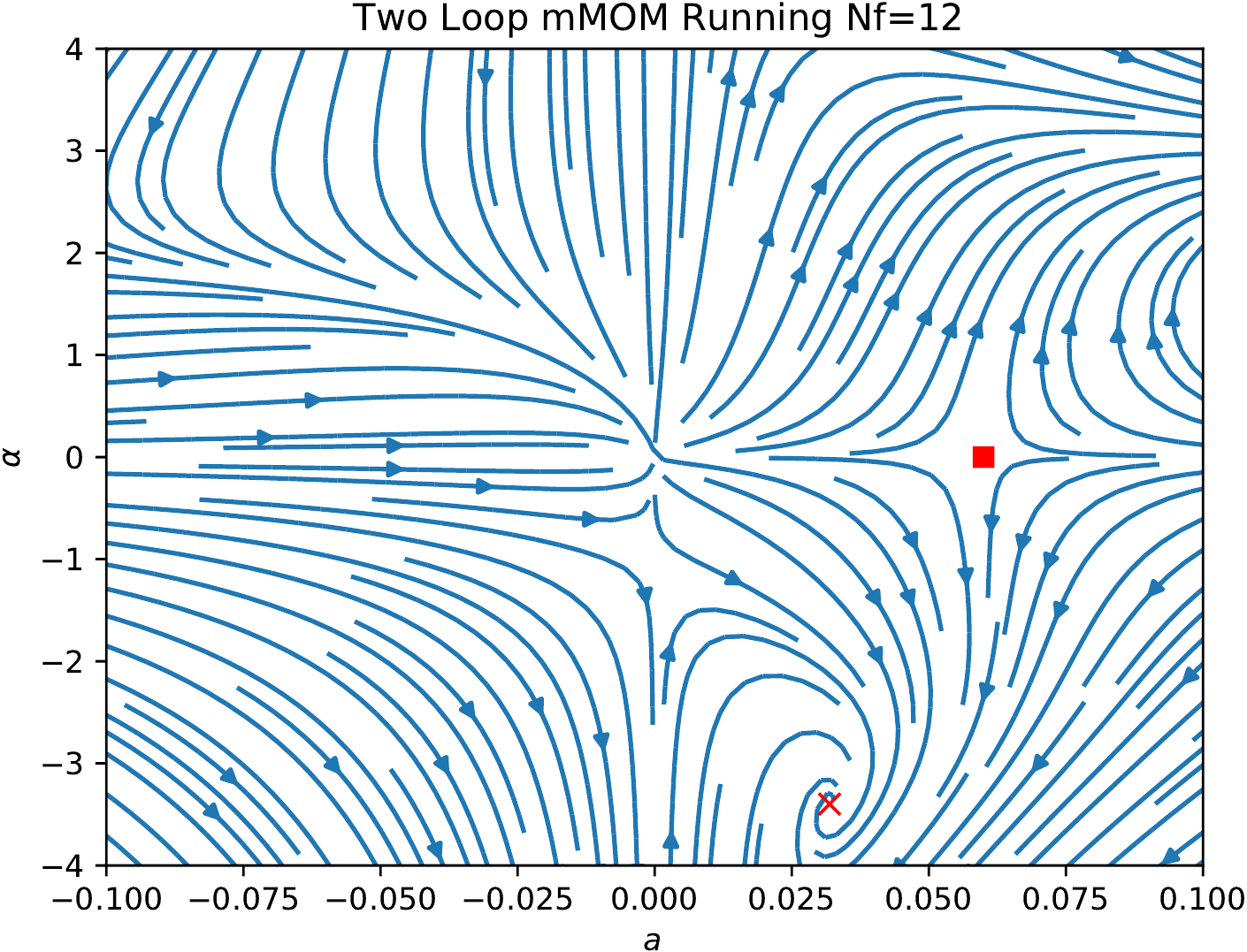}
\quad
\quad
\includegraphics[width=7.80cm,height=6cm]{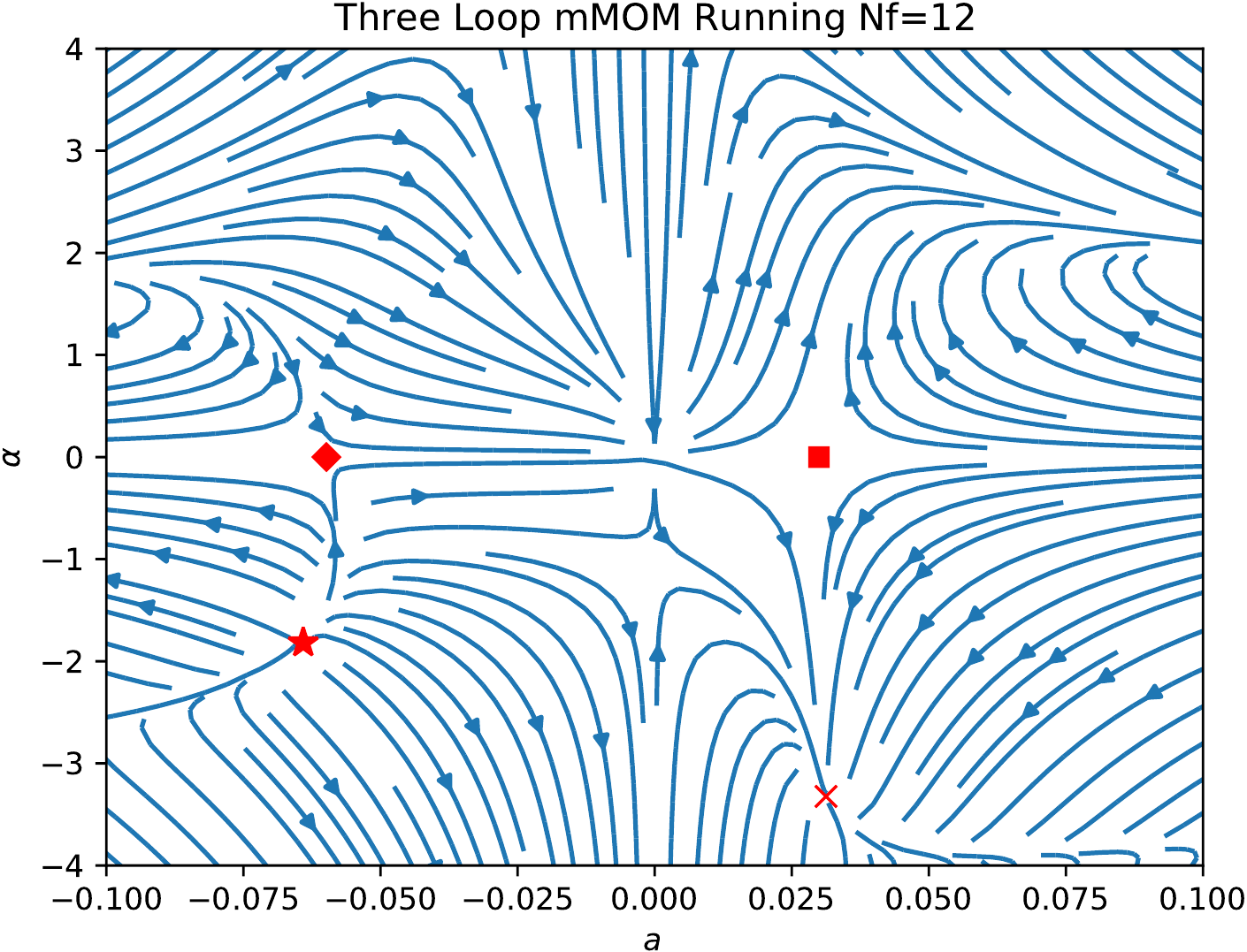}
\\ \\ \\ \\
\includegraphics[width=7.80cm,height=6cm]{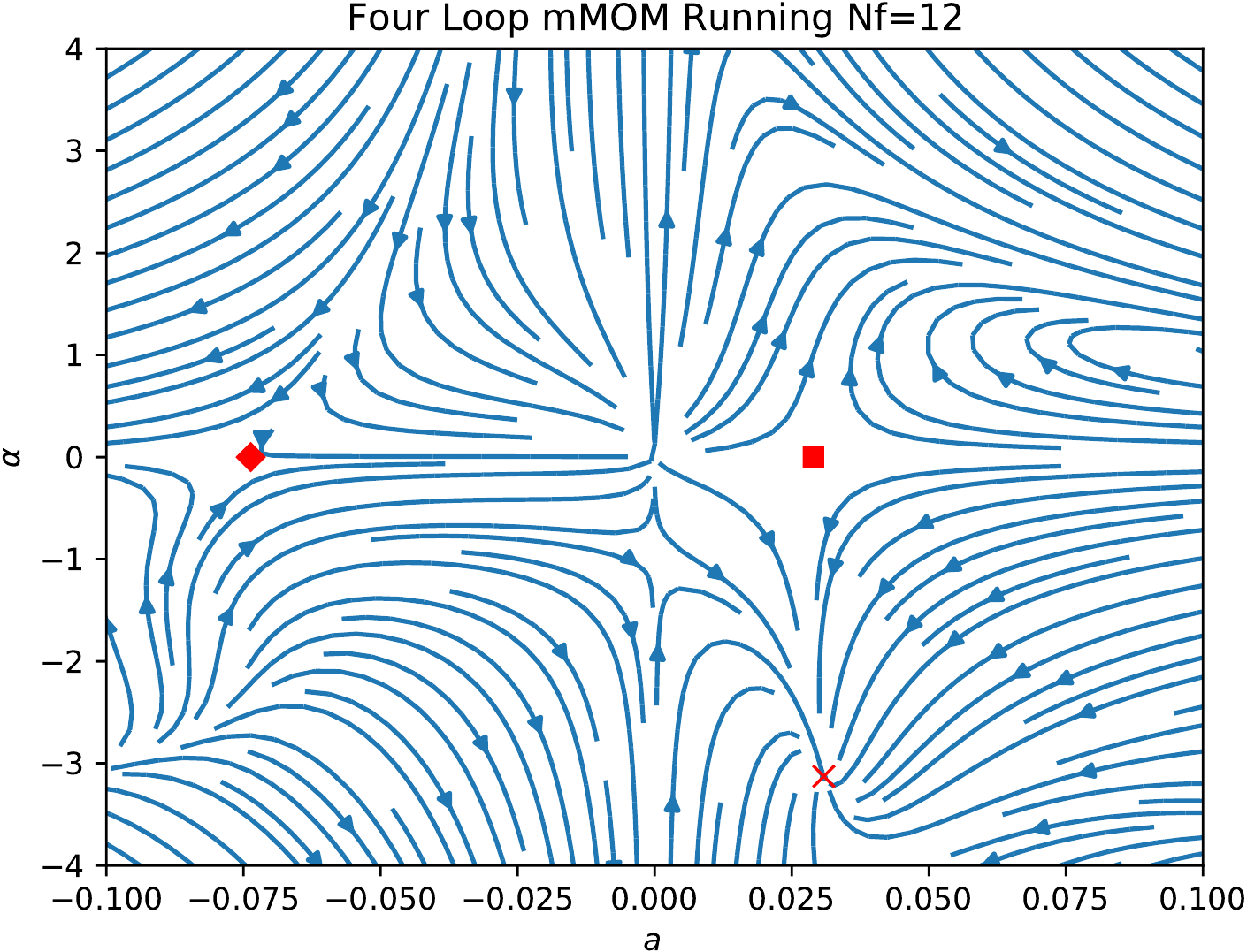}
\quad
\quad
\includegraphics[width=7.80cm,height=6cm]{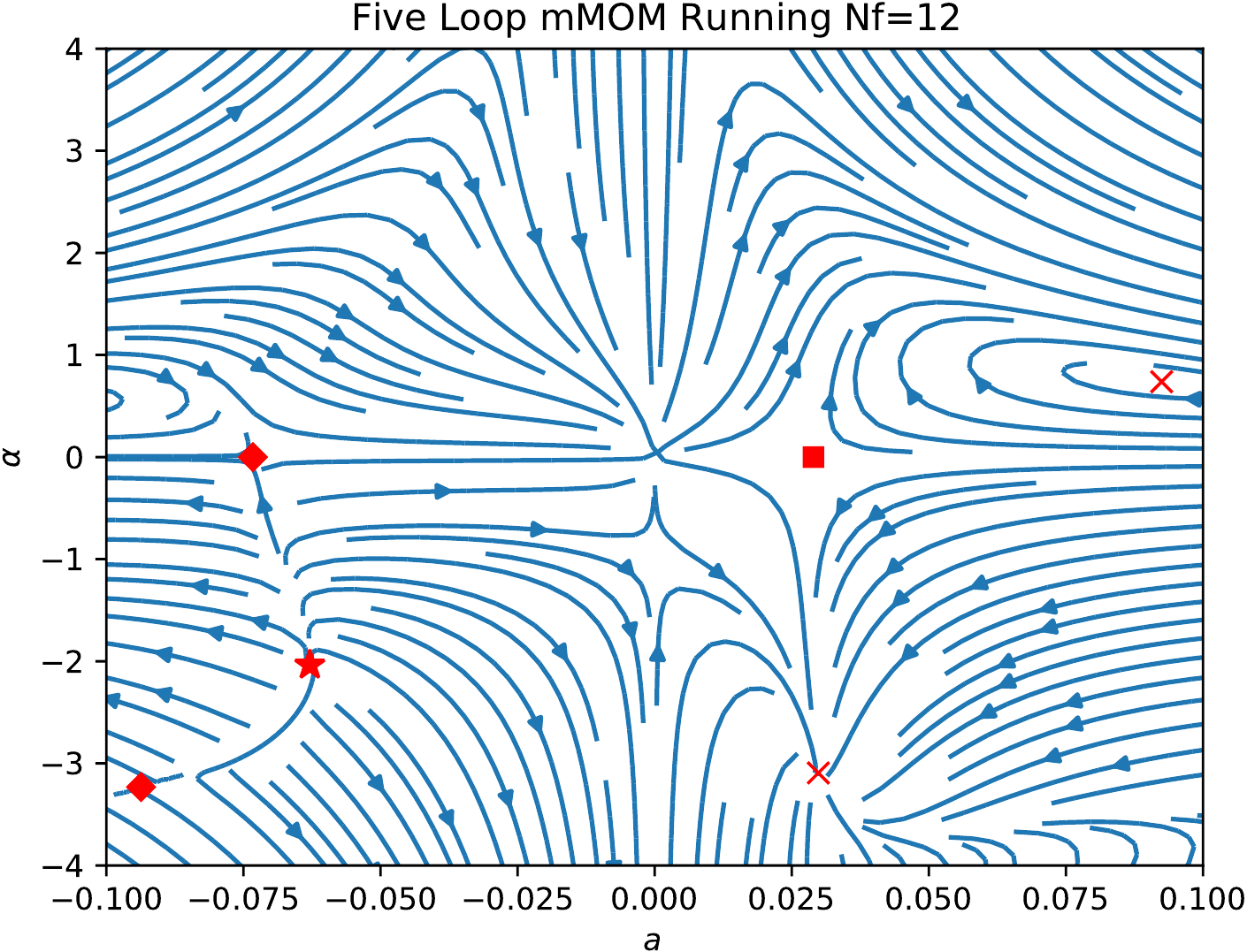}
\caption{Flow plane for the $\mMOM$ scheme $SU(3)$ linear gauge at two, three,
four and five loops when $\Nf$~$=$~$12$}
\label{flow3mmn12}
\end{figure}}

\clearpage 

{\begin{figure}[ht]
\includegraphics[width=7.80cm,height=6cm]{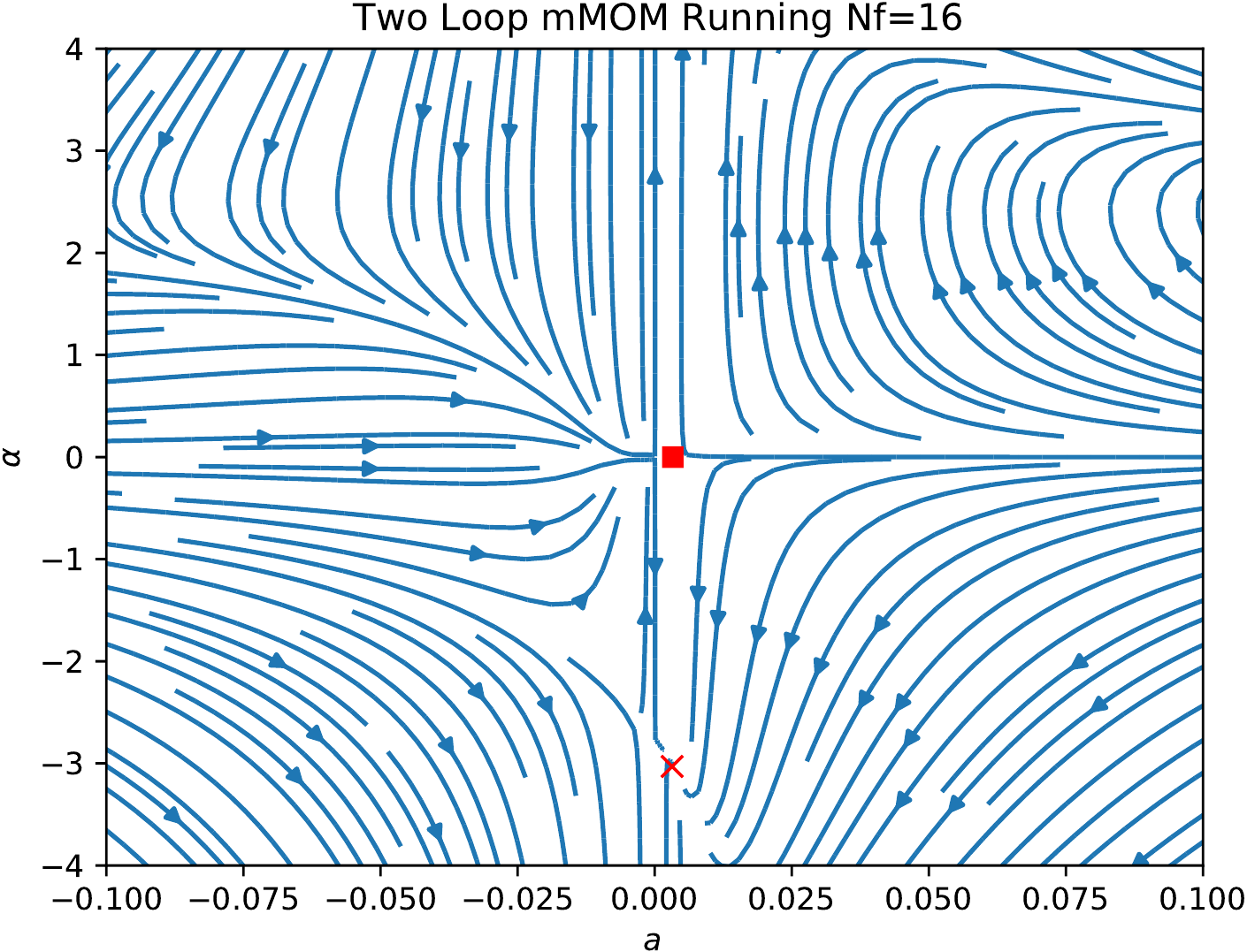}
\quad
\quad
\includegraphics[width=7.80cm,height=6cm]{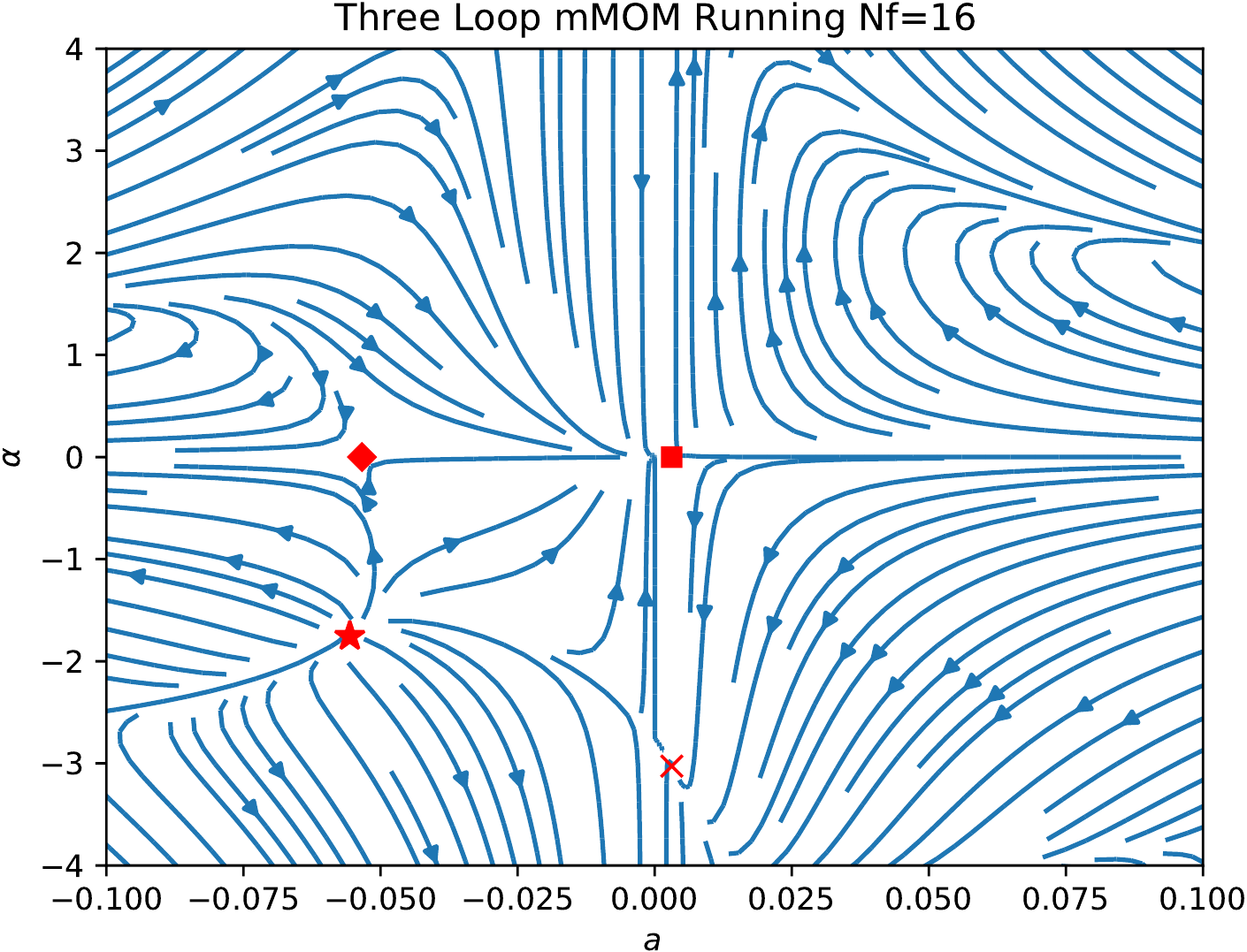}
\\ \\ \\ \\
\includegraphics[width=7.80cm,height=6cm]{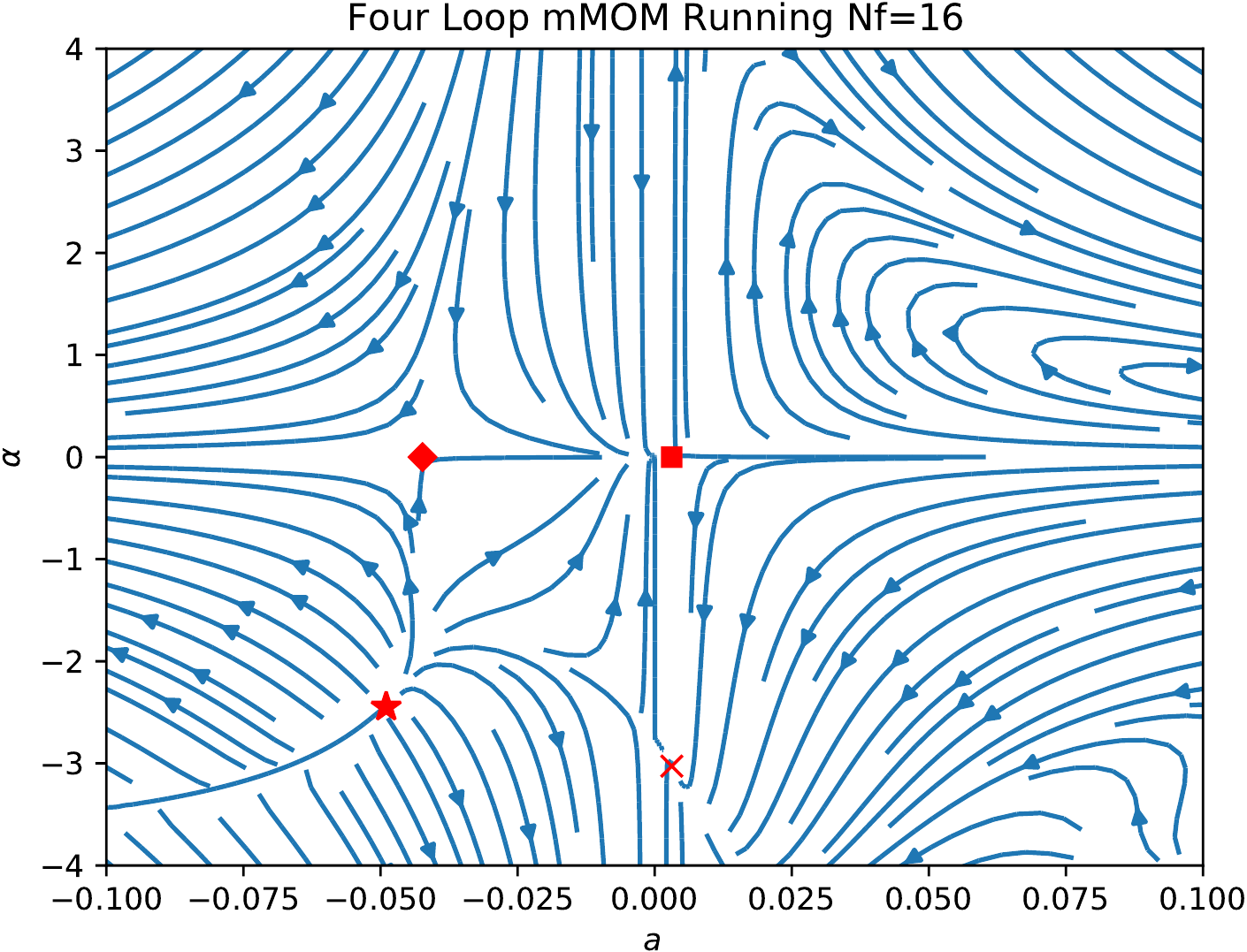}
\quad
\quad
\includegraphics[width=7.80cm,height=6cm]{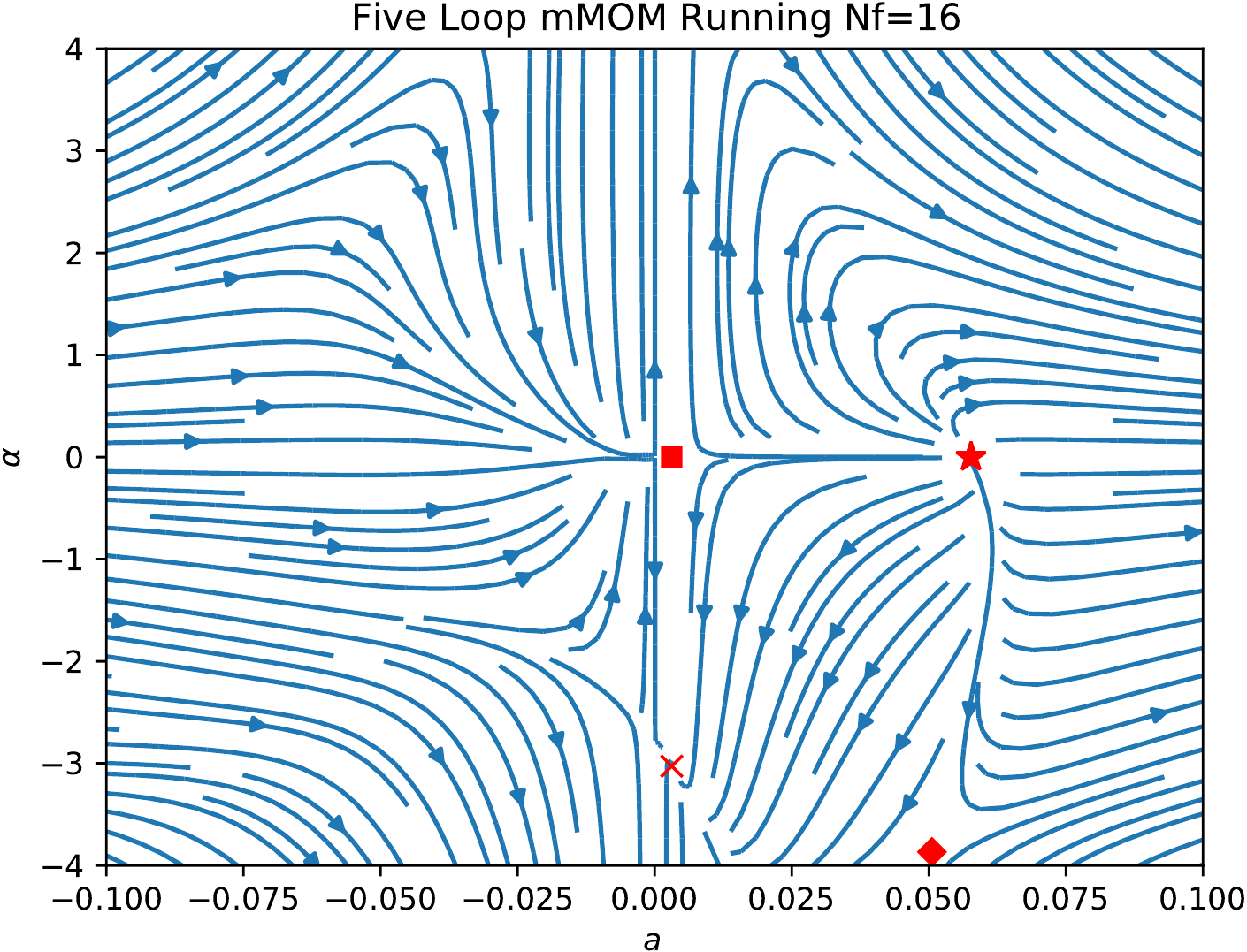}
\caption{Flow planes for the $\mMOM$ scheme $SU(3)$ linear gauge at two, three,
four and five loops when $\Nf$~$=$~$16$}
\label{flow3mmn16}
\end{figure}}

{\begin{figure}[ht]
\includegraphics[width=7.80cm,height=6cm]{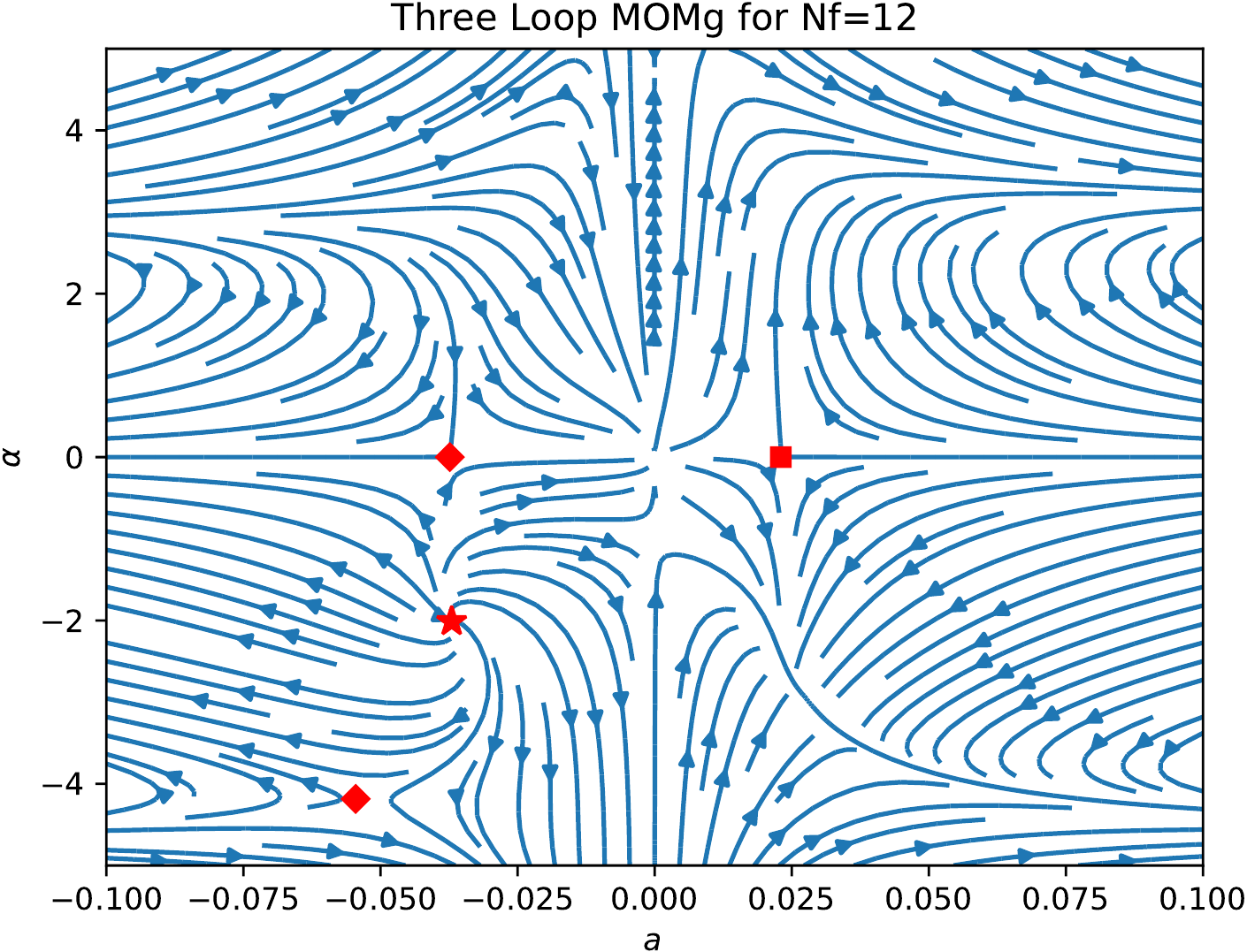}
\quad
\quad
\includegraphics[width=7.80cm,height=6cm]{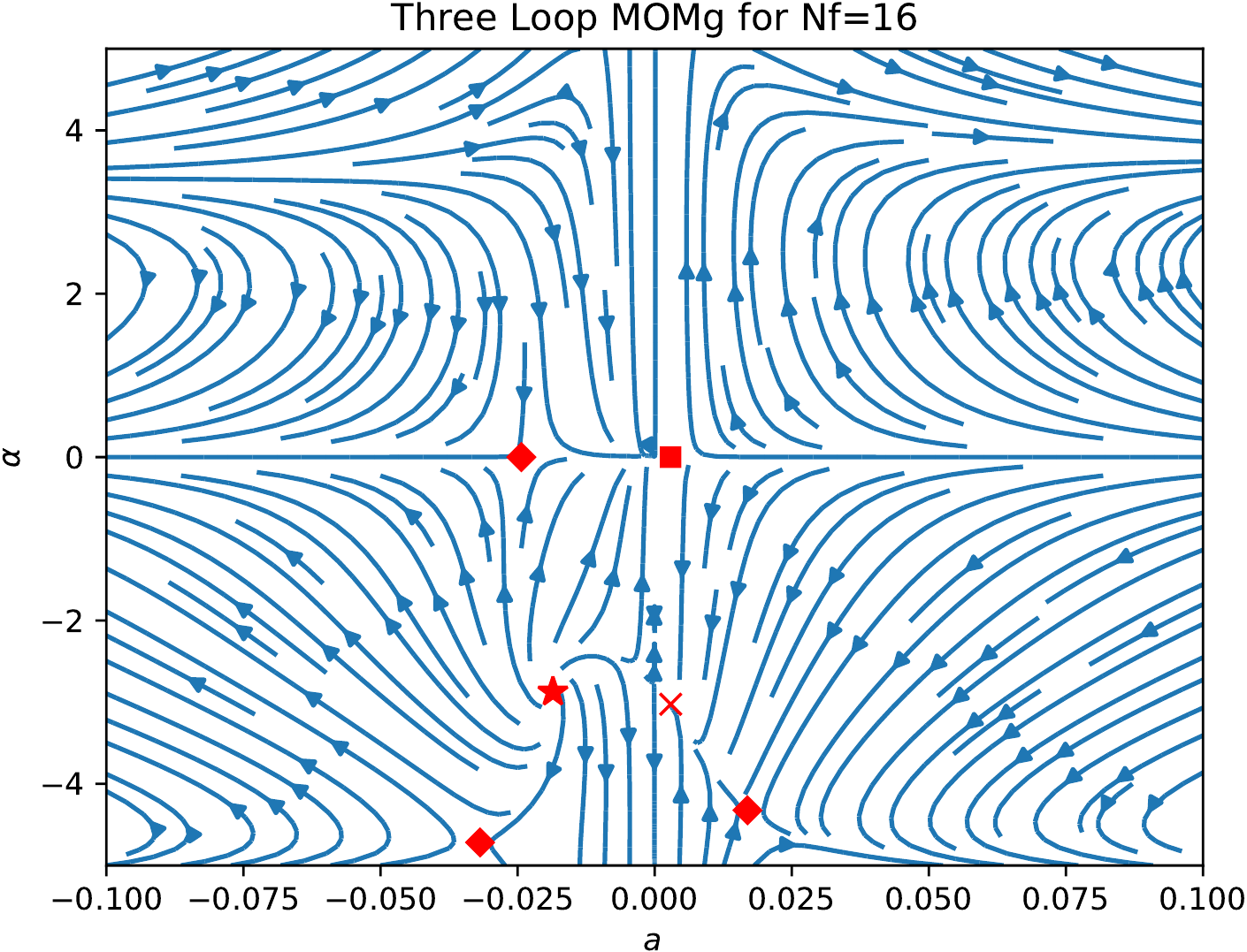}
\\ \\ \\ \\
\includegraphics[width=7.80cm,height=6cm]{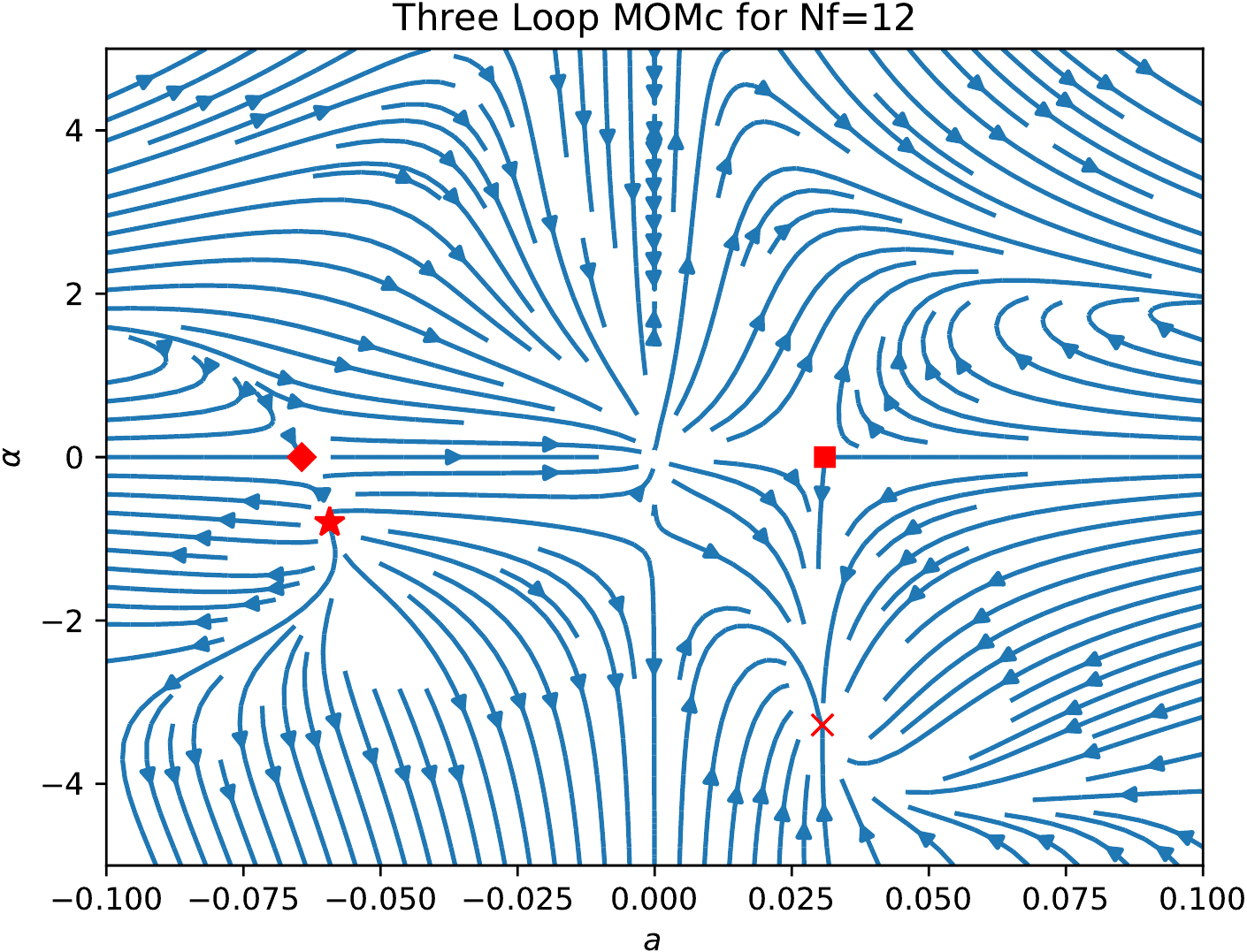}
\quad
\quad
\includegraphics[width=7.80cm,height=6cm]{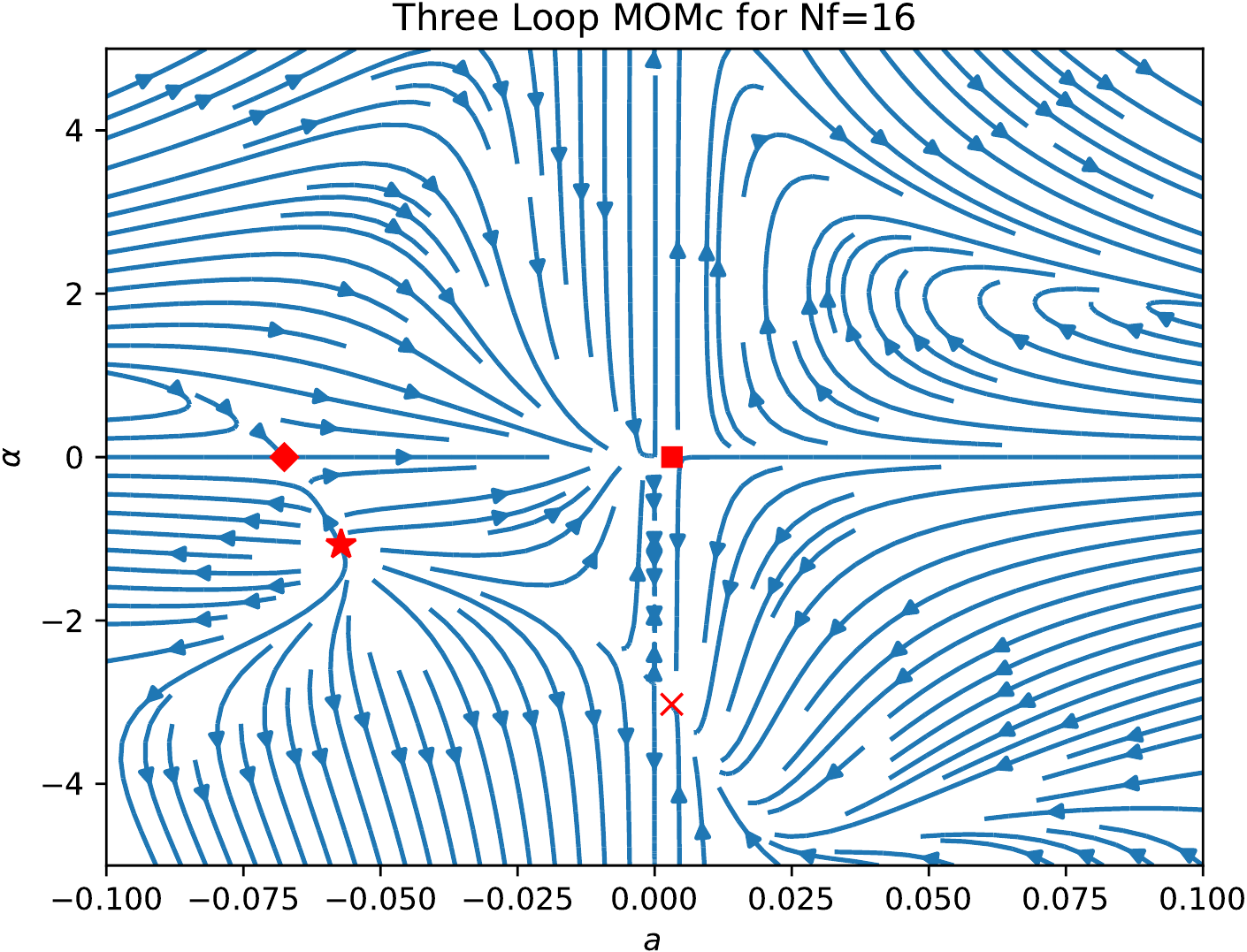}
\\ \\ \\ \\
\includegraphics[width=7.80cm,height=6cm]{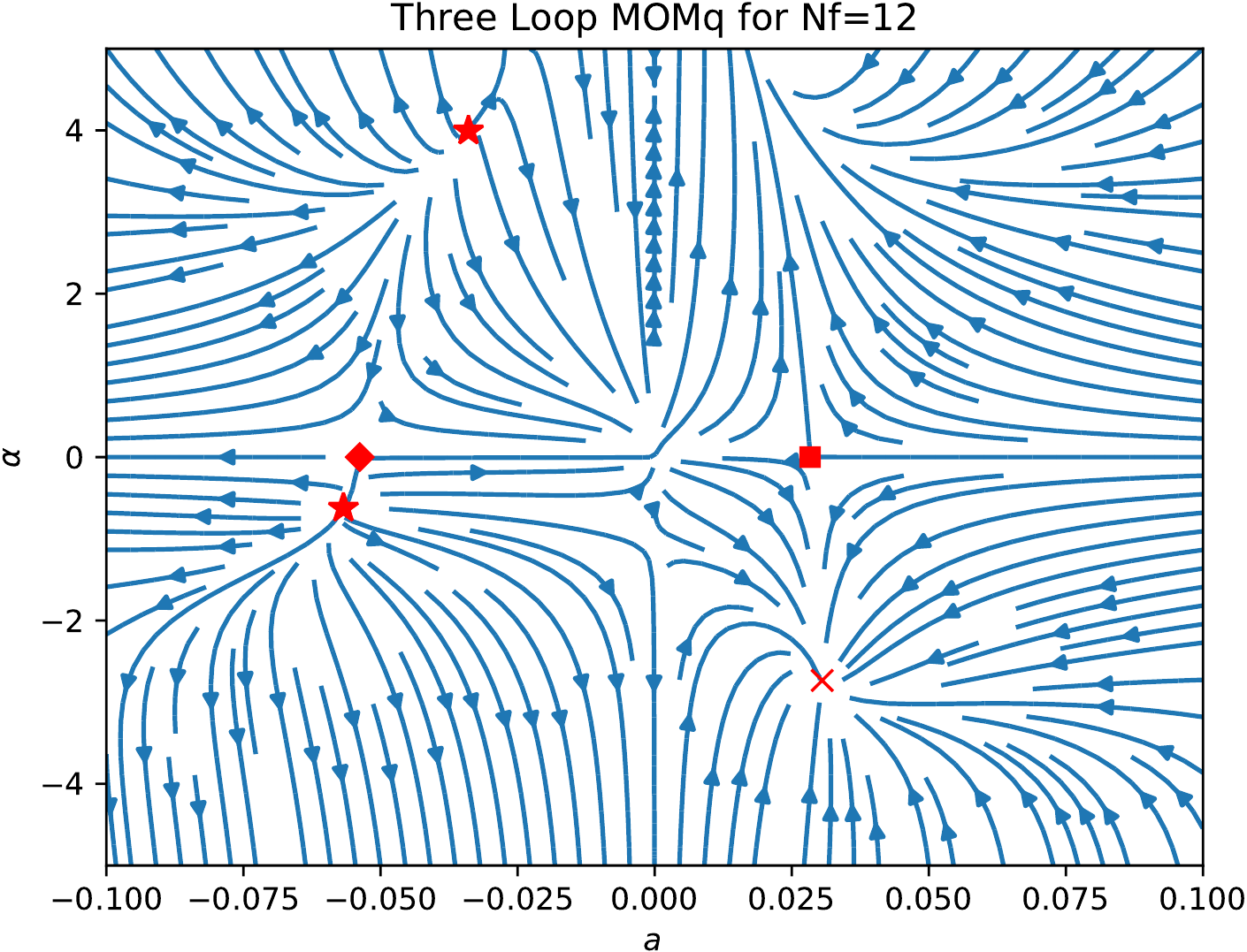}
\quad
\quad
\includegraphics[width=7.80cm,height=6cm]{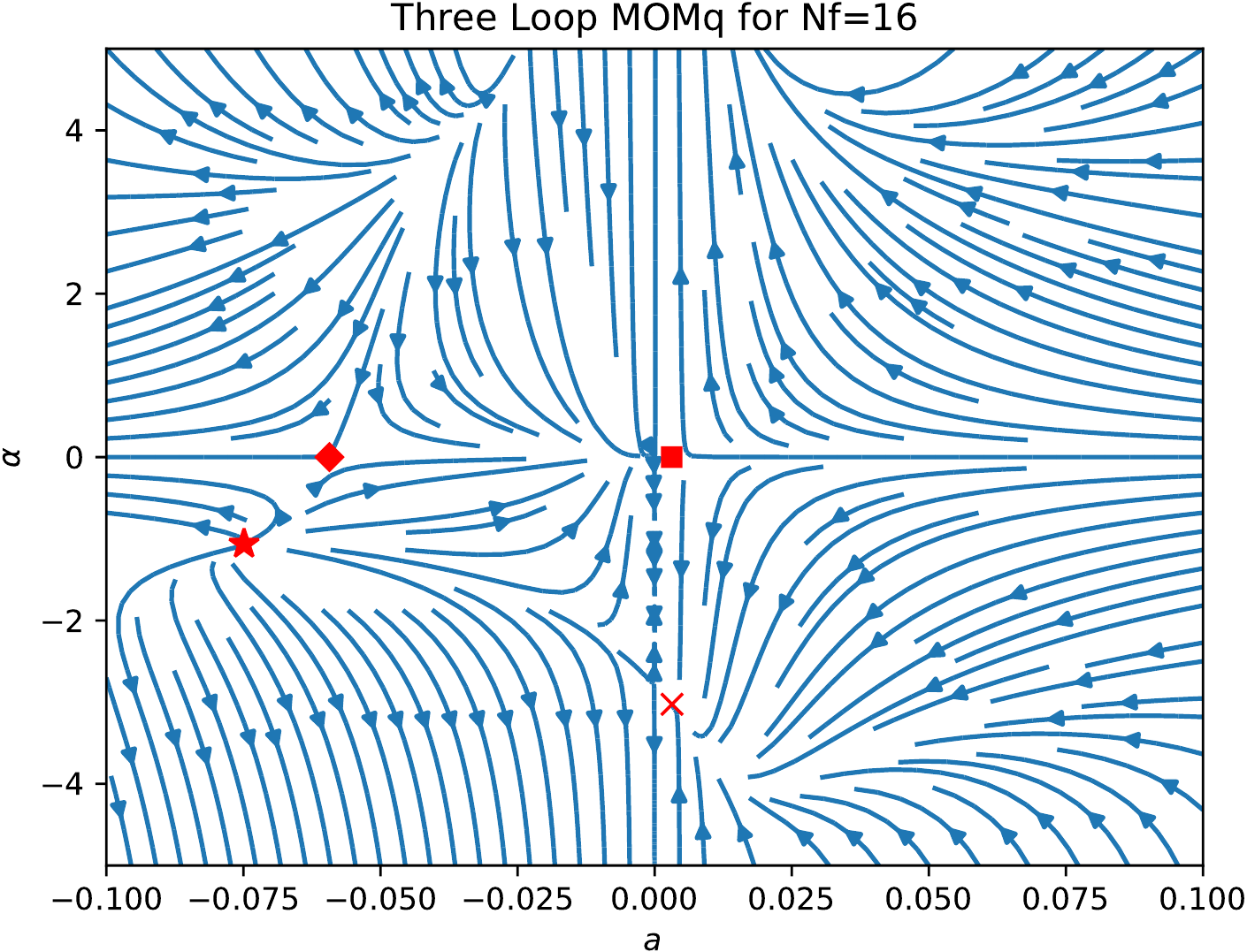}
\caption{Flow planes for the three MOM schemes in $SU(3)$ at three loops for 
$\Nf$~$=$~$12$ (left set) and $\Nf$~$=$~$16$ (right set).}
\label{flow3mom3n1216}
\end{figure}}

{\begin{figure}[ht]
\includegraphics[width=7.80cm,height=6cm]{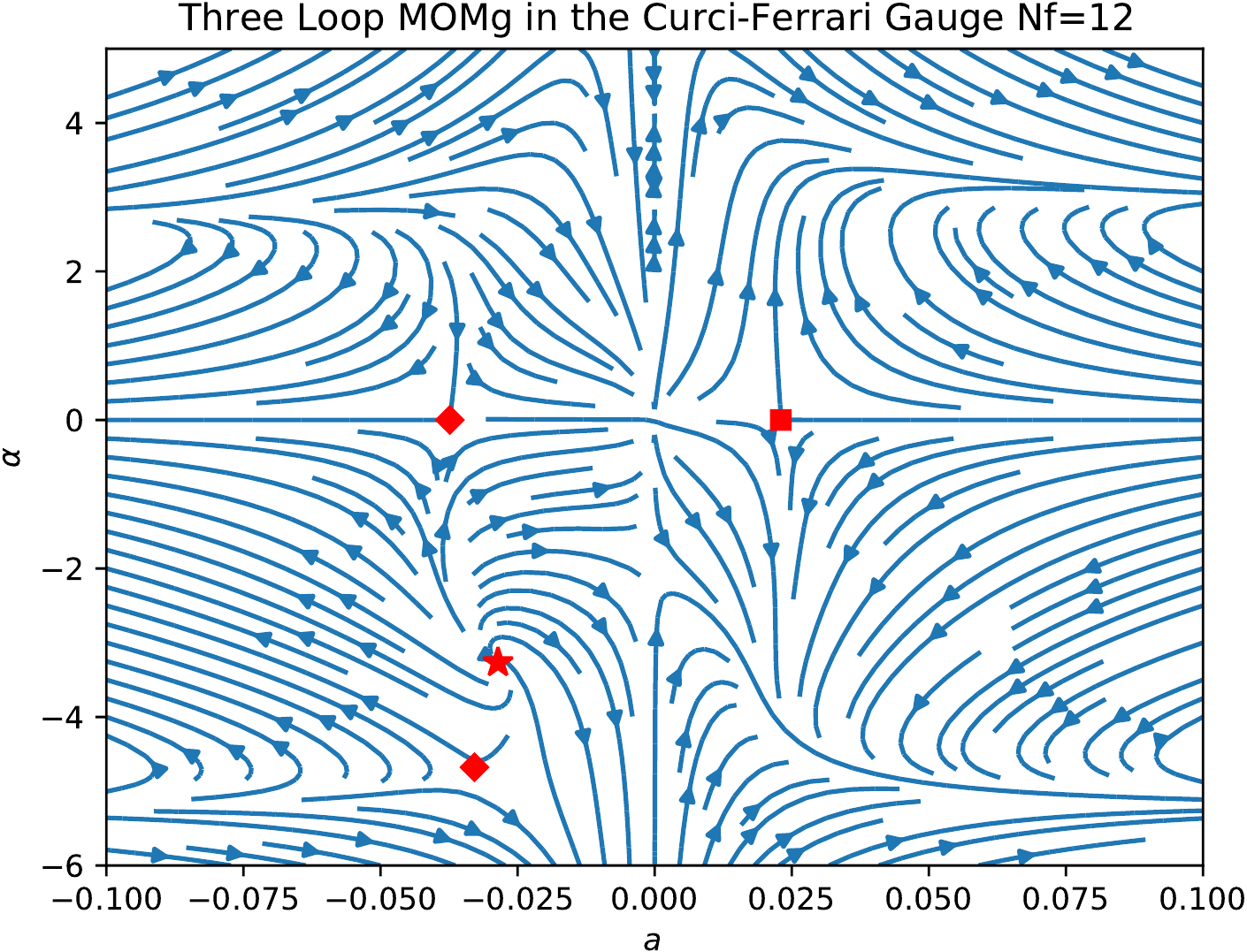}
\quad
\quad
\includegraphics[width=7.80cm,height=6cm]{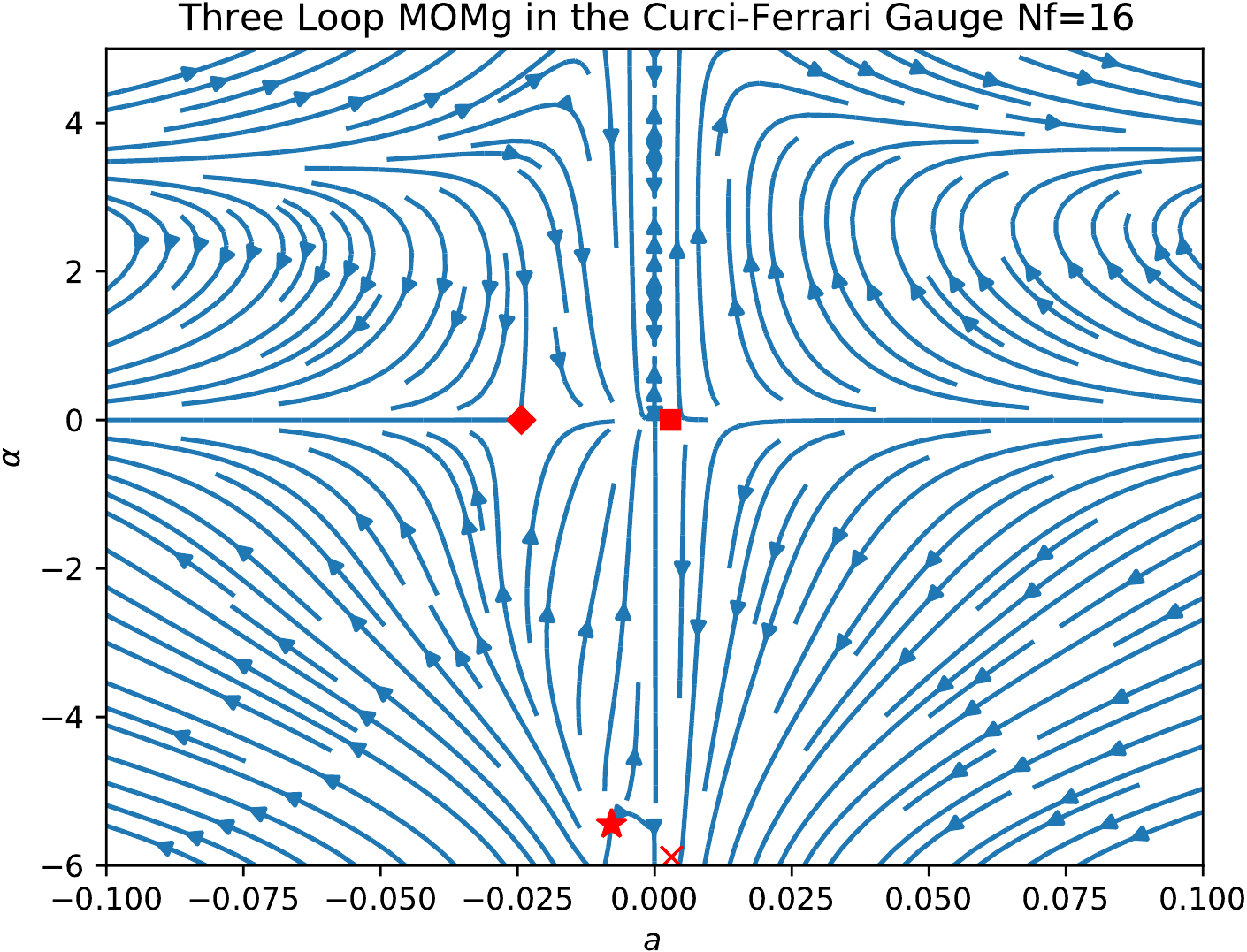}
\\ \\ \\ \\
\includegraphics[width=7.80cm,height=6cm]{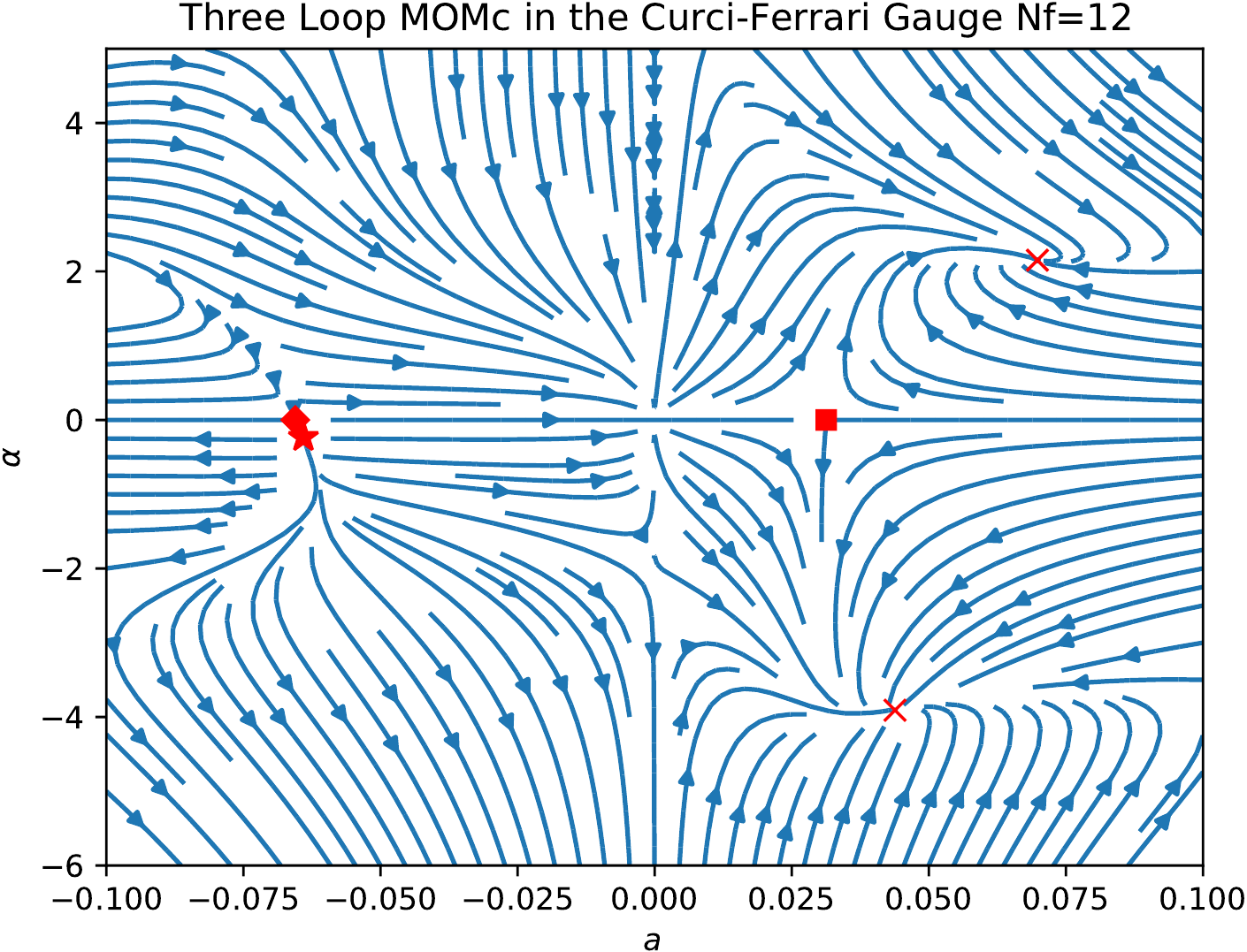}
\quad
\quad
\includegraphics[width=7.80cm,height=6cm]{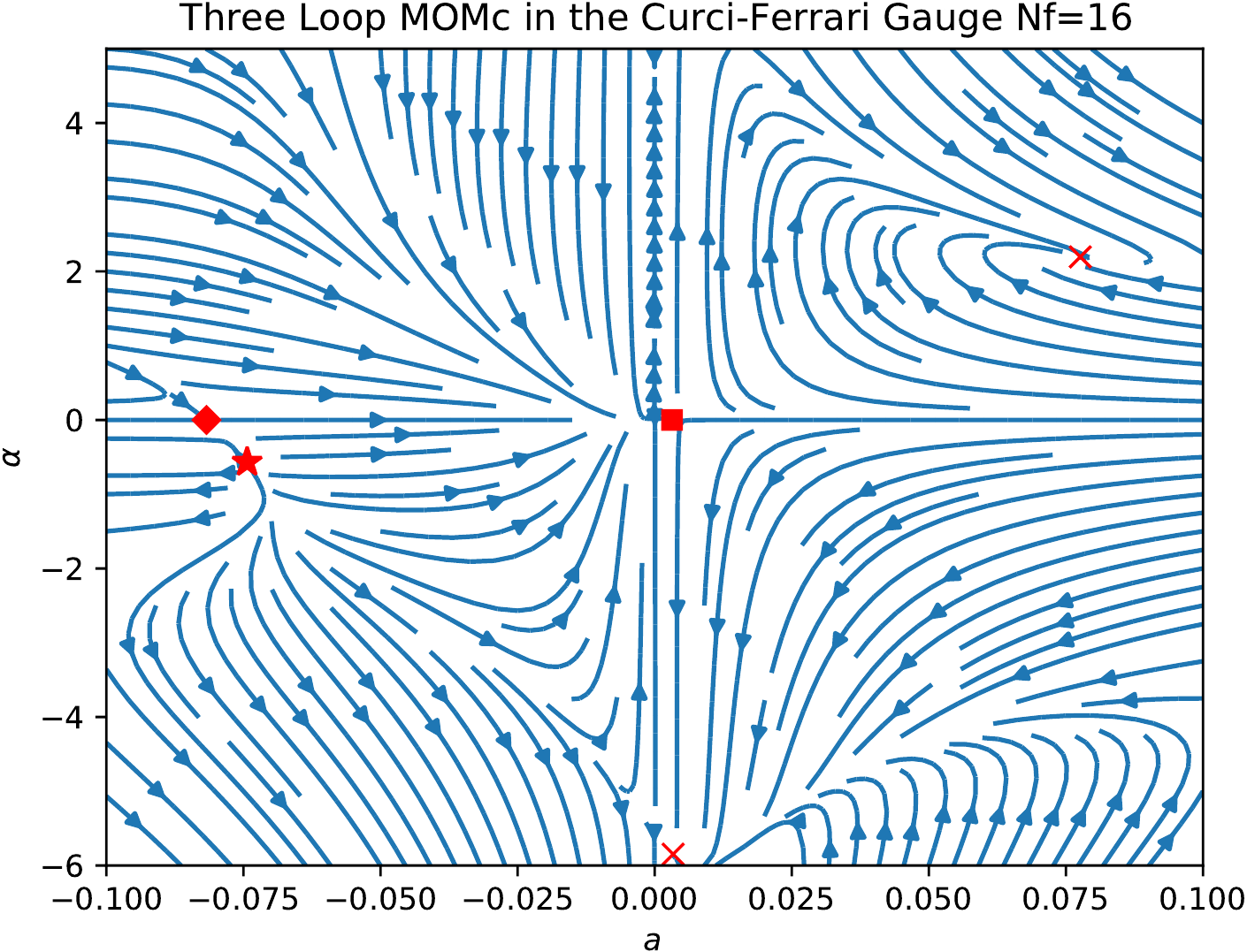}
\\ \\ \\ \\
\includegraphics[width=7.80cm,height=6cm]{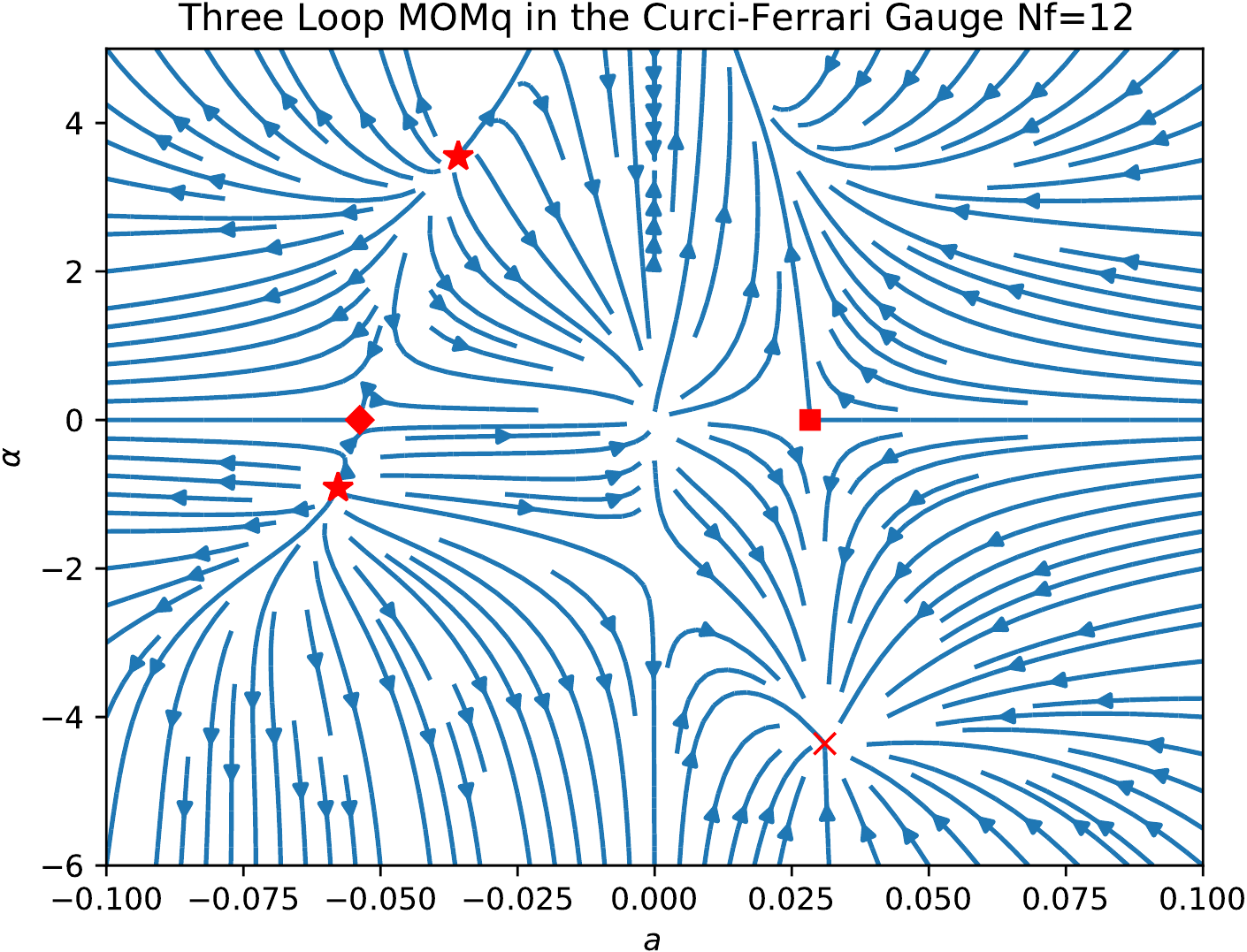}
\quad
\quad
\includegraphics[width=7.80cm,height=6cm]{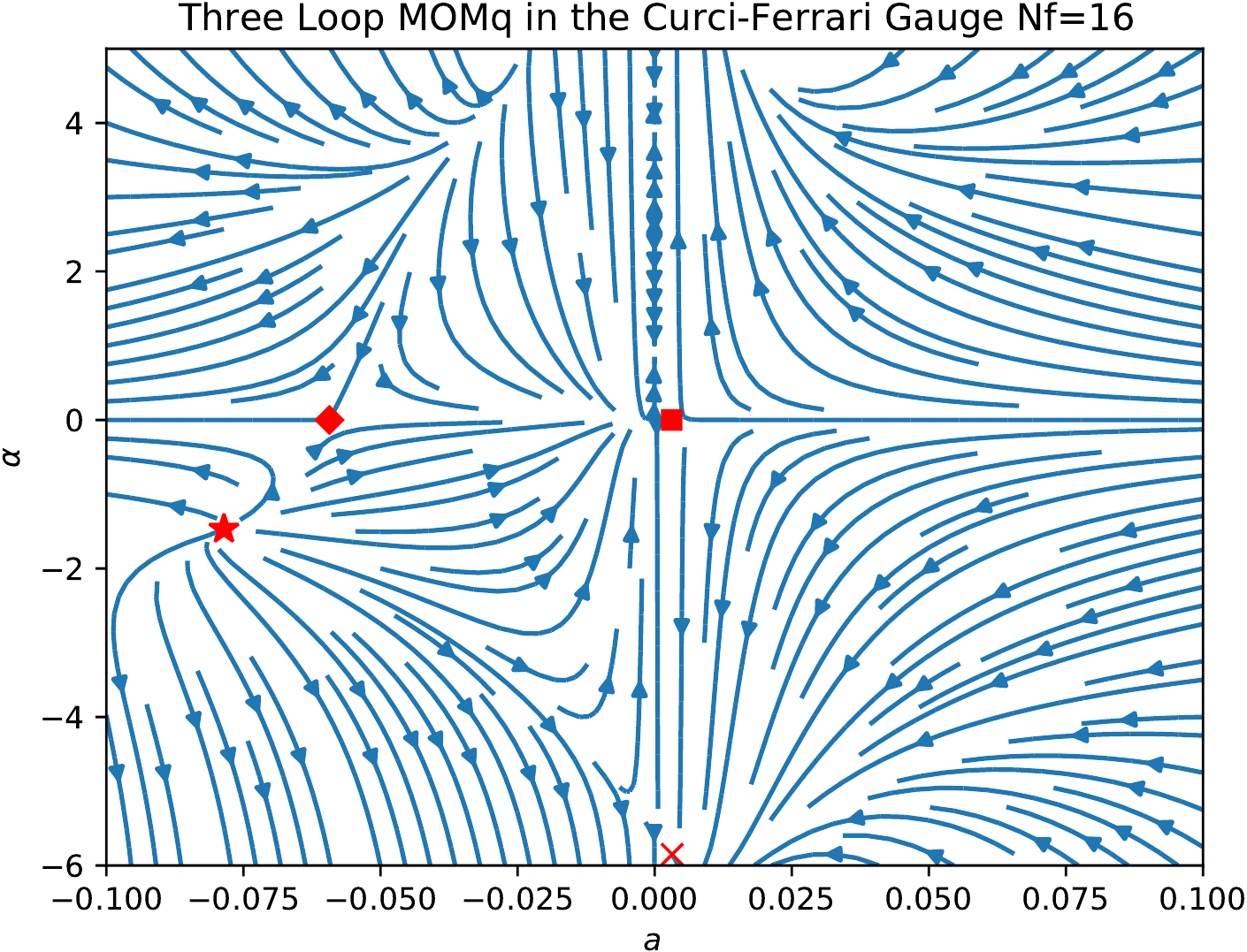}
\caption{Flow planes for the three MOM schemes in $SU(3)$ Curci-Ferrari at 
three loops for $\Nf$~$=$~$12$ (left set) and $\Nf$~$=$~$16$ (right set).}
\label{flow3momcf3n1216}
\end{figure}}

{\begin{figure}[ht]
\includegraphics[width=7.80cm,height=6cm]{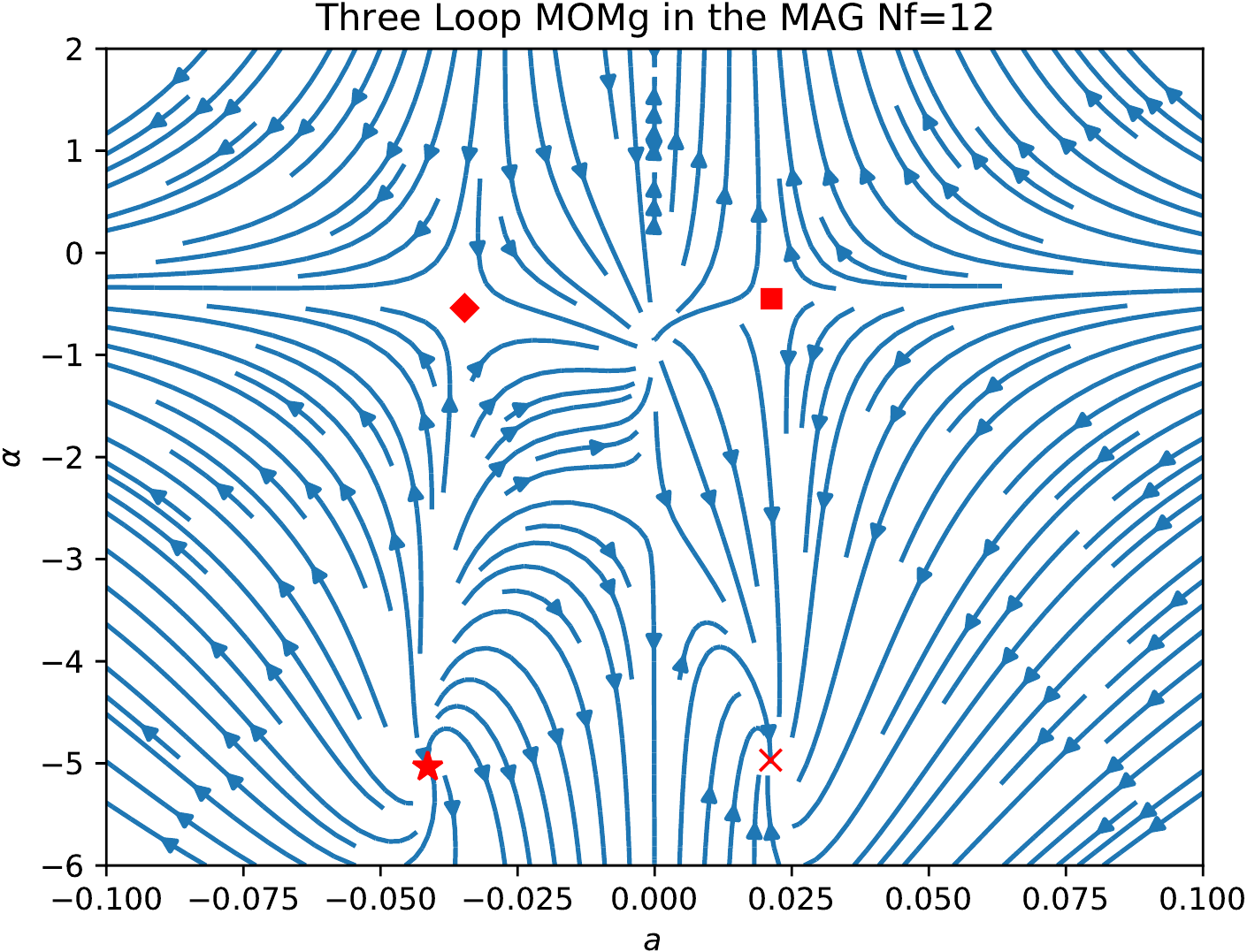}
\quad
\quad
\includegraphics[width=7.80cm,height=6cm]{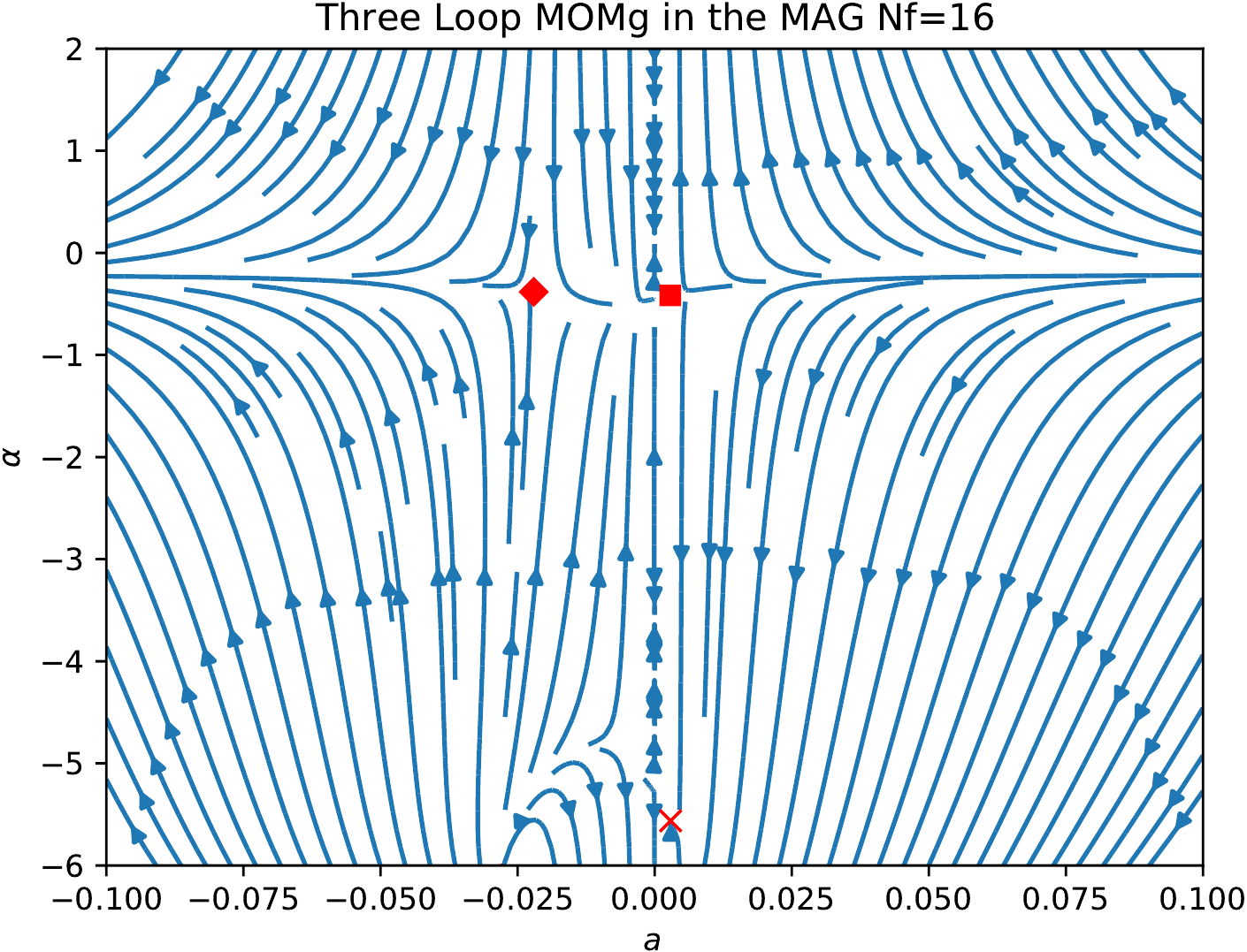}
\\ \\ \\ \\
\includegraphics[width=7.80cm,height=6cm]{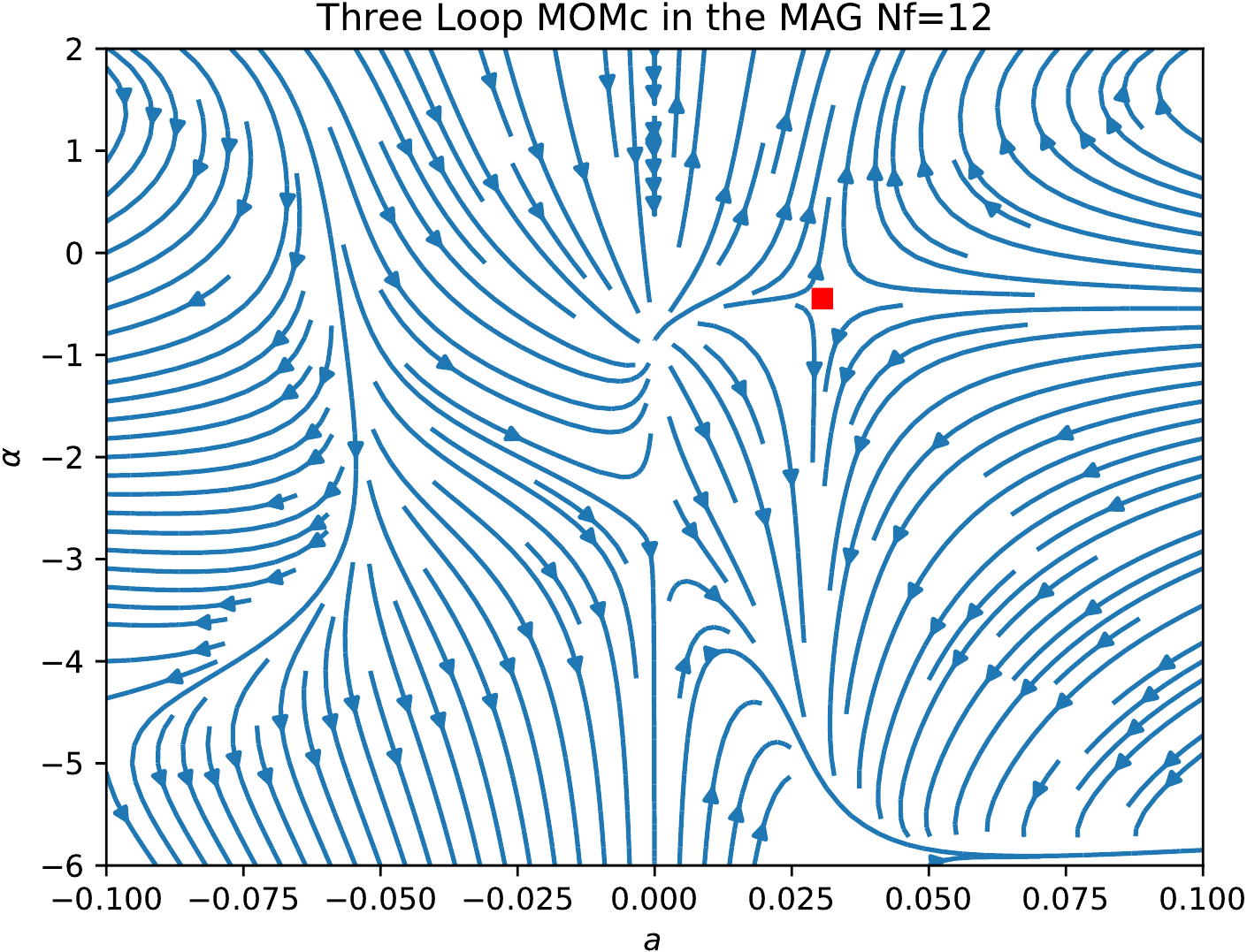}
\quad
\quad
\includegraphics[width=7.80cm,height=6cm]{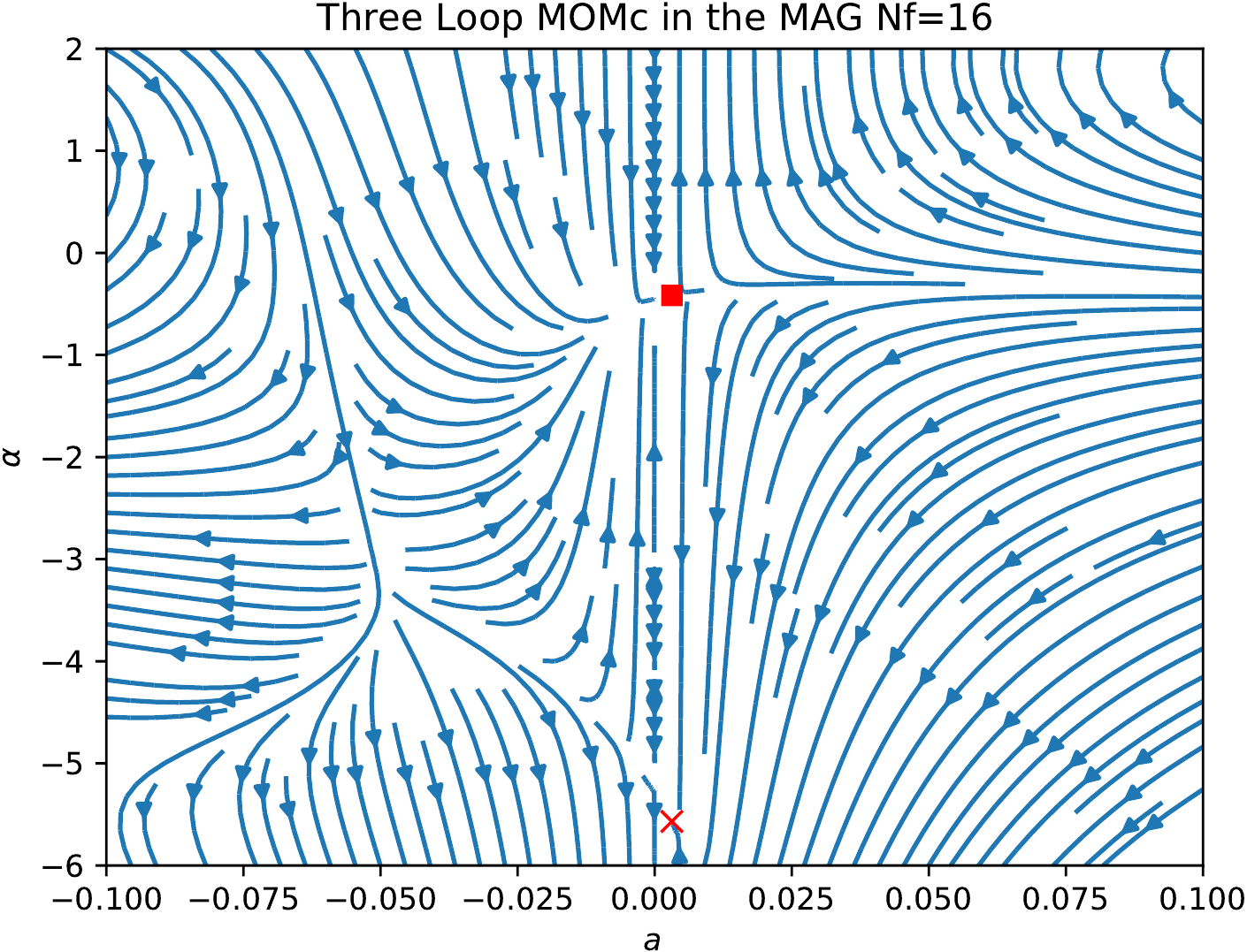}
\\ \\ \\ \\
\includegraphics[width=7.80cm,height=6cm]{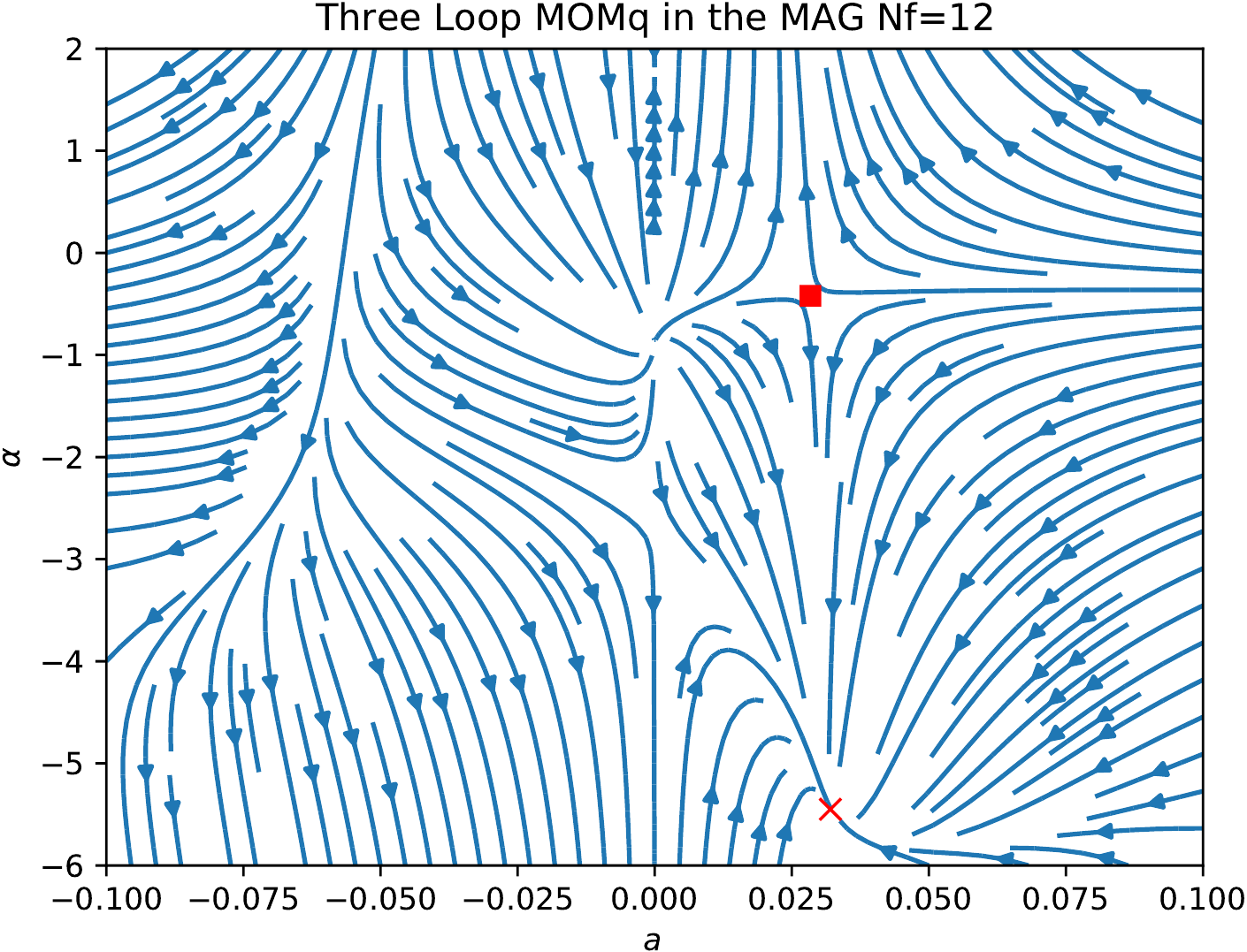}
\quad
\quad
\includegraphics[width=7.80cm,height=6cm]{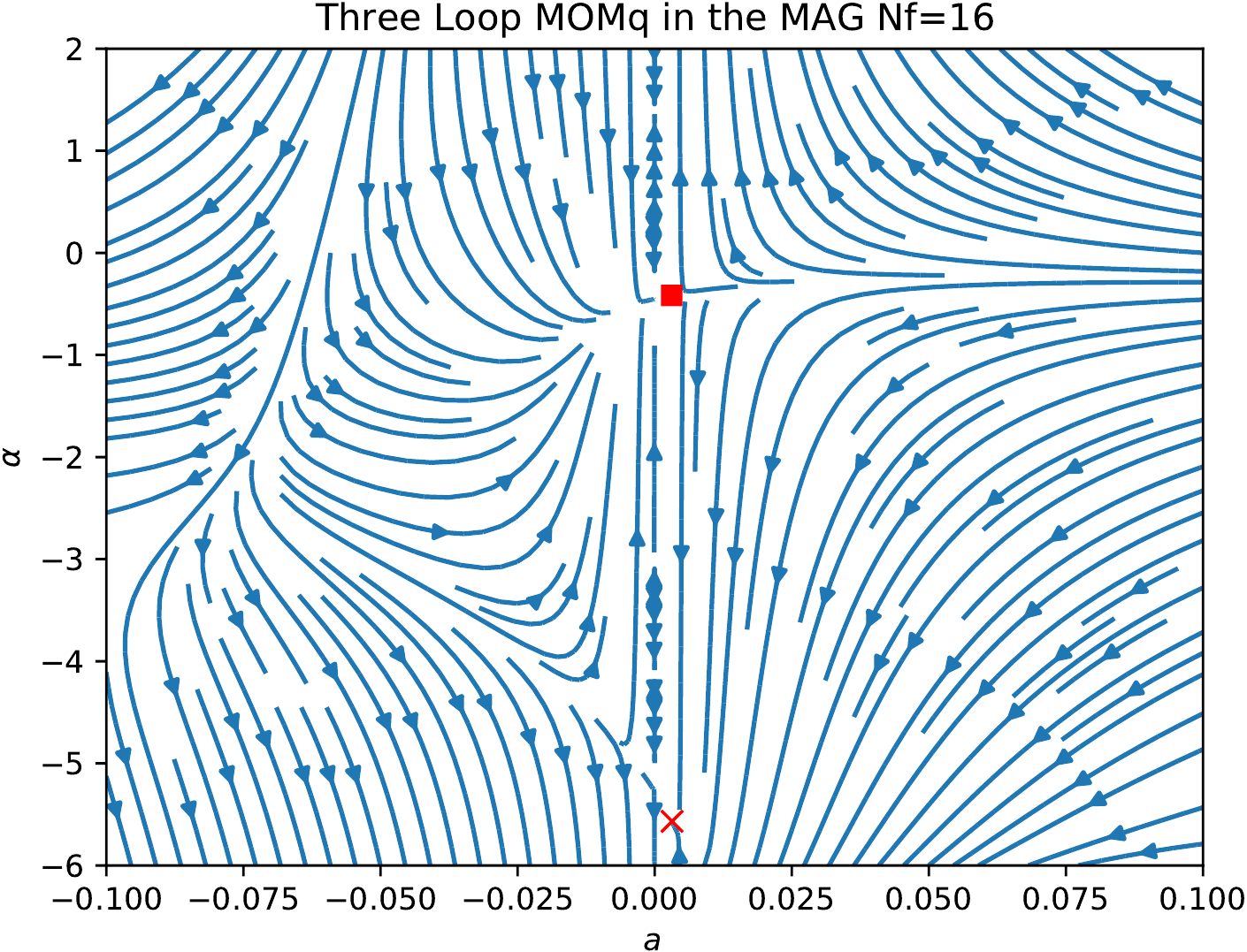}
\caption{Flow planes for the three MOM schemes in $SU(3)$ MAG  at three loops 
for $\Nf$~$=$~$12$ (left set) and $\Nf$~$=$~$16$ (right set).}
\label{flow3mommag3n1216}
\end{figure}}

\clearpage

{\begin{table}
\begin{center}
\begin{tabular}{ |c||c|c|c|c|c| } 
  \hline 
 $\Nf$ & $a_{\infty}$ &$\alpha_{\infty}$ &$\omega_1$&$\omega_2$& Infrared Stability \\ 
\hline
\hline
 9&$0.416667$&$0.000000$&$-4.153646$&$2.083333$& Saddle  \\ 
&$0.416667$&$-2.146730$&$2.353470$&$2.083333$& {\bf Stable}  \\ 
&$0.416667$&$-4.953270$&$-5.430293$&$2.083333$& Saddle  \\ 
\hline 
 10&$0.175676$&$0.000000$&$-1.118449$&$0.761261$& Saddle  \\ 
&$0.175676$&$-2.303972$&$0.749845$&$0.761261$& {\bf Stable}  \\ 
&$0.175676$&$-6.990900$&$-2.275239$&$0.761261$& Saddle  \\ 
\hline 
 11&$0.098214$&$0.000000$&$-0.520338$&$0.360119$& Saddle  \\ 
&$0.098214$&$-2.432707$&$0.391895$&$0.360119$& {\bf Stable}  \\ 
&$0.098214$&$-9.855172$&$-1.587611$&$0.360119$& Saddle  \\ 
\hline 
 12&$0.060000$&$0.000000$&$-0.290250$&$0.180000$& Saddle  \\ 
&$0.060000$&$-2.548046$&$0.237660$&$0.180000$& {\bf Stable}  \\ 
&$0.060000$&$-14.063065$&$-1.311685$&$0.180000$& Saddle  \\ 
\hline 
 13&$0.037234$&$0.000000$&$-0.171886$&$0.086879$& Saddle  \\ 
&$0.037234$&$-2.655694$&$0.149886$&$0.086879$& {\bf Stable}  \\ 
&$0.037234$&$-20.749068$&$-1.171065$&$0.086879$& Saddle  \\ 
\hline 
 14&$0.022124$&$0.000000$&$-0.099863$&$0.036873$& Saddle  \\ 
&$0.022124$&$-2.758253$&$0.091485$&$0.036873$& {\bf Stable}  \\ 
&$0.022124$&$-32.875081$&$-1.090390$&$0.036873$& Saddle  \\ 
\hline 
 15&$0.011364$&$0.000000$&$-0.050894$&$0.011364$& Saddle  \\ 
&$0.011364$&$-2.857085$&$0.048523$&$0.011364$& {\bf Stable}  \\ 
&$0.011364$&$-61.309582$&$-1.041235$&$0.011364$& Saddle  \\ 
\hline 
 16&$0.003311$&$0.000000$&$-0.014853$&$0.001104$& Saddle  \\ 
&$0.003311$&$-2.952985$&$0.014638$&$0.001104$& {\bf Stable}  \\ 
&$0.003311$&$-203.880349$&$-1.010609$&$0.001104$& Saddle  \\ 
\hline 
\end{tabular}
\end{center}
\begin{center}
\caption{Two loop $\MSbar$ scheme $SU(3)$ fixed points in the linear covariant
gauge in the conformal window.}
\label{2lsu3cfw}
\end{center}
\end{table}}

{\begin{table}
\begin{center}
 \begin{tabular}{ |c||c|c|c|c|c| } 
  \hline 
 $\Nf$ & $a_{\infty}$ &$\alpha_{\infty}$ &$\omega_1$&$\omega_2$& Infrared Stability \\ 
\hline
\hline
 9&$0.081803$&$0.000000$&$-0.193341$&$0.737728$& Saddle  \\
&$0.081803$&$-0.858773$&$0.172635$&$0.737728$& {\bf Stable}  \\
\hline
 10&$0.060824$&$0.000000$&$-0.175883$&$0.435887$& Saddle  \\ 
&$0.060824$&$-1.298293$&$0.154800$&$0.435887$& {\bf Stable}  \\ 
\hline 
 11&$0.046039$&$0.000000$&$-0.151892$&$0.258488$& Saddle  \\ 
&$0.046039$&$-1.692825$&$0.134329$&$0.258488$& {\bf Stable}  \\ 
\hline 
 12&$0.034607$&$0.000000$&$-0.125974$&$0.147759$& Saddle  \\ 
&$0.034607$&$-2.039860$&$0.113066$&$0.147759$& {\bf Stable}  \\ 
\hline 
 13&$0.025191$&$0.000000$&$-0.099007$&$0.077790$& Saddle  \\ 
&$0.025191$&$-2.339613$&$0.090641$&$0.077790$& {\bf Stable}  \\ 
\hline 
 14&$0.017070$&$0.000000$&$-0.071194$&$0.034949$& Saddle  \\ 
&$0.017070$&$-2.591922$&$0.066678$&$0.034949$& {\bf Stable}  \\ 
\hline 
 15&$0.009818$&$0.000000$&$-0.042753$&$0.011153$& Saddle  \\ 
&$0.009818$&$-2.795125$&$0.041052$&$0.011153$& {\bf Stable}  \\ 
\hline 
 16&$0.003162$&$0.000000$&$-0.014138$&$0.001102$& Saddle  \\ 
&$0.003162$&$-2.945839$&$0.013943$&$0.001102$& {\bf Stable}  \\ 
\hline 
\end{tabular}
\end{center}
\begin{center}
\caption{Three loop $\MSbar$ scheme $SU(3)$ fixed points in the linear
covariant gauge in the conformal window.}
\label{3lsu3cfw}
\end{center}
\end{table}}
 
{\begin{table}
\begin{center}
 \begin{tabular}{ |c||c|c|c|c|c| } 
  \hline 
 $\Nf$ & $a_{\infty}$ &$\alpha_{\infty}$ &$\omega_1$&$\omega_2$& Infrared Stability \\ 
\hline
\hline
 9&$0.085291$&$0.000000$&$0.017807$&$0.732193$& {\bf Stable}  \\ 
&$0.085291$&$0.078057$&$-0.018214$&$0.732193$& Saddle  \\ 
&$0.085291$&$-5.882389$&$-1.983191$&$0.732193$& Saddle  \\ 
&$0.962002$&$0.000000$&$-542.394739$&$3638.528050$& Saddle  \\ 
&$0.962002$&$2.372214$&$-12107.306162$&$-542.394739$& Unstable  \\ 
&$0.962002$&$-2.710372$&$-4877.065906$&$-542.394739$& Unstable  \\ 
\hline 
 10&$0.064860$&$0.000000$&$-0.096514$&$0.426703$& Saddle  \\ 
&$0.064860$&$-0.758788$&$0.084116$&$0.426703$& {\bf Stable}  \\ 
&$0.064860$&$-7.285851$&$-1.593756$&$0.426703$& Saddle  \\ 
&$0.446940$&$0.000000$&$-72.410221$&$216.607652$& Saddle  \\ 
&$0.446940$&$2.631336$&$-770.095520$&$-72.410221$& Unstable  \\ 
&$0.446940$&$-2.802990$&$-276.712319$&$-72.410221$& Unstable  \\ 
\hline 
 11&$0.049832$&$0.000000$&$-0.125835$&$0.249017$& Saddle  \\ 
&$0.049832$&$-1.444524$&$0.107670$&$0.249017$& {\bf Stable}  \\ 
&$0.049832$&$-9.164022$&$-1.495291$&$0.249017$& Saddle  \\ 
&$0.262159$&$0.000000$&$-18.805627$&$32.056161$& Saddle  \\ 
&$0.262159$&$2.797166$&$-114.552315$&$-18.805627$& Unstable  \\ 
&$0.262159$&$-3.009747$&$-40.855724$&$-18.805627$& Unstable  \\ 
\hline 
 12&$0.037434$&$0.000000$&$-0.121873$&$0.141164$& Saddle  \\ 
&$0.037434$&$-1.945284$&$0.106950$&$0.141164$& {\bf Stable}  \\ 
&$0.037434$&$-11.872619$&$-1.477787$&$0.141164$& Saddle  \\ 
&$0.182641$&$0.000000$&$-8.151111$&$9.044557$& Saddle  \\ 
&$0.182641$&$2.874023$&$-31.268962$&$-8.151111$& Unstable  \\ 
&$0.182641$&$-3.270103$&$-11.852903$&$-8.151111$& Unstable  \\ 
\hline 
\end{tabular}
\end{center}
\begin{center}
\caption{Four loop $\MSbar$ scheme $SU(3)$ fixed points in the linear covariant
gauge for $9$~$\leq$~$\Nf$~$\leq$~$12$.}
\label{4lsu3cfw912}
\end{center}
\end{table}}

{\begin{table}
\begin{center}
 \begin{tabular}{ |c||c|c|c|c|c| } 
  \hline 
 $\Nf$ & $a_{\infty}$ &$\alpha_{\infty}$ &$\omega_1$&$\omega_2$& Infrared Stability \\ 
\hline
\hline
 13&$0.026853$&$0.000000$&$-0.101310$&$0.074565$& Saddle  \\ 
&$0.026853$&$-2.312632$&$0.091708$&$0.074565$& {\bf Stable}  \\ 
&$0.026853$&$-16.124358$&$-1.476144$&$0.074565$& Saddle  \\ 
&$0.141687$&$0.000000$&$-4.907150$&$3.754144$& Saddle  \\ 
&$0.141687$&$2.881045$&$-12.298078$&$-4.907150$& Unstable  \\ 
&$0.141687$&$-3.553712$&$-5.175980$&$-4.907150$& Unstable  \\ 
\hline 
 14&$0.017793$&$0.000000$&$-0.073373$&$0.033906$& Saddle  \\ 
&$0.017793$&$-2.588296$&$0.068423$&$0.033906$& {\bf Stable}  \\ 
&$0.017793$&$-23.713794$&$-1.471549$&$0.033906$& Saddle  \\ 
&$0.117756$&$0.000000$&$-3.633611$&$1.980404$& Saddle  \\ 
&$0.117756$&$2.842775$&$-6.094849$&$-3.633611$& Unstable  \\ 
&$0.117756$&$-3.843479$&$-2.923578$&$-3.633611$& Unstable  \\ 
\hline 
 15&$0.010001$&$0.000000$&$-0.043478$&$0.010988$& Saddle  \\ 
&$0.010001$&$-2.795819$&$0.041706$&$0.010988$& {\bf Stable}  \\ 
&$0.010001$&$-41.152460$&$-1.462900$&$0.010988$& Saddle  \\ 
&$0.102368$&$0.000000$&$-3.051413$&$1.218110$& Saddle  \\ 
&$0.102368$&$2.778768$&$-3.516462$&$-3.051413$& Unstable  \\ 
&$0.102368$&$-4.129750$&$-1.950938$&$-3.051413$& Unstable  \\ 
\hline 
 16&$0.003170$&$0.000000$&$-0.014174$&$0.001099$& Saddle  \\ 
&$0.003170$&$-2.945935$&$0.013978$&$0.001099$& {\bf Stable}  \\ 
&$0.003170$&$-126.835296$&$-1.453369$&$0.001099$& Saddle  \\ 
&$0.091700$&$0.000000$&$-2.761435$&$0.828759$& Saddle  \\ 
&$0.091700$&$2.700828$&$-2.247803$&$-2.761435$& Unstable  \\ 
&$0.091700$&$-4.407403$&$-1.453956$&$-2.761435$& Unstable  \\ 
\hline 
\end{tabular}
\end{center}
\begin{center}
\caption{Four loop $\MSbar$ scheme $SU(3)$ fixed points in the linear covariant
gauge for $13$~$\leq$~$\Nf$~$\leq$~$16$.}
\label{4lsu3cfw1316}
\end{center}
\end{table}}

{\begin{table}
\begin{center}
 \begin{tabular}{ |c||c|c|c|c|c| } 
  \hline 
 $\Nf$ & $a_{\infty}$ &$\alpha_{\infty}$ &$\omega_1$&$\omega_2$& Infrared Stability \\ 
\hline
\hline
 13&$0.032315$&$0.000000$&$-0.122684$&$0.036682$& Saddle  \\ 
&$0.032315$&$-2.386609$&$0.109561$&$0.036682$& {\bf Stable}  \\ 
&$0.041343$&$0.000000$&$-0.151397$&$-0.072312$& Unstable  \\ 
&$0.041343$&$-2.470802$&$0.131351$&$-0.072312$& Saddle  \\ 
\hline 
 14&$0.018526$&$0.000000$&$-0.076782$&$0.031152$& Saddle  \\ 
&$0.018526$&$-2.603426$&$0.071510$&$0.031152$& {\bf Stable}  \\ 
&$0.046357$&$0.000000$&$-0.365291$&$-0.176646$& Unstable  \\ 
&$0.046357$&$-2.932818$&$0.152600$&$-0.365291$& Saddle  \\ 
\hline 
 15&$0.010078$&$0.000000$&$-0.043850$&$0.010828$& Saddle  \\ 
&$0.010078$&$-2.797897$&$0.042058$&$0.010828$& {\bf Stable}  \\ 
&$0.047107$&$0.000000$&$-0.587861$&$-0.190657$& Unstable  \\ 
&$0.047107$&$-3.326390$&$0.168639$&$-0.587861$& Saddle  \\ 
\hline 
 16&$0.003171$&$0.000000$&$-0.014178$&$0.001098$& Saddle  \\ 
&$0.003171$&$-2.945965$&$0.013982$&$0.001098$& {\bf Stable}  \\ 
&$0.046999$&$0.000000$&$-0.780464$&$-0.202969$& Unstable  \\ 
&$0.046999$&$-3.682521$&$0.185665$&$-0.780464$& Saddle  \\ 
\hline 
\end{tabular}
\end{center}
\begin{center}
\caption{Five loop $\MSbar$ scheme $SU(3)$ fixed points in the linear covariant
gauge for $13$~$\leq$~$\Nf$~$\leq$~$16$.}
\label{5lsu3cfw}
\end{center}
\end{table}}
 
{\begin{table}
\begin{center}
 \begin{tabular}{ |c||c|c|c|c|c| } 
  \hline 
 $\Nf$ & $a_{\infty}$ &$\alpha_{\infty}$ &$\omega_1$&$\omega_2$& Infrared Stability \\ 
\hline
\hline
 9&$0.416667$&$0.000000$&$-4.153646$&$2.083333$& Saddle  \\ 
&$0.416667$&$-2.387888$&$3.596809$&$2.083333$& {\bf Stable}  \\ 
&$0.416667$&$-17.812112$&$-26.829882$&$2.083333$& Saddle  \\ 
\hline 
 10&$0.175676$&$0.000000$&$-1.118449$&$0.761261$& Saddle  \\ 
&$0.175676$&$-2.981627$&$0.964118$&$0.761261$& {\bf Stable}  \\ 
&$0.175676$&$-21.608116$&$-6.987048$&$0.761261$& Saddle  \\ 
\hline 
 11&$0.098214$&$0.000000$&$-0.520338$&$0.360119$& Saddle  \\ 
&$0.098214$&$-3.548192$&$0.452028$&$0.360119$& {\bf Stable}  \\ 
&$0.098214$&$-27.027566$&$-3.443224$&$0.360119$& Saddle  \\ 
\hline 
 12&$0.060000$&$0.000000$&$-0.290250$&$0.180000$& Saddle  \\ 
&$0.060000$&$-4.078489$&$0.256566$&$0.180000$& {\bf Stable}  \\ 
&$0.060000$&$-35.143733$&$-2.210791$&$0.180000$& Saddle  \\ 
\hline 
 13&$0.037234$&$0.000000$&$-0.171886$&$0.086879$& Saddle  \\ 
&$0.037234$&$-4.569040$&$0.155606$&$0.086879$& {\bf Stable}  \\ 
&$0.037234$&$-48.240484$&$-1.642904$&$0.086879$& Saddle  \\ 
\hline 
 14&$0.022124$&$0.000000$&$-0.099863$&$0.036873$& Saddle  \\ 
&$0.022124$&$-5.020489$&$0.092924$&$0.036873$& {\bf Stable}  \\ 
&$0.022124$&$-72.246178$&$-1.337198$&$0.036873$& Saddle  \\ 
\hline 
 15&$0.011364$&$0.000000$&$-0.050894$&$0.011364$& Saddle  \\ 
&$0.011364$&$-5.435844$&$0.048748$&$0.011364$& {\bf Stable}  \\ 
&$0.011364$&$-128.897489$&$-1.155936$&$0.011364$& Saddle  \\ 
\hline 
 16&$0.003311$&$0.000000$&$-0.014853$&$0.001104$& Saddle  \\ 
&$0.003311$&$-5.819105$&$0.014644$&$0.001104$& {\bf Stable}  \\ 
&$0.003311$&$-413.847562$&$-1.041453$&$0.001104$& Saddle  \\ 
\hline 
\end{tabular}
\end{center}
\begin{center}
\caption{Two loop $\MSbar$ scheme $SU(3)$ Curci-Ferrari gauge fixed points for 
$9$~$\leq$~$\Nf$~$\leq$~$16$.}
\label{2lcfsu3cfw}
\end{center}
\end{table}}
 
{\begin{table}
\begin{center}
 \begin{tabular}{ |c||c|c|c|c|c| } 
  \hline 
 $\Nf$ & $a_{\infty}$ &$\alpha_{\infty}$ &$\omega_1$&$\omega_2$& Infrared Stability \\ 
\hline
\hline
 6&$1.012686$&$0.000000$&$40.841486$&$92.653913$& {\bf Stable}  \\ 
&$1.012686$&$0.773985$&$-101.745391$&$40.841486$& Saddle  \\ 
\hline 
 7&$0.195518$&$0.000000$&$0.161289$&$2.986261$& {\bf Stable}  \\ 
&$0.195518$&$0.134812$&$-0.163479$&$2.986261$& Saddle  \\ 
\hline 
 8&$0.116505$&$0.000000$&$-0.176683$&$1.329441$& Saddle  \\ 
&$0.116505$&$-0.525277$&$0.169160$&$1.329441$& {\bf Stable}  \\ 
\hline 
 9&$0.081803$&$0.000000$&$-0.193341$&$0.737728$& Saddle  \\ 
&$0.081803$&$-1.239118$&$0.177887$&$0.737728$& {\bf Stable}  \\ 
\hline 
 10&$0.060824$&$0.000000$&$-0.175883$&$0.435887$& Saddle  \\ 
&$0.060824$&$-1.988405$&$0.158258$&$0.435887$& {\bf Stable}  \\ 
\hline 
 11&$0.046039$&$0.000000$&$-0.151892$&$0.258488$& Saddle  \\ 
&$0.046039$&$-2.744498$&$0.135852$&$0.258488$& {\bf Stable}  \\ 
\hline 
 12&$0.034607$&$0.000000$&$-0.125974$&$0.147759$& Saddle  \\ 
&$0.034607$&$-3.479125$&$0.113468$&$0.147759$& {\bf Stable}  \\ 
\hline 
 13&$0.025191$&$0.000000$&$-0.099007$&$0.077790$& Saddle  \\ 
&$0.025191$&$-4.169158$&$0.090635$&$0.077790$& {\bf Stable}  \\ 
\hline 
 14&$0.017070$&$0.000000$&$-0.071194$&$0.034949$& Saddle  \\ 
&$0.017070$&$-4.796907$&$0.066620$&$0.034949$& {\bf Stable}  \\ 
\hline 
 15&$0.009818$&$0.000000$&$-0.042753$&$0.011153$& Saddle  \\ 
&$0.009818$&$-5.347954$&$0.041032$&$0.011153$& {\bf Stable}  \\ 
\hline 
 16&$0.003162$&$0.000000$&$-0.014138$&$0.001102$& Saddle  \\ 
&$0.003162$&$-5.808458$&$0.013942$&$0.001102$& {\bf Stable}  \\ 
\hline 
\end{tabular}
\end{center}
\begin{center}
\caption{Three loop $\MSbar$ scheme $SU(3)$ Curci-Ferrari gauge fixed points 
for $6$~$\leq$~$\Nf$~$\leq$~$16$.}
\label{3lcfsu3cfw}
\end{center}
\end{table}}
 
{\begin{table}
\begin{center}
 \begin{tabular}{ |c||c|c|c|c|c| } 
  \hline 
 $\Nf$ & $a_{\infty}$ &$\alpha_{\infty}$ &$\omega_1$&$\omega_2$& Infrared Stability \\ 
\hline
\hline
 9&$0.416667$&$0.373983$&$-14.338702$&$2.083333$& Saddle  \\ 
&$0.416667$&$-1.775685$&$12.032176$&$2.083333$& {\bf Stable}  \\ 
&$0.416667$&$-12.989602$&$-74.798981$&$2.083333$& Saddle  \\ 
\hline 
 10&$0.175676$&$0.246040$&$-2.897017$&$0.761261$& Saddle  \\ 
&$0.175676$&$-2.041475$&$2.432727$&$0.761261$& {\bf Stable}  \\ 
&$0.175676$&$-14.027308$&$-15.179420$&$0.761261$& Saddle  \\ 
\hline 
 11&$0.098214$&$0.113996$&$-1.082583$&$0.360119$& Saddle  \\ 
&$0.098214$&$-2.385628$&$0.909308$&$0.360119$& {\bf Stable}  \\ 
&$0.098214$&$-15.503072$&$-5.681145$&$0.360119$& Saddle  \\ 
\hline 
 12&$0.060000$&$-0.013878$&$-0.515865$&$0.180000$& Saddle  \\ 
&$0.060000$&$-2.822900$&$0.434197$&$0.180000$& {\bf Stable}  \\ 
&$0.060000$&$-17.757425$&$-2.742662$&$0.180000$& Saddle  \\ 
\hline 
 13&$0.037234$&$-0.131066$&$-0.275179$&$0.086879$& Saddle  \\ 
&$0.037234$&$-3.356225$&$0.233719$&$0.086879$& {\bf Stable}  \\ 
&$0.037234$&$-21.537553$&$-1.551277$&$0.086879$& Saddle  \\ 
\hline 
 14&$0.022124$&$-0.234163$&$-0.150092$&$0.036873$& Saddle  \\ 
&$0.022124$&$-3.968278$&$0.130471$&$0.036873$& {\bf Stable}  \\ 
&$0.022124$&$-28.797559$&$-0.998010$&$0.036873$& Saddle  \\ 
\hline 
 15&$0.011364$&$-0.322652$&$-0.073940$&$0.011364$& Saddle  \\ 
&$0.011364$&$-4.620179$&$0.067083$&$0.011364$& {\bf Stable}  \\ 
&$0.011364$&$-46.665865$&$-0.723403$&$0.011364$& Saddle  \\ 
\hline 
 16&$0.003311$&$-0.397788$&$-0.021258$&$0.001104$& Saddle  \\ 
&$0.003311$&$-5.263519$&$0.020511$&$0.001104$& {\bf Stable}  \\ 
&$0.003311$&$-138.990867$&$-0.584233$&$0.001104$& Saddle  \\ 
\hline 
\end{tabular}
\end{center}
\begin{center}
\caption{Two loop $\MSbar$ scheme $SU(3)$ MAG fixed points for
$9$~$\leq$~$\Nf$~$\leq$~$16$.}
\label{2lmagsu3cfw}
\end{center}
\end{table}}
 
{\begin{table}
\begin{center}
 \begin{tabular}{ |c||c|c|c|c|c| } 
  \hline 
 $\Nf$ & $a_{\infty}$ &$\alpha_{\infty}$ &$\omega_1$&$\omega_2$& Infrared Stability \\ 
\hline
\hline
 6&$1.012686$&$1.066181$&$-936.218875$&$40.841486$& Saddle  \\ 
&$1.012686$&$-0.302082$&$614.234192$&$40.841486$& {\bf Stable}  \\ 
&$1.012686$&$-3.789323$&$-1653.674943$&$40.841486$& Saddle  \\ 
&$1.012686$&$-14.751998$&$22323.155240$&$40.841486$& {\bf Stable}  \\ 
\hline 
 7&$0.195518$&$0.707879$&$-5.833518$&$2.986261$& Saddle  \\ 
&$0.195518$&$-0.392201$&$4.297884$&$2.986261$& {\bf Stable}  \\ 
&$0.195518$&$-4.599116$&$-14.711952$&$2.986261$& Saddle  \\
\hline 
\end{tabular}
\end{center}
\begin{center}
\caption{Three loop $\MSbar$ scheme $SU(3)$ MAG fixed points for
$6$~$\leq$~$\Nf$~$\leq$~$16$.}
\label{3lmagsu3cfw}
\end{center}
\end{table}}
 
{\begin{table}
\begin{center}
 \begin{tabular}{ |c||c|c|c|c|c| } 
  \hline 
 $\Nf$ & $a_{\infty}$ &$\alpha_{\infty}$ &$\omega_1$&$\omega_2$& Infrared Stability \\ 
\hline
\hline
 8&$0.052219$&$-3.795565$&$-0.151187 + 0.374262 i$&$\omega_1^*$& Unstable  \\ 
\hline 
 9&$0.416667$&$0.000000$&$2.083333$&$-6.171875$& Saddle  \\ 
&$0.048504$&$-3.705259$&$-0.074519 + 0.326283 i$&$\omega_1^*$& Unstable  \\ 
\hline 
 10&$0.175676$&$0.000000$&$0.761261$&$-1.508081$& Saddle  \\ 
&$0.043898$&$-3.608171$&$-0.013262 + 0.266097 i$&$\omega_1^*$& Unstable  \\ 
\hline 
 11&$0.098214$&$0.000000$&$0.360119$&$-0.651766$& Saddle  \\ 
&$0.038351$&$-3.505155$&$0.029014 + 0.199294 i$&$\omega_1^*$& {\bf Stable}  \\ 
\hline 
 12&$0.060000$&$0.000000$&$0.180000$&$-0.342900$& Saddle  \\ 
&$0.031919$&$-3.398484$&$0.050721 + 0.133051 i$&$\omega_1^*$& {\bf Stable}  \\ 
\hline 
 13&$0.037234$&$0.000000$&$0.086879$&$-0.193548$& Saddle  \\ 
&$0.024812$&$-3.292241$&$0.053562 + 0.074726 i$&$\omega_1^*$& {\bf Stable}  \\ 
\hline 
 14&$0.022124$&$0.000000$&$0.036873$&$-0.108001$& Saddle  \\ 
&$0.017389$&$-3.191979$&$0.042741 + 0.029067 i$&$\omega_1^*$& {\bf Stable}  \\ 
\hline 
 15&$0.011364$&$0.000000$&$0.011364$&$-0.053170$& Saddle  \\ 
&$0.010070$&$-3.103310$&$0.013247$&$0.037372$& {\bf Stable}  \\ 
\hline 
 16&$0.003311$&$0.000000$&$0.001104$&$-0.015057$& Saddle  \\ 
&$0.003200$&$-3.030182$&$0.001115$&$0.014086$& {\bf Stable}  \\ 
\hline 
\end{tabular}
\end{center}
\begin{center}
\caption{Two loop $\mMOM$ scheme $SU(3)$ linear covariant gauge fixed points 
for $8$~$\leq$~$\Nf$~$\leq$~$16$.}
\label{2lmmsu3cfw}
\end{center}
\end{table}}
 
{\begin{table}
\begin{center}
 \begin{tabular}{ |c||c|c|c|c|c| } 
  \hline 
 $\Nf$ & $a_{\infty}$ &$\alpha_{\infty}$ &$\omega_1$&$\omega_2$& Infrared Stability \\ 
\hline
\hline
 8&$0.089050$&$0.000000$&$1.014521$&$-0.712285$& Saddle  \\ 
&$0.247362$&$1.634268$&$4.420438$&$26.650609$& {\bf Stable}  \\ 
\hline 
 9&$0.064438$&$0.000000$&$0.594550$&$-0.406353$& Saddle  \\ 
&$0.188070$&$1.757395$&$3.267307$&$15.301919$& {\bf Stable}  \\ 
\hline 
 10&$0.049421$&$0.000000$&$0.368071$&$-0.275001$& Saddle  \\ 
&$0.162126$&$1.839694$&$2.847656$&$11.795069$& {\bf Stable}  \\ 
\hline 
 11&$0.038603$&$0.000000$&$0.227456$&$-0.199789$& Saddle  \\ 
&$0.042908$&$-3.513882$&$0.075548$&$0.453675$& {\bf Stable}  \\ 
&$0.093269$&$-3.787885$&$-0.107796$&$5.240065$& Saddle  \\ 
&$0.147705$&$1.893386$&$2.651692$&$10.179310$& {\bf Stable}  \\ 
&$0.481048$&$-3.475076$&$2.964235$&$647.235163$& {\bf Stable}  \\ 
\hline 
 12&$0.029962$&$0.000000$&$0.134887$&$-0.148283$& Saddle  \\ 
&$0.031269$&$-3.323533$&$0.087686$&$0.196713$& {\bf Stable}  \\ 
&$0.138858$&$1.925597$&$2.556433$&$9.295937$& {\bf Stable}  \\ 
\hline 
 13&$0.022535$&$0.000000$&$0.073340$&$-0.108285$& Saddle  \\ 
&$0.023188$&$-3.213214$&$0.065347$&$0.109665$& {\bf Stable}  \\ 
&$0.133279$&$1.940673$&$2.518423$&$8.776404$& {\bf Stable}  \\ 
\hline 
 14&$0.015786$&$0.000000$&$0.033847$&$-0.074248$& Saddle  \\ 
&$0.016146$&$-3.138534$&$0.033290$&$0.070371$& {\bf Stable}  \\ 
&$0.129903$&$1.941314$&$2.519382$&$8.470959$& {\bf Stable}  \\ 
\hline 
 15&$0.009383$&$0.000000$&$0.011019$&$-0.043354$& Saddle  \\ 
&$0.009537$&$-3.080626$&$0.011009$&$0.041625$& {\bf Stable}  \\ 
&$0.128198$&$1.929163$&$2.551400$&$8.310242$& {\bf Stable}  \\ 
\hline 
 16&$0.003118$&$0.000000$&$0.001100$&$-0.014158$& Saddle  \\ 
&$0.003138$&$-3.027421$&$0.001100$&$0.013960$& {\bf Stable}  \\ 
&$0.127908$&$1.905111$&$2.611967$&$8.260995$& {\bf Stable}  \\ 
\hline 
\end{tabular}
\end{center}
\begin{center}
\caption{Three loop $\mMOM$ scheme $SU(3)$ linear covariant gauge fixed points 
for $8$~$\leq$~$\Nf$~$\leq$~$16$.}
\label{3lmmsu3cfw}
\end{center}
\end{table}}
 
{\begin{table}
\begin{center}
 \begin{tabular}{ |c||c|c|c|c|c| } 
  \hline 
 $\Nf$ & $a_{\infty}$ &$\alpha_{\infty}$ &$\omega_1$&$\omega_2$& Infrared Stability \\ 
\hline
\hline
 8&$0.072796$&$0.000000$&$0.966044$&$-0.474778$& Saddle  \\ 
&$0.363423$&$0.712602$&$31.432482$&$87.465149$& {\bf Stable}  \\ 
&$0.060959$&$-3.485534$&$0.121193 + 0.476810 i$&$\omega_1^*$& {\bf Stable}  \\ 
\hline 
 9&$0.054935$&$0.000000$&$0.582834$&$-0.292463$& Saddle  \\ 
&$0.164446$&$0.956855$&$5.810601 + 4.718666 i$&$\omega_1^*$& {\bf Stable}  \\ 
&$0.055612$&$-3.212828$&$0.297571 + 0.228641 i$&$\omega_1^*$& {\bf Stable}  \\ 
\hline 
 10&$0.044230$&$0.000000$&$0.368153$&$-0.218958$& Saddle  \\
\hline 
\end{tabular}
\end{center}
\begin{center}
\caption{Four loop $\mMOM$ scheme $SU(3)$ linear covariant gauge fixed points 
for $8$~$\leq$~$\Nf$~$\leq$~$16$.}
\label{4lmmsu3cfw}
\end{center}
\end{table}}
 
{\begin{table}
\begin{center}
 \begin{tabular}{ |c||c|c|c|c|c| } 
  \hline 
 $\Nf$ & $a_{\infty}$ &$\alpha_{\infty}$ &$\omega_1$&$\omega_2$& Infrared Stability \\ 
\hline
\hline
 8&$0.050162$&$0.000000$&$0.928758$&$-0.249221$& Saddle  \\ 
&$0.047937$&$-3.192836$&$0.742629$&$0.261331$& {\bf Stable}  \\ 
&$0.178070$&$1.377827$&$4.732069$&$173.467310$& {\bf Stable}  \\ 
&$8.186013$&$1.470023$&$-7742265.147766$&$32959305743.666241$& Saddle  \\ 
\hline 
 9&$0.044015$&$0.000000$&$0.608228$&$-0.202835$& Saddle  \\ 
&$0.043123$&$-3.003521$&$0.510250$&$0.209211$& {\bf Stable}  \\ 
&$0.116454$&$1.315030$&$3.633119$&$18.970870$& {\bf Stable}  \\
\hline 
\end{tabular}
\end{center}
\begin{center}
\caption{Five loop $\mMOM$ scheme $SU(3)$ linear covariant gauge fixed points 
for $8$~$\leq$~$\Nf$~$\leq$~$16$.}
\label{5lmmsu3cfw}
\end{center}
\end{table}}
 
{\begin{table}
\begin{center}
 \begin{tabular}{ |c||c|c|c|c|c| } 
  \hline 
 $\Nf$ & $a_{\infty}$ &$\alpha_{\infty}$ &$\omega_1$&$\omega_2$& Infrared Stability \\ 
\hline
\hline
 9&$0.033833$&$-8.143603$&$-0.141127 + 0.284271 i$&$\omega_1^*$& Unstable  \\ 
\hline 
 10&$0.034117$&$-7.784169$&$-0.089868 + 0.260414 i$&$\omega_1^*$& Unstable  \\ 
&$0.323141$&$0.519306$&$1.041758$&$-4.497781$& Saddle  \\ 
\hline 
 11&$0.033655$&$-7.364723$&$-0.033540 + 0.222947 i$&$\omega_1^*$& Unstable  \\ 
&$0.123479$&$0.165298$&$0.406388$&$-1.137423$& Saddle  \\ 
&$5.519238$&$1.674527$&$188.897763$&$396.630816$& {\bf Stable}  \\ 
\hline 
 12&$0.031606$&$-6.879356$&$0.020789 + 0.167020 i$&$\omega_1^*$& {\bf Stable}  \\ 
&$0.067453$&$-0.046271$&$0.190327$&$-0.515078$& Saddle  \\ 
&$1.476988$&$1.916484$&$42.559782 + 20.144833 i$&$\omega_1^*$& {\bf Stable}  \\ 
\hline 
 13&$0.026761$&$-6.364145$&$0.055556 + 0.095535 i$&$\omega_1^*$& {\bf Stable}  \\ 
&$0.039716$&$-0.191430$&$0.089168$&$-0.273455$& Saddle  \\ 
&$0.806259$&$2.117233$&$18.914033 + 11.732178 i$&$\omega_1^*$& {\bf Stable}  \\ 
\hline 
 14&$0.019089$&$-5.932927$&$0.055911 + 0.026515 i$&$\omega_1^*$& {\bf Stable}  \\ 
&$0.022912$&$-0.293143$&$0.037277$&$-0.149419$& Saddle  \\ 
&$0.546940$&$2.294665$&$11.561464 + 8.159073 i$&$\omega_1^*$& {\bf Stable}  \\ 
\hline 
 15&$0.011554$&$-0.364086$&$0.011402$&$-0.073734$& Saddle  \\ 
&$0.010713$&$-5.675674$&$0.012346$&$0.056948$& {\bf Stable}  \\ 
&$0.412326$&$2.456372$&$8.208946 + 6.298732 i$&$\omega_1^*$& {\bf Stable}  \\ 
\hline 
 16&$0.003326$&$-0.412862$&$0.001104$&$-0.021234$& Saddle  \\ 
&$0.003263$&$-5.575657$&$0.001109$&$0.020108$& {\bf Stable}  \\ 
&$0.330726$&$2.606374$&$6.351019 + 5.181185 i$&$\omega_1^*$& {\bf Stable}  \\ 
\hline 
\end{tabular}
\end{center}
\begin{center}
\caption{Two loop MOMc scheme $SU(3)$ fixed points for the MAG for
$9$~$\leq$~$\Nf$~$\leq$~$16$.}
\label{2lsu3momcmagcfw}
\end{center}
\end{table}}

{\begin{table}
\begin{center}
 \begin{tabular}{ |c||c|c|c|c|c| } 
  \hline 
 $\Nf$ & $a_{\infty}$ &$\alpha_{\infty}$ &$\omega_1$&$\omega_2$& Infrared Stability \\ 
\hline
\hline
 9&$0.001323$&$-63.474435$&$-0.144700 + 0.059892 i$&$\omega_1^*$& Unstable  \\ 
&$0.060025$&$-0.761340$&$0.568218$&$-0.381005$& Saddle  \\ 
&$2.014127$&$-4.819139$&$387.341962$&$37185.984046$& {\bf Stable}  \\ 
\hline 
 10&$0.001338$&$-63.103585$&$-0.145138 + 0.059715 i$&$\omega_1^*$& Unstable  \\ 
&$0.048050$&$-0.606026$&$0.365311$&$-0.288998$& Saddle  \\ 
&$0.525666$&$-5.116521$&$25.061445$&$792.707980$& {\bf Stable}  \\ 
\hline 
 11&$0.001352$&$-62.729216$&$-0.145579 + 0.059527 i$&$\omega_1^*$& Unstable  \\
\hline 
\end{tabular}
\end{center}
\begin{center}
\caption{Three loop MOMc scheme $SU(3)$ fixed points for the MAG for 
$9$~$\leq$~$\Nf$~$\leq$~$16$.}
\label{3lsu3momcmagcfw}
\end{center}
\end{table}}
 
{\begin{table}
\begin{center}
 \begin{tabular}{ |c||c|c|c|c|c| } 
  \hline 
 $\Nf$ 	& Type 	&$\gamma_A$	&$\gamma_c$	&$\gamma_\psi$ 	&$\rho_m$ \\ 
\hline
  \hline
  8	&(IRS)/BZ	& $-0.595200$&$0.297600$&$-0.095887$&$-1.562280$   \\ \hline 
9	&BZ	&$-0.017807$&$0.008903$&$-0.048726$&$-0.143490$   \\\hline 
10	&BZ	& $0.096514$&$-0.048257$&$-0.026915$&$0.155885$  \\
	&IRS	&$ 0$	&$-0.070828$&$-0.098415$&$0.155885$  \\\hline 
11	&BZ	& $0.125835$&$-0.062918$&$-0.012892$&$0.249686$   \\
	&IRS	& $0$	&$-0.105421$&$-0.115454$&$0.249686$  \\\hline 
12	&BZ	& $0.121873$&$-0.060936$&$-0.005073$&$0.253328$  \\
	&IRS	& $ 0$	&$-0.109067$&$-0.108915$&$0.253328$    \\\hline 
13	&BZ	& $0.101310$&$-0.050655$&$-0.001559$&$0.209757$   \\
	&IRS	& $ 0$	&$-0.094220$&$-0.089777$&$0.209757$  \\\hline 
14	&BZ	& $0.073373$&$-0.036686$&$-0.000337$&$0.147421$   \\
	&IRS	& $ 0$	&$-0.070098$&$-0.064964$&$0.147421$   \\\hline 
15	&BZ	&  $0.043478$&$-0.021739$&$-0.000034$&$0.083600$  \\
	&IRS	& $ 0$	&$-0.042422$&$-0.038564$&$0.083600$    \\\hline 
16	&BZ	&$0.014174$&$-0.007087$&$0.000001$&$0.025895$  \\
	&IRS	& $ 0$	&$-0.014069$&$-0.012594$&$0.025895$     \\\hline 
\end{tabular}
\end{center}
\begin{center}
\caption{Four loop $\MSbar$ scheme $SU(3)$ linear covariant gauge critical 
exponents.}
\label{4lmssu3exp}
\end{center}
\end{table}}

{\begin{table}
\begin{center}
 \begin{tabular}{ |c||c|c|c|c|c| } 
  \hline 
 $\Nf$ 	&Type	&$\gamma_A$	&$\gamma_c$	&$\gamma_\psi$ 	&$\rho_m$ \\ 
\hline
  \hline 8	&  - 	& -	& - 	&- 	&-  \\\hline
  9	&  - 	& -	& - 	&- 	&-  \\\hline 
10	&  - 	& -	& - 	&- 	&-  \\\hline 
11	&  - 	& -	& - 	&- 	&-  \\\hline 
12	&  - 	& -	& - 	&- 	&-  \\\hline 
13	&BZ	& $0.122684$&$-0.061342$&$-0.001436$&$0.238701$  \\
	&IRS	& $ 0$	&$-0.115113$&$-0.109348$&$0.238701$   \\\hline 
14	&BZ	& $0.076782$&$-0.038391$&$-0.000170$&$0.154060$   \\
	&IRS	& $ 0$	&$-0.073395$&$-0.067884$&$0.154060$  \\\hline 
15	&BZ	& $0.043850$&$-0.021925$&$-0.000019$&$0.084340$  \\
	&IRS	& $ 0$	&$-0.042786$&$-0.038884$&$0.084340$   \\\hline 
16	&BZ	& $0.014178$&$-0.007089$&$0.000001$&$0.025903$   \\
	&IRS	& $ 0$	&$-0.014074$&$-0.012597$&$0.025903$ \\\hline 
\end{tabular}
\end{center}
\begin{center}
\caption{Five loop $\MSbar$ scheme $SU(3)$ linear covariant gauge critical
exponents.}
\label{5lmssu3exp}
\end{center}
\end{table}}

{\begin{table}
\begin{center}
 \begin{tabular}{ |c||c||c|c|c|c| } 
  \hline 
 $\Nf$ 	&Type	&$\gamma_A$	&$\gamma_c$	&$\gamma_\psi$ 	&$\rho_m$ \\ 
\hline
  \hline 
8	&BZ	& $0.474778$&$-0.237389$&$-0.017956$&$2.025675$   \\
	&(IRS)	& $0$	&$-2.258175$&$-60.188725$&$260.687360$ \\
	&IRS	& $0$	&$-0.265079$&$-0.226746$&$1.044736$   \\\hline 
9	&BZ	& $0.292463$&$-0.146232$&$-0.010391$&$0.979184$   \\
	&(IRS)	& $0$	&$0.001655$&$-2.579132$&$7.594993$  \\
	&IRS	& $0$	&$-0.222793$&$-0.215257$&$0.836272$   \\\hline 
10	&BZ	& $0.218958$&$-0.109479$&$-0.006569$&$0.620806$  \\
	&(IRS)	& $0$	&$0.117721$&$-1.072557$&$1.658957$   \\
	&IRS	& $ 0$	&$-0.184374$&$-0.182289$&$0.593885$  \\\hline 
11	&BZ	&$0.173654$&$-0.086827$&$-0.003834$&$0.436592$\\
	&(IRS)	& $ 0$	&$0.154522$&$-0.702656$&$0.071329$  \\
	&IRS	& $ 0$	&$-0.154669$&$-0.151401$&$0.434220$ \\\hline 
12	&BZ	& $0.137930$&$-0.068965$&$-0.001939$&$0.317156$   \\
	&(IRS)	& $0$	&$0.171764$&$-0.572375$&$-0.740644$  \\
	&IRS	& $ 0$	&$-0.127538$&$-0.122541$&$0.320300$ \\\hline 
13	&BZ	& $0.105460$&$-0.052730$&$-0.000786$&$0.226367$  \\
	&(IRS)	& $0$	&$0.180910$&$-0.539644$&$-1.365889$   \\
	&IRS	& $ 0$	&$-0.099963$&$-0.094153$&$0.229135$  \\\hline 
14	&BZ	& $0.074090$&$-0.037045$&$-0.000220$&$0.150241$   \\
	&(IRS)	& $0$	&$0.184904$&$-0.573180$&$-2.017939$   \\
	&IRS	& $ 0$	&$-0.071427$&$-0.066024$&$0.151374$ \\\hline 
15	&BZ	&$0.043538$&$-0.021769$&$-0.000027$&$0.083816$  \\
	&IRS	& $0$	&$-0.042586$&$-0.038701$&$0.084007$  \\\hline 
16	&BZ	&$0.014174$&$-0.007087$&$0.000001$&$0.025896$ \\
	&IRS	& $0$	&$-0.014072$&$-0.012595$&$0.025899$  \\\hline 
\end{tabular}
\end{center}
\begin{center}
\caption{Four loop $\mMOM$ scheme $SU(3)$ linear covariant gauge critical
exponents.}
\label{4lmmsu3exp}
\end{center}
\end{table}}

{\begin{table}
\begin{center}
\begin{tabular}{ |c||c||c|c|c|c| } 
  \hline 
 $\Nf$ 	&Type	&$\gamma_A$	&$\gamma_c$	&$\gamma_\psi$ 	&$\rho_m$ \\ 
\hline
  \hline
8	&BZ	& $0.249221$&$-0.124610$&$-0.001137$&$1.005585$ \\
	&(IRS)	& $0$	&$2.952013$&$6.488203$&$20.134644$  \\
	&IRS 	& $0$	&$-0.189463$&$-0.182394$&$0.692361$  \\\hline 
9	&BZ	& $0.202835$&$-0.101418$&$-0.005582$&$0.672233$   \\
	&(IRS)	& $0$	&$0.575398$&$0.816390$&$0.424152$  \\
	&IRS	& $0$	&$-0.166196$&$-0.162738$&$0.551290$  \\\hline 
10	&BZ	& $0.178382$&$-0.089191$&$-0.005186$&$0.503784$ \\
	&(IRS)	& $ 0$	&$0.414674$&$0.449910$&$-1.079122$ \\
	&IRS	& $0$	&$-0.151697$&$-0.148790$&$0.450991$  \\\hline 
11	&BZ	& $0.158660$&$-0.079330$&$-0.003476$&$0.395544$  \\
	&(IRS)	& $0$	&$0.325593$&$0.276743$&$-1.666897$  \\
	&IRS	& $ 0$	&$-0.139217$&$-0.135229$&$0.373631$  \\\hline 
12	&BZ	&  $0.135918$&$-0.067959$&$-0.001749$&$0.309177$   \\
	&(IRS)	& $0$	&$0.227143$&$0.131599$&$-2.007590$  \\
	&IRS	& $0$	&$-0.123246$&$-0.117907$&$0.302656$ \\\hline 
13	&BZ	& $0.107263$&$-0.053632$&$-0.000633$&$0.228539$  \\
	&IRS	& $0$	&$-0.100243$&$-0.094248$&$0.227759$  \\\hline 
14	&BZ	& $0.075242$&$-0.037621$&$-0.000158$&$0.152168$  \\
	&IRS	& $ 0$	&$-0.071998$&$-0.066518$&$0.152228$  \\\hline 
15	&BZ	& $0.043759$&$-0.021879$&$-0.000019$&$0.084206$ \\
	&IRS	& $0$	&$-0.042706$&$-0.038808$&$0.084220$  \\\hline 
16	&BZ	& $0.014178$&$-0.007089$&$0.000001$&$0.025902$  \\
	&IRS	& $ 0$	&$-0.014073$&$-0.012597$&$0.025902$  \\\hline 
\end{tabular}
\end{center}
\begin{center}
\caption{Five loop $\mMOM$ scheme $SU(3)$ linear covariant gauge critical
exponents.}
\label{5lmmsu3exp}
\end{center}
\end{table}}

{\begin{table}
\begin{center}
 \begin{tabular}{ |c||c||c|c|c|c| } 
  \hline 
 $\Nf$ 	& Type 	&$\gamma_A$	&$\gamma_c$	&$\gamma_\psi$ 	&$\rho_m$ \\ 
\hline
  \hline 
8	&BZ	& $6.471485$&$-0.141493$&$-0.216872$&$10.793654$ \\
	&(IRS)	& $0$	&$-2.424696$&$3.827038$&$19.536414$  \\
\hline 
9	&BZ	& $1.758067$&$0.145285$&$-0.189162$&$1.462828$  \\
	&(IRS)	& $0$	&$-0.419659$&$0.935252$&$3.663383$   \\
\hline 
10	&BZ	&  $0.655285$&$0.059319$&$-0.099279$&$0.342432$  \\
	&(IRS)	& $0$	&$-0.121571$&$0.388783$&$1.145796$   \\
\hline 
11	&BZ	& $0.287141$&$-0.011344$&$-0.042769$&$0.221298$  \\
	&(IRS)	& $0$ 	&$-0.056209$&$0.264999$&$0.589514$   \\
	&IRS	& $ 0$	&$-0.020924$&$-0.095855$&$0.186593$  \\ \hline 
12	&BZ	& $0.159018$&$-0.042487$&$-0.014862$&$0.224515$  \\
	&(IRS)	& $ 0$	&$-0.025747$&$0.250893$&$0.442411$   \\
	&IRS	& $ 0$	&$-0.082487$&$-0.104959$&$0.209696$   \\ \hline 
13	&BZ	& $0.107236$&$-0.045861$&$-0.003945$&$0.199534$  \\
	&(IRS)	& $ 0$	&$0.006053$&$0.249136$&$0.348713$   \\
	&IRS	& $0$	&$-0.087072$&$-0.087682$&$0.196374$ \\ \hline 
14	&BZ	& $0.073867$&$-0.035894$&$-0.000711$&$0.145441$  \\
	&(IRS)	& $0$	&$0.033897$&$0.227024$&$0.245013$   \\
	&IRS	& $ 0$	&$-0.068745$&$-0.064264$&$0.144981$   \\ \hline 
15	&BZ	& $0.043484$&$-0.021684$&$-0.000059$&$0.083448$  \\
	&(IRS)	& $0$	&$0.049227$&$0.181022$&$0.139471$   \\
	&IRS	& $ 0$	&$-0.042316$&$-0.038493$&$0.083421$  \\ \hline 
16	&BZ	&$0.014174$&$-0.007087$&$0.000001$&$0.025894$   \\
	&(IRS)	& $0$	&$0.043509$&$0.105463$&$0.043333$  \\
	&IRS	& $ 0$	&$-0.014069$&$-0.012593$&$0.025894$  \\ \hline 
\end{tabular}
\end{center}
\begin{center}
\caption{Four loop $\RI$ scheme $SU(3)$ linear covariant gauge critical
exponents.}
\label{4lrisu3exp}
\end{center}
\end{table}}
 
{\begin{table}
\begin{center}
 \begin{tabular}{ |c||c||c|c|c|c| } 
  \hline 
 $\Nf$ 	& Type 	&$\gamma_A$	&$\gamma_c$	&$\gamma_\psi$ 	&$\gamma_m$ \\ 
\hline
  \hline 
8	&  - 	&- 	&- 	&- 	&- \\\hline 
9	&  - 	&- 	&- 	&- 	&- \\\hline 
10	& - 	&- 	&- 	&- 	&- \\\hline 
11	& - 	&- 	&- 	&- 	&- \\\hline 
12	& - 	&- 	&- 	&- 	&- \\\hline 
13	&BZ	& $0.070465$&$-0.051662$&$-0.004040$&$0.187491$  \\
	&(IRS)	& $0$	&$-0.033365$&$0.181740$&$0.313061$   \\
	&IRS	& $ 0$	&$-0.081046$&$-0.060264$&$0.187045$   \\ \hline 
14	&BZ	&$0.073887$&$-0.037954$&$-0.000270$&$0.151600$  \\
	&(IRS)	& $0$	&$0.016256$&$0.203936$&$0.249187$  \\
	&IRS	& $ 0$	&$-0.071499$&$-0.064958$&$0.151500$  \\
\hline 
15	&BZ	& $0.043744$&$-0.021913$&$-0.000020$&$0.084266$  \\
	&(IRS)	& $ 0$	&$0.038957$&$0.163924$&$0.138931$  \\
	&IRS	& $ 0$	&$-0.042724$&$-0.038782$&$0.084269$  \\
\hline 
16	&BZ	& $0.014178$&$-0.007089$&$0.000001$&$0.025902$ \\
	&(IRS)	& $ 0$	&$0.038776$&$0.097100$&$0.043049$  \\
	&IRS	& $ 0$	&$-0.014074$&$-0.012597$&$0.025902$   \\
\hline 
\end{tabular}
\end{center}
\begin{center}
\caption{Five loop $\RI$ scheme $SU(3)$ linear covariant gauge critical
exponents.}
\label{5lrisu3exp}
\end{center}
\end{table}}

{\begin{table}
\begin{center}
 \begin{tabular}{ |c||c||c|c|c|c| } 
  \hline 
 $\Nf$ 	& Type 	&$\gamma_A$	&$\gamma_c$	&$\gamma_\psi$ 	&$\rho_m$ \\ 
\hline
  \hline
8	&BZ	& $0.276082$&$-0.413130$&$0.099159$&$1.331235$  \\\hline 
9	&BZ	& $0.224312$&$-0.262619$&$0.042710$&$0.954324$  \\
\hline 
10	&BZ	& $0.193650$&$-0.182608$&$0.020019$&$0.701362$  \\
\hline 
11	&BZ	& $0.166770$&$-0.131130$&$0.009440$&$0.516777$  \\
\hline 
12	&BZ	& $0.138971$&$-0.094074$&$0.004212$&$0.374024$  \\
\hline 
13	&BZ	& $0.109064$&$-0.065442$&$0.001649$&$0.259356$  \\
	&IRS	& $0$	&$-0.155825$&$-0.139024$&$0.343730$  \\\hline 
14	&BZ	& $0.077336$&$-0.042319$&$0.000497$&$0.165375$  \\
	&IRS	& $0$	&$-0.093956$&$-0.084584$&$0.196204$  \\\hline 
15	&BZ	& $0.045066$&$-0.023188$&$0.000086$&$0.088262$   \\
	&IRS	& $0$	&$-0.048493$&$-0.043719$&$0.095293$   \\
\hline 
16	&BZ	& $0.014301$&$-0.007163$&$0.000003$&$0.026154$  \\
	&IRS	& $0$	&$-0.014374$&$-0.012860$&$0.026451$   \\ \hline 
\end{tabular}
\end{center}
\begin{center}
\caption{Three loop $\MOMg$ scheme $SU(3)$ linear covariant gauge critical
exponents.}
\label{3lmgsu3exp}
\end{center}
\end{table}}
 
{\begin{table}
\begin{center}
 \begin{tabular}{ |c||c||c|c|c|c| } 
  \hline 
 $\Nf$ 	&Type	&$\gamma_A$	&$\gamma_c$	&$\gamma_\psi$ 	&$\rho_m$ \\ 
\hline
  \hline
9	&BZ	& $0.329837$&$-0.149574$&$0.017904$&$0.375534$   \\
	&(IRS)	& $0$	&$0.391474$&$-0.086183$&$-9.428263$  \\
	&IRS	& $ 0$	&$-0.347993$&$-0.304238$&$0.754394$  \\\hline 
10	&BZ	& $0.233644$&$-0.110954$&$0.006969$&$0.377682$  \\
	&(IRS)	& $ 0$	&$0.277219$&$0.102546$&$-4.979258$  \\
	&IRS	& $ 0$	&$-0.241319$&$-0.224706$&$0.544544$  \\
\hline 
11	&BZ	& $0.176390$&$-0.086067$&$0.002517$&$0.330763$ \\
	&(IRS)	& $0$	&$0.228272$&$0.166763$&$-3.142147$  \\
	&IRS	& $ 0$	&$-0.179505$&$-0.169441$&$0.408303$  \\\hline 
12	&BZ	& $0.135538$&$-0.067109$&$0.000730$&$0.270097$  \\
	&(IRS)	& $0$	&$0.202150$&$0.193153$&$-2.170112$  \\
	&IRS	& $ 0$&$-0.136076$&$-0.128086$&$0.305591$  \\\hline 
13	&BZ	& $0.102106$&$-0.050907$&$0.000115$&$0.206167$   \\
	&(IRS)	& $ 0$	&$0.186388$&$0.203886$&$-1.568987$   \\
	&IRS	& $ 0$	&$-0.101443$&$-0.094591$&$0.220930$  \\\hline 
14	&BZ	& $0.071889$&$-0.035931$&$-0.000017$&$0.142818$   \\
	&(IRS)	& $0$	&$0.176148$&$0.206352$&$-1.150713$  \\
	&IRS	& $ 0$	&$-0.071000$&$-0.065379$&$0.147737$  \\\hline 
15	&BZ	& $0.042821$&$-0.021412$&$-0.000007$&$0.082094$  \\
	&(IRS)	& $0$	&$0.169193$&$0.203158$&$-0.827549$  \\
	&IRS	& $ 0$&$-0.042282$&$-0.038392$&$0.083079$   \\\hline 
16	&BZ	& $0.014138$&$-0.007069$&$0.000001$&$0.025826$  \\
	&(IRS)	& $0$	&$0.164339$&$0.194794$&$-0.550086$  \\
	&IRS	& $ 0$	&$-0.014052$&$-0.012578$&$0.025860$ \\\hline 
\end{tabular}
\end{center}
\begin{center}
\caption{Three loop $\MOMc$ scheme $SU(3)$ linear covariant gauge critical
exponents.}
\label{3lmcsu3exp}
\end{center}
\end{table}}
 
{\begin{table}
\begin{center}
 \begin{tabular}{ |c||c||c|c|c|c| } 
  \hline 
 $\Nf$ 	&Type 	&$\gamma_A$	&$\gamma_c$	&$\gamma_\psi$ 	&$\rho_m$ \\ 
\hline
  \hline
9	&BZ	& $0.189612$&$-0.170966$&$0.020033$&$0.553462$  \\
	&IRS	& $ 0$	&$-0.225997$&$-0.144117$&$0.482308$  \\\hline 
10	&BZ	&$0.173917$&$-0.125391$&$0.009356$&$0.452897$  \\
	&IRS	& $ 0$	&$-0.180608$&$-0.134742$&$0.406464$  \\\hline 
11	&BZ	& $0.151840$&$-0.094592$&$0.004063$&$0.364656$  \\
	&IRS	& $ 0$	&$-0.146612$&$-0.120352$&$0.336490$  \\\hline 
12	&BZ	& $0.126653$&$-0.071608$&$0.001526$&$0.285218$  \\
	&IRS	& $ 0$	&$-0.118634$&$-0.103174$&$0.269817$   \\\hline 
13	&BZ	& $0.099677$&$-0.052935$&$0.000439$&$0.212345$  \\
	&IRS	& $ 0$	&$-0.093224$&$-0.083658$&$0.205116$  \\\hline 
14	&BZ	&$0.071577$&$-0.036630$&$0.000074$&$0.144860$   \\
	&IRS	& $ 0$	&$-0.067999$&$-0.061735$&$0.142206$   \\ \hline 
15	&BZ	& $0.042870$&$-0.021550$&$0.000005$&$0.082504$  \\
	&IRS	& $ 0$	&$-0.041642$&$-0.037698$&$0.081920$  \\ \hline 
16	&BZ	& $0.014144$&$-0.007073$&$0.000002$&$0.025840$ \\
	&IRS	& $ 0$	&$-0.014030$&$-0.012556$&$0.025819$   \\
\hline 
\end{tabular}
\end{center}
\begin{center}
\caption{Three loop $\MOMq$ scheme $SU(3)$ linear covariant gauge critical
exponents.}
\label{3lmqsu3exp}
\end{center}
\end{table}}
 
{\begin{table}
\begin{center}
 \begin{tabular}{ |c||c||c|c|c|c| } 
  \hline 
 $\Nf$ 	&Type 	& $\gamma_A$	&$\gamma_c$	&$\gamma_\psi$ 	&$\rho_m$ \\ 
\hline
  \hline
9	&BZ	& $0.224312$&$-0.262619$&$0.042710$&$0.954324$  \\
	&(IRS)	&$1.452883$&$-0.120220$&$2.190266$&$2.552738$   \\
\hline 
10	&BZ	& $0.193650$&$-0.182608$&$0.020019$&$0.701362$ \\
	&(IRS)	&$5.227316$&$0.353456$&$7.616440$&$7.175322$  \\
\hline 
11	&BZ	& $0.166770$&$-0.131130$&$0.009440$&$0.516777$  \\
\hline 
12	&BZ	& $0.138971$&$-0.094074$&$0.004212$&$0.374024$  \\
\hline 
13	&BZ	& $0.109064$&$-0.065442$&$0.001649$&$0.259356$  \\
\hline 
14	&BZ	& $0.077336$&$-0.042319$&$0.000497$&$0.165375$  \\
\hline 
15	&BZ	& $0.045066$&$-0.023188$&$0.000086$&$0.088262$  \\ \hline 
16	&BZ	& $0.014301$&$-0.007163$&$0.000003$&$0.026154$  \\
	&IRS	& $-0.017872$&$-0.026734$&$-0.032502$&$0.033489$  \\ \hline 
\end{tabular}
\end{center}
\begin{center}
\caption{Three loop $\MOMg$ scheme $SU(3)$ Curci-Ferrari gauge critical
exponents.}
\label{3lmgcfsu3exp}
\end{center}
\end{table}}
 
{\begin{table}
\begin{center}
 \begin{tabular}{ |c||c||c|c|c|c| } 
  \hline 
 $\Nf$ 	& Type 	&$\gamma_A$	&$\gamma_c$	&$\gamma_\psi$ 	&$\rho_m$ \\ 
\hline
  \hline
9	&BZ	& $-0.240091$&$-0.133464$&$0.002704$&$0.306029$  \\
	&(IRS)	& $ 0$	&$0.003384$&$0.147923$&$-0.471734$  \\
	&IRS	& $ 0$	&$-0.541331$&$-0.439621$&$0.684557$  \\ \hline 
10	&BZ	& $-0.188899$&$-0.102092$&$0.000514$&$0.326369$  \\
	&(IRS)	& $0$	&$0.028004$&$0.173992$&$-0.254634$  \\
	&IRS	& $ 0$	&$-0.399200$&$-0.360987$&$0.582751$   \\ \hline 
11	&BZ	& $-0.153163$&$-0.080911$&$-0.000309$&$0.298706$  \\
	&(IRS)	& $0$	&$0.044654$&$0.192948$&$-0.142850$   \\
	&IRS	& $ 0$	&$-0.312859$&$-0.306670$&$0.490657$  \\ \hline 
12	&BZ	& $-0.123892$&$-0.064194$&$-0.000459$&$0.252326$   \\
	&(IRS)	& $0$	&$0.057411$&$0.209116$&$-0.074662$  \\
	&IRS	& $ 0$	&$-0.252137$&$-0.263491$&$0.407208$  \\ \hline 
13	&BZ	& $-0.096909$&$-0.049431$&$-0.000326$&$0.197780$   \\
	&(IRS)	& $0$	&$0.068224$&$0.224669$&$-0.025628$  \\
	&IRS	& $ 0$	&$-0.202447$&$-0.223580$&$0.326803$  \\ \hline 
14	&BZ	& $-0.070077$&$-0.035343$&$-0.000140$&$0.139805$   \\
	&(IRS)	& $0$	&$0.078333$&$0.241159$&$0.016759$  \\
	&IRS	& $ 0$	&$-0.151857$&$-0.177302$&$0.237937$   \\ \hline 
15	&BZ 	& $-0.042456$&$-0.021275$&$-0.000025$&$0.081477$  \\
	&(IRS)	& $0$	&$0.088848$&$0.260182$&$0.062111$   \\
	&IRS	& $ 0$	&$-0.074893$&$-0.092871$&$0.106324$   \\ \hline 
16	&BZ	& $-0.014126$&$-0.007064$&$0.000001$&$0.025806$  \\
	&(IRS) 	& $0$	&$0.101281$&$0.283939$&$0.123768$   \\
	&IRS	& $ 0$	&$-0.020939$&$-0.025335$&$0.026249$  \\ \hline 
\end{tabular}
\end{center}
\begin{center}
\caption{Three loop $\MOMc$ scheme $SU(3)$ Curci-Ferrari gauge critical
exponents.}
\label{3lmccfsu3exp}
\end{center}
\end{table}}
 
{\begin{table}
\begin{center}
 \begin{tabular}{ |c||c||c|c|c|c| } 
  \hline 
 $\Nf$ 	& Type	&$\gamma_A$	&$\gamma_c$	&$\gamma_\psi$ 	&$\rho_m$ \\ 
\hline
  \hline
9	&BZ	& $0.189612$&$-0.170966$&$0.020033$&$0.553462$   \\
	&(IRS)	&$0.312662$&$-0.008175$&$0.475178$&$0.589581$  \\
	&IRS	& $-0.153980$&$-0.292270$&$-0.245203$&$0.493722$   \\ \hline 
10	&BZ	& $0.173917$&$-0.125391$&$0.009356$&$0.452897$  \\
	&(IRS)	& $0.256557$&$0.063582$&$0.393010$&$0.363224$  \\
	&IRS	& $-0.130485$&$-0.231574$&$-0.221890$&$0.412805$  \\
\hline 
11	&BZ	& $0.151840$&$-0.094592$&$0.004063$&$0.364656$   \\
	&(IRS)	& $0.177706$&$0.088548$&$0.280144$&$0.169961$   \\
	&IRS	& $-0.110305$&$-0.186018$&$-0.198121$&$0.338608$  \\ \hline 
12	&BZ	& $0.126653$&$-0.071608$&$0.001526$&$0.285218$  \\
	&IRS	& $-0.092896$&$-0.150331$&$-0.174062$&$0.269299$  \\ \hline 
13	&BZ	& $0.099677$&$-0.052935$&$0.000439$&$0.212345$  \\
	&IRS	&$-0.076738$&$-0.120097$&$-0.147352$&$0.203649$  \\ \hline 
14	&BZ	& $0.071577$&$-0.036630$&$0.000074$&$0.144860$   \\
	&IRS	& $-0.059450$&$-0.090785$&$-0.114343$&$0.141085$   \\
\hline 
15	&BZ	& $0.042870$&$-0.021550$&$0.000005$&$0.082504$  \\
	&IRS	& $-0.038683$&$-0.058270$&$-0.072841$&$0.081552$ \\ \hline 
16	&BZ	& $0.014144$&$-0.007073$&$0.000002$&$0.025840$   \\ 
	&IRS	& $-0.013724$&$-0.020589$&$-0.024909$&$0.025802$  \\
\hline 
\end{tabular}
\end{center}
\begin{center}
\caption{Three loop $\MOMq$ scheme $SU(3)$ Curci-Ferrari gauge critical
exponents.}
\label{3lmqcfsu3exp}
\end{center}
\end{table}}
 
{\begin{table}
\begin{center}
 \begin{tabular}{ |c||c||c|c|c|c| } 
  \hline 
 $\Nf$ 	& Type 	&$\gamma_A$	&$\gamma_c$	&$\gamma_\psi$ 	&$\rho_m$ \\ 
\hline
  \hline
9	&(BZ)	& $0.335178$&$-0.440643$&$-0.034855$&$1.079969$   \\
	&IRS	& $-0.009456$&$-0.193145$&$-0.197174$&$0.506658$  \\\hline 
10	&(BZ)	& $0.275999$&$-0.306410$&$-0.027619$&$0.758291$  \\
	&IRS	& $0.001070$&$-0.171865$&$-0.185892$&$0.450244$  \\\hline 
11	&(BZ)	& $0.228961$&$-0.220004$&$-0.021271$&$0.544096$   \\
	&IRS	& $0.005716$&$-0.145500$&$-0.164339$&$0.367052$  \\\hline 
12	&(BZ)	& $0.185274$&$-0.157883$&$-0.015932$&$0.387345$   \\
	&IRS	& $0.007434$&$-0.119473$&$-0.139186$&$0.284884$   \\\hline 
13	&(BZ)	& $0.142121$&$-0.109950$&$-0.011408$&$0.265692$   \\
	&IRS	& $0.007438$&$-0.094446$&$-0.111933$&$0.210196$  \\\hline 
14	&(BZ)	&$0.099005$&$-0.071251$&$-0.007537$&$0.168103$  \\
	&IRS	& $0.006239$&$-0.069643$&$-0.082647$&$0.142984$   \\\hline 
15	&(BZ)	& $0.056867$&$-0.039180$&$-0.004211$&$0.089115$  \\
	&IRS	& $0.004116$&$-0.043831$&$-0.051081$&$0.081838$ \\\hline 
16	&(BZ)	& $0.017803$&$-0.012188$&$-0.001338$&$0.026215$   \\
	&IRS	& $0.001392$&$-0.015520$&$-0.017328$&$0.025816$ \\\hline 
\end{tabular}
\end{center}
\begin{center}
\caption{Three loop $\MOMg$ scheme $SU(3)$ MAG critical exponents.}
\label{3lmgmagsu3exp}
\end{center}
\end{table}}
 
{\begin{table}
\begin{center}
 \begin{tabular}{ |c||c||c|c|c|c| } 
  \hline 
 $\Nf$ 	& Type 	&$\gamma_A$	&$\gamma_c$	&$\gamma_\psi$ 	&$\rho_m$ \\ 
\hline
  \hline
9	&(BZ)	& $0.295832$&$-0.240449$&$-0.034696$&$0.409975$  \\
	&IRS	& $-2715.990473$&$-15974.919978$&$-7524.837593$&$9467.421651$  \\\hline 
10	&(BZ)	&$0.234670$&$-0.175561$&$-0.025879$&$0.372691$  \\
	&IRS	& $-56.221412$&$-301.361636$&$-170.525780$&$231.933695$  \\ \hline 
11	&(BZ)	& $0.191903$&$-0.135957$&$-0.019136$&$0.320416$   \\
	&IRS	& $-11.249839$&$-57.735060$&$-38.212415$&$54.336768$   \\ \hline 
12	&(BZ)	& $0.155944$&$-0.106659$&$-0.014011$&$0.262295$   \\
	&IRS	& $-3.920648$&$-19.973268$&$-15.156930$&$22.089447$   \\ \hline 
13	&(BZ)	& $0.122059$&$-0.081811$&$-0.009984$&$0.201901$  \\
	&IRS	& $-1.731560$&$-9.032689$&$-7.739827$&$11.403803$   \\ \hline 
14	&(BZ)	& $0.088049$&$-0.058618$&$-0.006705$&$0.141149$ \\
	&IRS	& $0.006603$&$-0.090235$&$-0.113243$&$0.185858$   \\
	&(IRS)	& $-0.849158$&$-4.683999$&$-4.483094$&$6.606901$   \\ \hline 
15	&(BZ)	&$0.053091$&$-0.035578$&$-0.003909$&$0.081733$ \\
	&IRS	& $0.004338$&$-0.046830$&$-0.055569$&$0.087995$  \\
	&(IRS)	& $-0.414965$&$-2.533893$&$-2.698436$&$3.938682$   \\\hline 
16	&(BZ)	& $0.017550$&$-0.011991$&$-0.001317$&$0.025814$  \\
	&IRS	& $0.001402$&$-0.015624$&$-0.017464$&$0.026006$ \\ \hline 
\end{tabular}
\end{center}
\begin{center}
\caption{Three loop $\MOMc$ scheme $SU(3)$ MAG critical exponents.}
\label{3lmcmagsu3exp}
\end{center}
\end{table}}
 
{\begin{table}
\begin{center}
 \begin{tabular}{ |c||c|c|c|c|c|c|c|c| } 
  \hline 
 $\Nf$ 	& Type 	&$\gamma_A$	&$\gamma_c$	&$\gamma_\psi$ 	&$\rho_m$ \\ 
\hline
  \hline 
9	&(BZ)	& $0.297538$&$-0.279798$&$-0.027555$&$0.592685$   \\
	&IRS	&$-0.077144$&$-0.495177$&$-0.299364$&$0.530421$  \\ \hline 
10	&BZ	& $0.250981$&$-0.204323$&$-0.020839$&$0.479055$  \\
	&IRS	&$-0.041489$&$-0.313878$&$-0.239936$&$0.434601$   \\
\hline 
11	&(BZ)	& $0.207877$&$-0.154160$&$-0.016380$&$0.380314$  \\
	&IRS	& $-0.018655$&$-0.209733$&$-0.193922$&$0.351356$   \\ \hline 
12	&(BZ)	&$0.167371$&$-0.117002$&$-0.012761$&$0.293629$  \\
	&IRS	& $-0.004711$&$-0.145483$&$-0.155019$&$0.276501$   \\ \hline 
13	&(BZ)	& $0.128545$&$-0.086897$&$-0.009551$&$0.216209$  \\
	&IRS	& $0.002606$&$-0.102882$&$-0.119598$&$0.207414$  \\ \hline 
14	&(BZ)	& $0.090759$&$-0.060566$&$-0.006621$&$0.146218$  \\
	&IRS	& $0.004944$&$-0.071411$&$-0.085559$&$0.142727$  \\ \hline 
15	&(BZ)	& $0.053726$&$-0.036016$&$-0.003913$&$0.082779$  \\
	&IRS	& $0.003965$&$-0.043956$&$-0.051686$&$0.081985$  \\
\hline 
16	&(BZ)	& $0.017573$&$-0.012008$&$-0.001319$&$0.025849$   \\
	&IRS 	& $0.001391$&$-0.015517$&$-0.017339$&$0.025820$    \\ \hline 
\end{tabular}
\end{center}
\begin{center}
\caption{Three loop $\MOMq$ scheme $SU(3)$ MAG critical exponents.}
\label{3lmqmagsu3exp}
\end{center}
\end{table}}
 
{\begin{table}
\begin{center}
\begin{tabular}{|c||c|c||c|c||c|c|}
\hline 
\rule{0pt}{14pt}
$\Nf$&$a_\infty^{\MOMqs}$&$\omega$&$a_\infty^{\MOMcs}$&$\omega$&$a_\infty^{\MOMgs}$&$\omega$\\ \hline
\hline
8 & $0.074165$ & $0.688880$ & $0.109629$ & $0.889557$& $0.040112$ & $0.596453$ \\
9 & $0.055773$ & $0.480084$ & $0.067722$ & $0.594048$& $0.032173$ & $0.393796$ \\
10& $0.045054$ & $0.322398$ & $0.052237$ & $0.371608$& $0.026493$ & $0.263778$ \\
11& $0.036815$ & $0.209040$ & $0.041634$ & $0.229043$& $0.022025$ & $0.173972$ \\
12& $0.029594$ & $0.127769$ & $0.032764$ & $0.134706$& $0.018232$ & $0.109698$ \\
13& $0.022785$ & $0.070870$ & $0.024643$ & $0.072715$& $0.014765$ & $0.063414$ \\
14& $0.016132$ & $0.033179$ & $0.017001$ & $0.033501$& $0.011320$ & $0.031089$ \\
15& $0.009571$ & $0.010922$ & $0.009835$ & $0.010946$& $0.007536$ & $0.010668$ \\
16& $0.003143$ & $0.001099$ & $0.003165$ & $0.001099$& $0.002882$ & $0.001097$ \\
\hline
\end{tabular}
\end{center}
\begin{center}
\caption{Four loop Landau gauge Banks-Zaks fixed points and $\omega$ in the
$\MOMq$, $\MOMc$ and $\MOMg$ schemes.}
\label{momilan4}
\end{center}
\end{table}}

\begin{table}
\begin{center}
\rotatebox{90}{
\begin{tabular}{ |c||c|c|c||c|c|c||c|c|c| } 
 \hline $\Nf$ & $a_{[3,1]}$ & $\alpha_{[3,1]}$ & $\rho_{m{[3,1]}}$ & $a_{[2,2]}$ & $\alpha_{[2,2]}$ & $\rho_{m{[2,2]}}$ & $a_{[1,3]}$ & $\alpha_{[1,3]}$ & $\rho_{m{[1,3]}}$  \\ 
 \hline 
\hline
$9$ & $0.081552 {\zp} $ & $0.000000$ & $-0.355655$ & $-$ & $-$ & $-$ & $-$ & $-$ & $-$ \\ 
& $0.081552 {\zp} $ & $-0.009802$ & $-0.355655$ & $-$ & $-$ & $-$ & $-$ & $-$ & $-$ \\ 
& $0.081552 {\zp} $ & $-4.117330$ & $-0.355655$ & $-$ & $-$ & $-$ & $-$ & $-$ & $-$ \\ 
\hline 
$10$ & $0.060179 {\zp} $ & $0.000000$ & $0.211071$ & $-$ & $-$ & $-$ & $-$ & $-$ & $-$ \\ 
& $0.060179 {\zp} $ & $-0.708479$ & $0.211071$ & $-$ & $-$ & $-$ & $-$ & $-$ & $-$ \\ 
& $0.060179 {\zp} $ & $-3.428390$ & $0.211071$ & $-$ & $-$ & $-$ & $-$ & $-$ & $-$ \\ 
\hline 
$11$ & $0.044777 {\zp} $ & $0.000000$ & $0.293997$ & $-$ & $-$ & $-$ & $-$ & $-$ & $-$ \\
\hline 
\end{tabular}} 
\caption{Values of $a$ and $\alpha$ in the $\MSbar$ scheme from the Pad\'{e} 
analysis.}
\label{padems} 
\end{center}
\end{table}

\begin{table}
\begin{center}
\rotatebox{90}{
\begin{tabular}{ |c||c|c|c||c|c|c||c|c|c| } 
 \hline $\Nf$ & $a_{[3,1]}$ & $\alpha_{[3,1]}$ & $\rho_{m{[3,1]}}$ & $a_{[2,2]}$ & $\alpha_{[2,2]}$ & $\rho_{m{[2,2]}}$ & $a_{[1,3]}$ & $\alpha_{[1,3]}$ & $\rho_{m{[1,3]}}$  \\ 
 \hline
\hline
$9$ & $0.081552 {\zp} $ & $0.000000$ & $-2.193550$ & $-$ & $-$ & $-$ & $-$ & $-$ & $-$ \\  
\hline 
$10$ & $0.060179 {\zp} $ & $0.000000$ & $-0.635482$ & $-$ & $-$ & $-$ & $-$ & $-$ & $-$ \\ 
\hline 
$11$ & $0.044777 {\zp} $ & $0.000000$ & $0.056006$ & $-$ & $-$ & $-$ & $-$ & $-$ & $-$ \\ 
\hline 
$12$ & $0.032429 {\zp} $ & $0.000000$ & $0.209582$ & $0.050422$ & $0.000000$ & $-0.296439$ & $0.048898$ & $0.000000$ & $-0.209082$ \\ 
& $0.032429 {\zp} $ & $2.511522$ & $0.445578$ & $-$ & $-$ & $-$ & $-$ & $-$ & $-$ \\ 
& $0.032429 {\zp} $ & $-1.557388$ & $0.204125$ & $-$ & $-$ & $-$ & $-$ & $-$ & $-$ \\ 
& $0.032429 {\zp} $ & $-6.482723$ & $0.617028$ & $-$ & $-$ & $-$ & $-$ & $-$ & $-$ \\
\hline 
\end{tabular} } 
\caption{Values of $a$ and $\alpha$ in the $\RI$ scheme from the Pad\'{e} 
analysis.}
\label{paderi} 
\end{center}
\end{table}

\begin{table}
\begin{center}
\rotatebox{90}{
\begin{tabular}{ |c||c|c|c||c|c|c||c|c|c| } 
 \hline $\Nf$ & $a_{[3,1]}$ & $\alpha_{[3,1]}$ & $\rho_{m{[3,1]}}$ & $a_{[2,2]}$ & $\alpha_{[2,2]}$ & $\rho_{m{[2,2]}}$ & $a_{[1,3]}$ & $\alpha_{[1,3]}$ & $\rho_{m{[1,3]}}$  \\ 
 \hline
\hline
$9$ & $0.068777 {\zp} $ & $0.000000$ & $1.555108$ & $0.034658$ & $0.000000$ & $0.459046$ & $0.015650$ & $0.000000$ & $0.156758$ \\
& $0.037028$ & $-3.586442$ & $0.411340$ & $0.021048$ & $-3.613442$ & $0.207376$ & $0.016895$ & $-0.385698$ & $0.172742$ \\
\hline 
$10$ & $0.019275 {\zp} $ & $0.000000$ & $0.199491$ & $0.035297$ & $0.000000$ & $0.438099$ & $0.022986$ & $0.000000$ & $0.248950$ \\
& $0.021361 {\zp} $ & $-1.236002$ & $0.227161$ & $0.027313$ & $-3.724616$ & $0.273199$ & $0.023299$ & $-1.634361$ & $0.251958 $ \\
& $0.035930$ & $-3.625784$ & $0.376564$ & $-$ & $-$ & $-$ & $-$ & $-$ & $-$ \\ 
\hline 
$11$ & $0.026800 {\zp} $ & $0.000000$ & $0.292618$ & $0.033149$ & $0.000000$ & $0.379081$ & $0.027364$ & $0.000000$ & $0.300178$ \\
& $0.115245$ & $1.079306$ & $-5.582485$ & $-$ & $-$ & $-$ & $0.028914$ & $-2.681089$ & $0.307396$ \\
\hline 
\end{tabular} } 
\caption{Values of $a$ and $\alpha$ in the $\mMOM$ scheme from the Pad\'{e} 
analysis.}
\label{pademmom}
\end{center}
\end{table}

\end{document}